\documentclass[aip,amsmath,amssymb,reprint]{revtex4-2}

\usepackage{graphicx}
\usepackage{dcolumn}
\usepackage{bm}

\usepackage[utf8]{inputenc}
\usepackage[T1]{fontenc}
\usepackage{mathptmx}
\usepackage{etoolbox}

\makeatletter
\def\@email#1#2{%
 \endgroup
 \patchcmd{\titleblock@produce}
  {\frontmatter@RRAPformat}
  {\frontmatter@RRAPformat{\produce@RRAP{*#1\href{mailto:#2}{#2}}}\frontmatter@RRAPformat}
  {}{}
}%
\makeatother
\begin{document}

\preprint{AIP/123-QED}

\title{Giant overreflection of magnetohydrodynamic waves from inhomogeneous plasmas with nonuniform shear flows}
\author{Seulong Kim}
\affiliation{Research Institute for Basic Sciences, Ajou University, Suwon 16499, Korea}

\author{Kihong Kim*}%
\email{khkim@ajou.ac.kr}
\affiliation{Department of Physics, Ajou University, Suwon 16499, Korea}%
\affiliation{School of Physics, Korea Institute for Advanced Study, Seoul 02455, Korea}

\date{\today}

\begin{abstract}
We study theoretically mode conversion and resonant overreflection of magnetohydrodynamic waves in an inhomogeneous plane-stratified plasma in
the presence of a nonuniform shear flow, using precise numerical calculations of the reflection and transmission coefficients and the field distributions based on
the invariant imbedding method.
The cases where the flow velocity and the external magnetic field are directed
perpendicularly to the inhomogeneity direction
and both the flow velocity and the plasma density vary arbitrarily along it are considered.
When there is a shear flow, the wave frequency is modulated locally by the Doppler shift and
resonant amplification and overreflection occur where the modulated frequency is negative and its absolute
value matches the local Alfv\'en or slow frequency.
For many different types of the density and flow velocity profiles, we find that, especially when the parameters are such that the incident waves are totally reflected, there arises a giant overreflection where the reflectance is much larger than 10
in a fairly broad range of the incident angles, the frequency,
and the plasma $\beta$ and its maximum attains values larger than $10^5$.
In a finite $\beta$ plasma, both incident fast and slow magnetosonic waves are found to cause strong overreflection
and there appear multiple positions exhibiting both Alfv\'en and slow resonances inside the plasma.
We explain the mechanism of overreflection in terms of the formation of inhomogeneous and open
cavities close to the resonances and the strong enhancement of the wave energy due to
the occurrence of semi-bound states there.
We give discussions of the observational consequences in magnetized terrestrial and solar plasmas.
\end{abstract}

\maketitle

\section{introduction}\label{sec:intro}

Understanding the processes of electromagnetic wave propagation in inhomogeneous complex plasmas plays a central role in various areas of plasma physics and astrophysics \cite{1Keiling2016,2Walker2005,3Roberts2019}. In this paper, we investigate theoretically the propagation of magnetohydrodynamic (MHD) waves in the
presence of a {\it nonuniform} shear flow of plasma.
Mode conversion and the associated resonant absorption of electromagnetic waves
in inhomogeneous plasmas have been studied extensively for many decades \cite{4Swanson1998, 5Forslund1975, 6Mjolhus1990, 7Hinkel1992, 8Kim2005, 9Kim2006, 10Kim2007, 11McDougall2007, 12Lee2008, 13Yu2016}. In the
low-frequency regime where magnetohydrodynamics is applicable,
there can exist three kinds of wave modes in uniform plasmas, namely, shear Alfv\'en waves and fast and slow magnetosonic waves,
among which slow magnetosonic waves are excited only in finite temperature plasmas.
When fast or slow magnetosonic waves with frequencies in the Alfv\'en or slow continuum are incident onto inhomogeneous stratified
MHD plasmas, the wave frequency can match
that of the local Alfv\'en or slow (or cusp) oscillations at some resonance positions,
which causes the localized Alfv\'en or slow resonance and the resonant absorption of the wave energy.
This phenomenon, especially the Alfv\'en resonance that has often been called the field line resonance (FLR),
has been suggested to be very important
for the heating of a plasma and the transport of the wave energy \cite{14Chen1974, 15Southwood1974, 16Leonovich2013, 17Lee2002, 18Lysak2022}.
There have been proposals that the FLR is relevant
for the excitation of ultra-low-frequency (ULF) magnetic pulsations
in the terrestrial magnetosphere and solar coronal heating \cite{19Chen1974,20Nakariakov2016, 21Nakariakov2020, 22VanDoorsselaere2020}.

In the theories of ULF pulsations based on the FLR,
there has been emphasis on two important features, which are
the cavity- and waveguide-like effects of the inhomogeneous boundary region between the magnetosheath
and the magnetosphere and the shear flow
due to the abrupt change of the flow velocity from the magnetosheath into the magnetosphere \cite{23Walker1998, 24Walker2000, 25Mazur2013}.
In many theoretical and observational studies, it has been argued that the excitation and properties of
Pc 3--5 pulsations can be explained from such considerations.

When there is a shear flow of plasma, the FLR can occur in a significantly different manner.
The effective frequency of MHD waves is modulated by the Doppler effect due to plasma flow and
a negative effective frequency can arise when the flow speed is sufficiently high.
In such a case, waves gain energy from the shear flow and the phenomenon of overreflection
with the reflectance larger than 1 occurs \cite{26McKenzie1970, 28Mann1999}.
In plane-stratified plasmas,
resonant amplification and overreflection occur at the planes where the modulated frequency is negative and
its absolute value is equal to the local Alf\'ven or slow frequency.

The overreflection phenomena are expected to take place generally in
magnetized plasmas with structured inhomogeneity and strong flow shear.
The examples in space physics and solar physics include magnetopause and magnetotail regions in planetary magnetospheres and coronal loops, plumes, and magnetic reconnection regions in the solar atmosphere.
Linear MHD waves in the presence of mean flow have been studied previously
in nonuniform cylindrical and planar systems
\cite{48Goossens1992,29Csik1998,30Csik2000,46Andries2000,49Andries2001a,50Andries2001b}.
Cs\'ik et al. have studied theoretically the effects of a stationary mass flow on
resonant absorption and overreflection in
low $\beta$ solar plasmas \cite{29Csik1998,30Csik2000}. They have considered a model where the Alfv\'en velocity varies linearly while
the shear flow velocity has a step-function profile and calculated the absorption coefficient
as a function of flow speed and frequency, when waves are incident
from a high density region with no flow to a low density region with a uniform flow.
The incident angle has been fixed to a single value and only the configurations where the incident wave is totally reflected
have been considered.
Their results have demonstrated that for both fast and slow magnetosonic waves,
overreflection as well as resonant absorption can be obtained depending on the parameters.
The numerically obtained maximum values of the absorption coefficient and the reflectance
have been found to be about 0.7 and 3.5 respectively.
The results have been compared with the parameters for the solar atmosphere.

The influence of shear flows on the generation and propagation of MHD waves has also been
studied extensively in the context of the Kelvin-Helmholtz instability. There have been many studies
suggesting that this instability plays an important role in the generation of ULF pulsations
in the planetary magnetospheres. In the presence of plasma resonances, the Kelvin-Helmholtz instability can be combined with the FLR
to produce a phenomenon termed resonant flow instability \cite{46Andries2000,49Andries2001a,47Turkakin2013,delamere,45Kim2022}. Though our method can be generalized to such situations,
we restrict our interest to the case where the waves propagate in a stationary manner
in the presence of a steady flow.

In this paper, we generalize the study of Cs\'ik et al. to the case where both the equilibrium flow velocity
and the plasma density vary arbitrarily and continuously along the inhomogeneity direction.
The energy that causes the overreflection comes from the nonuniform equilibrium flow.
Based on the invariant imbedding method (IIM) \cite{31Bellman1976, 32Klyatskin2005, 33GOLBERG19751, 8Kim2005, 9Kim2006, 34Kim2016}, we develop an efficient numerical method to solve the MHD wave equation
and calculate the reflection and transmission coefficients and the field distributions in a numerically precise manner
for an {\it entire range of the incident angles}.
For many different types of the density and flow velocity profiles, we find that, especially when the parameters are such that the incident waves are totally reflected, giant overreflection where the reflectance is much larger than 10
in a fairly broad range of the incident angles, the frequency,
and the plasma $\beta$ occurs and its maximum attains values larger than $10^5$.
We find that in a finite $\beta$ plasma, both fast and slow magnetosonic waves cause strong overreflection
and there appear multiple positions exhibiting both Alfv\'en and slow resonances inside the plasma.
Our results also include the overreflection phenomena occurring when the incident wave propagates in the transmitted region with negative energy \cite{27Cairns1979,37Ostrovskii1986, 38Joarder1997}.
We explain the mechanism of giant overreflection in terms of the formation of inhomogeneous and open
cavities close to the resonances and the strong enhancement of the wave energy due to the formation
of semi-bound states there. We also give some discussions on the possible consequences of our theory
in magnetized terrestrial and solar plasmas.
Giant overreflection with a similar mechanism has been previously studied in the context of
amplification of circularly-polarized electromagnetic waves in chiral optical media with gain \cite{35Kim2016}.

It is worth mentioning the merits of the IIM used in the present work.
Using this method, it is possible to solve the MHD wave equation in the general cases where the external
magnetic field as well as the plasma density and the flow velocity vary arbitrarily along the inhomogeneity direction.
Although the wave equation becomes singular in the presence of resonances,
we are able to solve it in a numerically exact manner
without using any approximation such as the WKB approximation.
We propose that the IIM can be a very useful tool in the study
of wave propagation problems in a variety of laboratory and astrophysical plasmas.

The rest of this paper is organized as follows. In Sec.~\ref{sec_we}, we develop the basic MHD theory and
derive the wave equations used in this study. In Sec.~\ref{sec_iim} and Appendix, we discuss the IIM and derive the invariant imbedding equations. A simplified version of the theory in the case of zero $\beta$ plasmas is given in Sec.~\ref{sec_c}. In Sec.~\ref{sec_num},
we present extensive numerical results obtained by using the IIM for various configurations of the plasma density
and the flow velocity.  In Sec.~\ref{sec_or}, we provide an explanation
of the mechanism of giant overreflection. Some discussions on the observational consequences
of the theory are given in Sec.~\ref{sec_dc}. Finally, we conclude the paper in Sec.~\ref{sec_con}.

\section{Wave equation}
\label{sec_we}

We start with the standard set of ideal MHD equations
\begin{eqnarray}
&&\frac{\partial\rho}{\partial t}+\nabla\cdot\left(\rho{\bf v}\right)=0,\nonumber\\
&&\frac{\partial {\bf v}}{\partial t}+\left({\bf v}\cdot\nabla\right) {\bf v}
=-\frac{\nabla p}{\rho}+\frac{1}{\mu_0 \rho}\left(\nabla\times{\bf B}\right)\times{\bf B},\nonumber\\
&&\frac{\partial p}{\partial t}+\left({\bf v}\cdot\nabla\right) p=\frac{\gamma p}{\rho}\left[\frac{\partial \rho}{\partial t}+\left({\bf v}\cdot\nabla\right) \rho\right],\nonumber\\
&&\frac{\partial {\bf B}}{\partial t}=\nabla\times\left({\bf v}\times{\bf B}\right),\nonumber\\
&&\nabla\cdot {\bf B}=0,
\label{eq:bwe}
\end{eqnarray}
where $\bf v$, $\bf B$, $\rho$, and $p$ are the fluid velocity, the magnetic field,
the mass density, and the pressure respectively. $\gamma$ is the ratio of the specific heats
and $\mu_0$ is the vacuum magnetic permeability. We have ignored the gravitational field here.
We consider the situation where a nonuniform plasma layer is placed in the region $0\le z\le L$ and is surrounded by
two semi-infinite uniform plasma layers. The nonuniform plasma is stratified
and its medium parameters are assumed to depend only on the coordinate $z$.
Plane waves are assumed to be incident from the region $z>L$
and transmitted to the region $z<0$.

We linearize the MHD equations by substituting $\rho=\rho_0+\rho_1$, ${\bf v}={\bf v}_0+{\bf v}_1$,
$p=p_0+p_1$, and ${\bf B}={\bf B}_0+{\bf B}_1$, where $\rho_0$, ${\bf v}_0$, $p_0$, and ${\bf B}_0$ are
the zeroth-order background quantities.
Inside the nonuniform layer, these quantities generally depend on $z$ and are time-independent.
Then the zeroth-order equation of continuity takes the form
\begin{eqnarray}
\frac{d}{dz}\left(\rho_0 v_{0z}\right)=0,
\end{eqnarray}
where $v_{0z}$ is the $z$ component of ${\bf v}_0$. Barring the improbable situation where $\rho_0(z)$ is strictly proportional to
${[v_{0z}(z)]}^{-1}$,
this equation can be satisfied only when $v_{0z}=0$.
That is, there is no equilibrium flow in the direction of inhomogeneity.
Furthermore, when $v_{0z}=0$, the zeroth-order forms of Faraday's equation and $\nabla\cdot {\bf B}=0$ lead to
\begin{eqnarray}
\frac{d}{dz}\left(v_{0x}B_{0z}\right)=\frac{d}{dz}\left(v_{0y}B_{0z}\right)=\frac{d B_{0z}}{dz}=0.
\end{eqnarray}
Since the flow velocity ${\bf v}_0$ is not uniform in the whole space, we obtain the condition $B_{0z}=0$.
We note that if ${\bf v}_0$ were uniform in the whole space, we could have chosen a reference frame moving with the same velocity and removed ${\bf v}_0$ from the equations.
From these considerations, we assume, with no loss of generality, that
${\bf B}_0=B_0(z) \hat{\bf x}$ and ${\bf v}_0=U_x(z) \hat{\bf x}+U_y(z) \hat{\bf y}$.
The $z$ dependence of $U_x$ and $U_y$ implies that there is a nonuniform shear flow in the direction perpendicular to that of inhomogeneity.

From the zeroth-order momentum equation, we obtain
\begin{eqnarray}
\frac{d}{dz}\left(p_0+\frac{{B_0}^2}{2\mu_0}\right)=0,
\label{eq:tp}
\end{eqnarray}
where the expression $p_0+{B_0}^2/(2\mu_0)$ is the equilibrium total pressure which is a constant.
If the external magnetic field is uniform in the whole space, the equilibrium pressure $p_0$ is also uniform.

We linearize Eq.~(\ref{eq:bwe}) to first order in the perturbed quantities.
In the uniform regions, we can Fourier analyze the resulting equations by making the substitutions
$\nabla\rightarrow i{\bf k}$ and $\partial/\partial t\rightarrow -i\omega$.
We assume that the wave vector is in an arbitrary direction and write its components as
\begin{eqnarray}
k_x=k\sin\theta\cos\phi,~
k_y=k\sin\theta\sin\phi,~
k_z=k\cos\theta,
\end{eqnarray}
where $\theta$ ($0\le \theta<\pi/2$) is defined as the angle between $\bf k$ and the {\it negative} $z$ direction
and $k_z$ is the {\it negative} $z$ component of the wave vector.
The angle $\phi$ is the azimuthal angle of the wave vector ($0\le \phi<2\pi$).
Following the standard procedure for deriving the dispersion relations for MHD waves, we obtain
\begin{eqnarray}
&&\frac{\Omega^2}{k^2}=\frac{1}{2}\left({v_A}^2+{v_s}^2\right)\nonumber\\
&&~+\frac{1}{2}\left[\left({v_A}^2+{v_s}^2\right)^2-4{v_A}^2{v_s}^2\sin^2\theta\cos^2\phi\right]^{1/2},\label{eq:fw}\\
&&\frac{\Omega^2}{k^2}=\frac{1}{2}\left({v_A}^2+{v_s}^2\right)\nonumber\\
&&~-\frac{1}{2}\left[\left({v_A}^2+{v_s}^2\right)^2-4{v_A}^2{v_s}^2\sin^2\theta\cos^2\phi\right]^{1/2},\label{eq:sw}\\
&&\frac{\Omega^2}{k^2}={v_A}^2\sin^2\theta\cos^2\phi, \label{eq:aw}
\end{eqnarray}
where the Alfv\'en velocity $v_A$ and the speed of sound $v_s$ are defined by
\begin{equation}
v_A=\frac{B_0}{\sqrt{\mu_0\rho_0}},~~v_s=\sqrt{\frac{\gamma p_0}{\rho_0}}
\end{equation}
and $\Omega$ is the wave frequency shifted by the Doppler effect in the presence of a mass flow and is given by
\begin{equation}
\Omega=\omega-U_xk_x-U_yk_y.
\end{equation}
We note that the quantity $\sin\theta\cos\phi$ is equal to the cosine of the angle between ${\bf k}$ and ${\bf B}_0$.
The first two dispersion relations, Eqs.~(\ref{eq:fw}) and (\ref{eq:sw}), correspond to the fast and slow magnetosonic modes respectively, while
the third dispersion relation, Eq.~(\ref{eq:aw}), corresponds to the transverse Alfv\'en mode. These dispersion relations differ from the usual ones
only in that the frequency $\omega$ is replaced by $\Omega$.

In the nonuniform region where the medium parameters depend on $z$, we substitute
\begin{eqnarray}
&&\rho_1={\tilde \rho}\left(z\right)e^{i\left(k_xx+k_yy-\omega t\right)},\nonumber\\
&&{\bf v}_1={\tilde {\bf v}}\left(z\right)e^{i\left(k_xx+k_yy-\omega t\right)},\nonumber\\
&&p_1={\tilde p}\left(z\right)e^{i\left(k_xx+k_yy-\omega t\right)},\nonumber\\
&&{\bf B}_1={\tilde {\bf B}}\left(z\right)e^{i\left(k_xx+k_yy-\omega t\right)}
\end{eqnarray}
into the linearized equations and derive the ordinary differential equations satisfied by the variables $\tilde \rho$, $\tilde {\bf v}$,
$\tilde p$, and $\tilde {\bf B}$.
It turns out that we can combine these equations into two first-order differential equations of the form
\begin{eqnarray}
\frac{dP}{dz}=A\xi_z,~~\frac{d\xi_z}{dz}=-\frac{C}{D}P,
\label{eq:fode}
\end{eqnarray}
where
\begin{eqnarray}
&&A=\rho_0\left(\Omega^2-{\omega_A}^2\right),\nonumber\\
&&C=\left(\Omega^2-{\omega_A}^2\right)\left(\Omega^2-{\omega_s}^2\right)-\Omega^2{v_A}^2{k_y}^2,\nonumber\\
&&D=\rho_0\left({v_A}^2+{v_s}^2\right)\left(\Omega^2-{\omega_A}^2\right)\left(\Omega^2-{\omega_c}^2\right).
\label{eq:abc}
\end{eqnarray}
The parameters $\omega_A$, $\omega_s$, and $\omega_c$ are defined by
\begin{eqnarray}
&&{\omega_A}^2={k_x}^2{v_A}^2,~~{\omega_s}^2=\left({k_x}^2+{k_y}^2\right){v_s}^2,\nonumber\\
&&{\omega_c}^2={k_x}^2{v_c}^2,
\label{eq:omdef}
\end{eqnarray}
where the cusp velocity $v_c$ is given by
\begin{eqnarray}
{v_c}=\frac{{v_A}{v_s}}{\sqrt{{v_A}^2+{v_s}^2}}.
\end{eqnarray}
Note that the cusp frequency $\omega_c$ is unrelated to the cyclotron frequency.
The variables $P$ and $\xi_z$ are defined by
\begin{eqnarray}
P={\tilde p}+\frac{B_0}{\mu_0}{\tilde B}_x,~~\xi_z=i\frac{{\tilde v}_z}{\Omega}.
\label{eq:ppp}
\end{eqnarray}
$P$ corresponds to the first-order perturbation of the total pressure $p+B^2/(2\mu_0)$ and $\xi_z$ is the $z$ component of the Lagrangian displacement vector.
From Eq.~(\ref{eq:fode}), we can derive the second-order differential equations satisfied by $P$ and $\xi_z$:
\begin{eqnarray}
&&\frac{d}{dz}\left(\frac{1}{A}\frac{dP}{dz}\right)+\frac{C}{D}P=0,\label{eq:wee1}\\
&&\frac{d}{dz}\left(\frac{D}{C}\frac{d\xi_z}{dz}\right)+A\xi_z=0.\label{eq:wee2}
\end{eqnarray}
We can solve either of these equations to obtain $P(z)$ or $\xi_z(z)$,
and then calculate all other variables $\tilde \rho$, $\tilde {\bf v}$,
$\tilde p$, $\tilde {\bf B}$, and the perturbed electric field $\tilde{\bf E}$ using simple relationships among the variables.
For instance, we can express ${\tilde E}_y$ in terms of $\xi_z$ by
\begin{eqnarray}
{\tilde E}_y=i\left(\omega-U_yk_y\right)B_0\xi_z.
\end{eqnarray}
In the case where the flow velocity is parallel to ${\bf B}_0$ (that is, when $U_y=0$), equations equivalent to Eq.~(\ref{eq:fode}) were derived in Ref.~\onlinecite{36Cadez1997}.
The formalism and the equations similar to those given in this
section have also been developed and analyzed in many previous papers \cite{48Goossens1992,29Csik1998,30Csik2000,46Andries2000,49Andries2001a,50Andries2001b}.

From the form of the wave equation, Eq.~(\ref{eq:wee1}), we notice that it becomes {\it singular}
at the positions $z=z_A$ and $z=z_c$ satisfying
\begin{eqnarray}
\Omega\left(z_A\right)=\pm \omega_A\left(z_A\right),~~
\Omega\left(z_c\right)=\pm \omega_c\left(z_c\right).
\label{eq:resa}
\end{eqnarray}
The resonances at $z_A$ and $z_c$ are called the Alfv\'en resonance and the slow (or cusp) resonance respectively.
We observe that the shifted frequency $\Omega$ can be either positive or negative. When it is positive, we have
\begin{eqnarray}
&&\omega=U_xk_x+U_yk_y+\omega_A,\nonumber\\&& \omega=U_xk_x+U_yk_y+\omega_c,
\label{eq:resb}
\end{eqnarray}
which implies that the energy of the incident wave is converted to that of the local Alfv\'en mode or slow MHD mode
and the resonant wave absorption occurs.
In contrast, when $\Omega$ is negative, we have
\begin{eqnarray}
&&U_xk_x+U_yk_y=\omega+\omega_A,\nonumber\\&& U_xk_x+U_yk_y=\omega+\omega_c,
\label{eq:resc}
\end{eqnarray}
which implies that the flow supplies energy to the wave
and the resonant amplification occurs. In this case, the reflectance becomes greater than one, resulting in the phenomenon of  overreflection.

The parameter $C$ in Eq.~(\ref{eq:abc}) can be factorized as
\begin{eqnarray}
C=\left(\Omega^2-{\omega_{\rm I}}^2\right)\left(\Omega^2-{\omega_{\rm II}}^2\right),
\end{eqnarray}
where
\begin{eqnarray}
&&{\omega_{\rm I}}^2=\frac{1}{2}\left({v_A}^2+{v_s}^2\right)\left({k_x}^2+{k_y}^2\right)\nonumber\\
&&~~\times\left\{1-\left[1-\frac{4{\omega_c}^2}{\left({v_A}^2+{v_s}^2\right)\left({k_x}^2+{k_y}^2\right)}\right]^{1/2}\right\},\nonumber\\
&&{\omega_{\rm II}}^2=\frac{1}{2}\left({v_A}^2+{v_s}^2\right)\left({k_x}^2+{k_y}^2\right)\nonumber\\
&&~~\times\left\{1+\left[1-\frac{4{\omega_c}^2}{\left({v_A}^2+{v_s}^2\right)\left({k_x}^2+{k_y}^2\right)}\right]^{1/2}\right\}.
\end{eqnarray}
In the uniform regions, we can derive
\begin{eqnarray}
{k_z}^2=\frac{AC}{D}=\frac{\left(\Omega^2-{\omega_{\rm I}}^2\right)\left(\Omega^2-{\omega_{\rm II}}^2\right)}
{\left({v_A}^2+{v_s}^2\right)\left(\Omega^2-{\omega_c}^2\right)}
\label{eq:kz}
\end{eqnarray}
from Eq.~(\ref{eq:wee1}) or (\ref{eq:wee2}).
Using the condition $k_z=0$ in Eqs.~(\ref{eq:fw}) and (\ref{eq:sw}), we can show that $\Omega^2={\omega_{\rm I}}^2$ and $\Omega^2={\omega_{\rm II}}^2$ correspond to
the cutoff conditions for the slow and fast magnetosonic waves in uniform media respectively.
It is straightforward to prove the inequalities
\begin{eqnarray}
{\omega_c} \le {\omega_{\rm I}} \le {\omega_A} \le {\omega_{\rm II}}.
\label{eq:ineq}
\end{eqnarray}
When $\Omega^2>{\omega_{\rm II}}^2$, fast magnetosonic waves can propagate in the medium, while, when ${\omega_c}^2<\Omega^2<{\omega_{\rm I}}^2$,
slow magnetosonic waves can.
Waves become evanescent in the frequency region where $\Omega^2<{\omega_c}^2$ or ${\omega_{\rm I}}^2<\Omega^2<{\omega_{\rm II}}^2$, which includes
$\Omega^2={\omega_A}^2$.

By taking the derivative of Eq.~(\ref{eq:kz}) with respect to $\Omega$, we can derive the relationship between the $z$ component of the phase velocity, $V_{pz}$ ($=\Omega/k_z$), and that of the group
velocity, $V_{gz}$ ($=\partial\Omega/\partial k_z$), for magnetosonic waves in uniform media of the form
\begin{equation}
V_{pz}V_{gz}=\frac{\left({v_A}^2+{v_s}^2\right)\left(\Omega^2-{\omega_c}^2\right)^2}{\left(\Omega^2-{\omega_c}^2\right)^2-\left({\omega_{\rm I}}^2-{\omega_c}^2\right)\left({\omega_{\rm II}}^2-{\omega_c}^2\right)}.
\end{equation}
We easily verify that for fast magnetosonic waves with $\Omega^2>{\omega_{\rm II}}^2$, this expression is positive and therefore $V_{pz}$ and $V_{gz}$ have the same signs.
On the other hand, for slow magnetosonic waves with ${\omega_c}^2<\Omega^2<{\omega_{\rm I}}^2$, $V_{pz}V_{gz}$ is negative and $V_{pz}$ and $V_{gz}$ have the opposite signs.

\section{Invariant imbedding method}
\label{sec_iim}

Using the IIM, we can accurately and efficiently solve the wave equation, Eq.~(\ref{eq:wee1}), in the cases where the plasma density $\rho_0$, the flow velocity components $U_x$ and $U_y$,
and the external magnetic field $B_0$ depend
on $z$ in an {\it arbitrary} manner in the region $0\le z\le L$. We point out that, in general, the quantities $v_A$, $v_s$, $v_c$,
$\omega_A$, $\omega_s$, $\omega_c$, $\Omega$, $A$, $C$, and $D$ also depend on $z$.
Alternatively, we may choose to solve Eq.~(\ref{eq:wee2}). We have verified explicitly that the final numerical results are the same regardless of the choice of wave equation.

In the IIM, we
consider plane waves
launched towards the nonuniform region with a given frequency.
We are interested in calculating the reflection and transmission coefficients $r$ and $t$ defined
by the wave functions in the incident and transmitted regions:
\begin{eqnarray}
P\left(z;L\right)=\left\{\begin{array}{ll}
  e^{ik_{z1}\left(L-z\right)}+r(L)e^{ik_{z1}\left(z-L\right)}, & z>L \\
  t(L)e^{-ik_{z2} z}, & z<0
  \end{array},\right.
  \label{eq:rt}
\end{eqnarray}
where $k_{z1}$ and $k_{z2}$ are the negative $z$ components of the wave vector in the incident and transmitted regions respectively
and $r$ and $t$ are regarded as functions of $L$. The wave function $P(z;L)$ represents the value of $P$
at the position $z$ when the thickness of the inhomogeneous medium is $L$.
We assume that the waves are not evanescent in the incident region and therefore $k_{z1}$ is a real number. Since $V_{pz}$ and $V_{gz}$ have the same (opposite) signs for fast (slow) waves,
we have to choose $k_{z1}$ to be positive (negative) for fast (slow) waves. We obtain
\begin{eqnarray}
k_{z1}=\left\{\begin{array}{ll} \sqrt{\frac{A_1C_1}{D_1}} &~~\mbox{for fast waves } \\
-\sqrt{\frac{A_1C_1}{D_1}} &~~\mbox{for slow waves } \end{array}\right.,
\end{eqnarray}
where $A_1$, $C_1$, and $D_1$ are the values of $A$, $C$, and $D$ in the incident region and $A_1C_1/D_1$ is assumed to be positive.
In the transmitted region where $z<0$, the waves can be either propagative or evanescent depending on the sign of $A_2C_2/D_2$,
where $A_2$, $C_2$, and $D_2$ are the values of $A$, $C$, and $D$ in the transmitted region.
We define $k_{z2}$ by
\begin{eqnarray}
k_{z2}=\left\{\begin{array}{ll} \sqrt{\frac{A_2C_2}{D_2}} &~\mbox{for fast waves, $\frac{A_2C_2}{D_2}>0$, $\Omega_2>0$} \\
-\sqrt{\frac{A_2C_2}{D_2}} &~\mbox{for fast waves, $\frac{A_2C_2}{D_2}>0$, $\Omega_2<0$} \\
-\sqrt{\frac{A_2C_2}{D_2}} &~\mbox{for slow waves, $\frac{A_2C_2}{D_2}>0$, $\Omega_2>0$} \\
\sqrt{\frac{A_2C_2}{D_2}} &~\mbox{for slow waves, $\frac{A_2C_2}{D_2}>0$, $\Omega_2<0$} \\
i\sqrt{\big\vert\frac{A_2C_2}{D_2}\big\vert} &~\mbox{for $\frac{A_2C_2}{D_2}\le 0$} \end{array}\right.,\nonumber\\
\end{eqnarray}
where $\Omega_2$ is the value of $\Omega$ in the transmitted region. The choices of sign for the propagative cases are made to make sure that the
group velocity in the region $z<0$ is always in the negative $z$ direction.

Following the procedure given in Ref.~\onlinecite{34Kim2016},
we derive the invariant imbedding equations satisfied by $r$ and $t$:
\begin{eqnarray}
\frac{dr}{dl}&=&2ik_{z1}\frac{A}{A_1}r+\frac{i}{2k_{z1}}\left(\frac{CA_1}{D}-\frac{A}{A_1}{k_{z1}}^2\right)\left(1+r\right)^2,\nonumber\\
\frac{dt}{dl}&=&ik_{z1}\frac{A}{A_1}t+\frac{i}{2k_{z1}}\left(\frac{CA_1}{D}-\frac{A}{A_1}{k_{z1}}^2\right)\left(1+r\right)t.
\label{eq:mie}
\end{eqnarray}
The details of the IIM and the derivation of Eq.~(\ref{eq:mie}) are given in Appendix.
We notice that the parameter $D$ appears in the denominator and the singularity at the resonances corresponding to $D=0$ is clearly exhibited in the invariant imbedding equations.
The resonance conditions for the incident fast and slow waves given by Eqs.~(\ref{eq:resb}) and (\ref{eq:resc})
can be expressed more explicitly as
\begin{widetext}
\begin{eqnarray}
&&f_1(z)=F-\frac{U_x(z)}{v_{A1}}\sin\theta\cos\phi-\frac{U_y(z)}{v_{A1}}\sin\theta\sin\phi-\sqrt{\frac{\rho_{01}}{\rho_0(z)}}\left\vert\sin\theta\cos\phi\right\vert=0,\nonumber\\
&&f_2(z)=F-\frac{U_x(z)}{v_{A1}}\sin\theta\cos\phi-\frac{U_y(z)}{v_{A1}}\sin\theta\sin\phi+\sqrt{\frac{\rho_{01}}{\rho_0(z)}}\left\vert\sin\theta\cos\phi\right\vert=0,\nonumber\\
&&f_3(z)=F-\frac{U_x(z)}{v_{A1}}\sin\theta\cos\phi-\frac{U_y(z)}{v_{A1}}\sin\theta\sin\phi
-\sqrt{\frac{\gamma\beta(z)}{2+\gamma\beta(z)}}\sqrt{\frac{\rho_{01}}{\rho_0(z)}}\left\vert\sin\theta\cos\phi\right\vert=0,\nonumber\\
&&f_4(z)=F-\frac{U_x(z)}{v_{A1}}\sin\theta\cos\phi-\frac{U_y(z)}{v_{A1}}\sin\theta\sin\phi
+\sqrt{\frac{\gamma\beta(z)}{2+\gamma\beta(z)}}\sqrt{\frac{\rho_{01}}{\rho_0(z)}}\left\vert\sin\theta\cos\phi\right\vert=0,
\label{eq:resw}
\end{eqnarray}
\end{widetext}
where we have used the definitions of $\omega_A$ and and $\omega_c$ given in Eq.~(\ref{eq:omdef}) and the relationship $v_s=\sqrt{\gamma\beta/2}~v_A$. When the flow velocity in the incident region is zero, the parameter $F$ [$\equiv \omega/(kv_{A1})$]
is expressed as
\begin{widetext}
\begin{eqnarray}
F=\left\{\begin{array}{ll} \frac{1}{2}\sqrt{2+\gamma\beta_1+\sqrt{\left(2+\gamma\beta_1\right)^2-8\gamma\beta_1\sin^2\theta\cos^2\phi}} &~~\mbox{for fast-wave incidence } \\
\frac{1}{2}\sqrt{2+\gamma\beta_1-\sqrt{\left(2+\gamma\beta_1\right)^2-8\gamma\beta_1\sin^2\theta\cos^2\phi}} &~~\mbox{for slow-wave incidence } \end{array}\right.,
\end{eqnarray}
\end{widetext}
using Eqs.~(\ref{eq:fw}) and (\ref{eq:sw}).
The plasma $\beta$ parameter is defined by
\begin{eqnarray}
\beta(z)=\frac{2\mu_0 p_0(z)}{\left[{B_0}(z)\right]^2}
\end{eqnarray}
and is generally a function of $z$.
The parameters $\rho_{01}$, $v_{A1}$, and $\beta_1$ are the mass density, the Alfv\'en velocity, and the plasma $\beta$ in the incident region respectively.
The condition $f_1=0$ ($f_2=0$) corresponds to the resonant absorption (amplification) due to the Alfv\'en resonance with $\Omega=\omega_A$ ($\Omega=-\omega_A$),
whereas $f_3=0$ ($f_4=0$) does to the resonant absorption (amplification) due to the slow resonance with $\Omega=\omega_c$ ($\Omega=-\omega_c$).

We calculate $r(L)$ and $t(L)$ by integrating Eq.~(\ref{eq:mie}) numerically from $l=0$ to $l=L$ using
the initial conditions
\begin{equation}
r(0)=\frac{k_{z1}A_2-k_{z2} A_1}{k_{z1}A_2+k_{z2} A_1},~~t(0)=\frac{2k_{z1}A_2}{k_{z1}A_2+k_{z2} A_1}.
\end{equation}
The reflectance $R$ and the transmittance $T$ are obtained using
\begin{eqnarray}
R=\vert r\vert^2,~~T=\left\{\begin{matrix} \frac{k_{z2} A_1}{k_{z1}A_2}\vert t\vert^2 &~~\mbox{if $\frac{A_2C_2}{D_2}>0$} \\
0 &~~\mbox{if $\frac{A_2C_2}{D_2}\le 0$} \end{matrix}\right..
\end{eqnarray}
As we have mentioned already, we can assume that the flow velocity is zero in the incident region.
Then $\Omega$ ($=\omega$) is positive in that region.
If $\Omega_2$ is also positive in the propagative case, $k_{zi}$ ($i=1,2$) is positive for fast waves and negative for slow waves, but at the same time $A_i$ has the same sign as $k_{zi}$ due to the inequalities in Eq.~(\ref{eq:ineq}). Therefore the transmittance $T$ is positive. In contrast, when $\Omega_2$ is negative, $k_{z2}$ and $A_2$ have the opposite signs for both fast and slow waves and {\it $T$ becomes negative}.
This latter case corresponding to the negative transmittance can be considered as an example of a wave with a  negative energy and is caused by the Doppler shift of $\Omega$
to a negative value due to a fast flow of the plasma \cite{27Cairns1979,37Ostrovskii1986, 38Joarder1997}. In such a case, the signs of the frequencies of the incident and transmitted waves are opposite to each other and overreflection with $R>1$ can occur due to the energy exchange between positive and negative energy waves.
However, the giant overreflection phenomenon which is the main focus of this paper is caused primarily not by negative energy waves, but by the mode conversion between the incident wave and resonant local oscillations when the transmitted wave is evanescent
and the transmittance is zero.

In realistic plasmas, there always exists some amount of dissipation due to collisions. In the simplest approximation,
the effects of dissipation can be incorporated into the formalism by replacing $\omega$
with $\omega+i\nu$, where $\nu$ ($>0$) is the collision frequency. The inclusion of the imaginary part of the frequency removes the singularities at the resonances
and allows us to solve Eq.~(\ref{eq:mie}) numerically for any spatial configurations of $\rho_0$, $U_x$, $U_y$, and $B_0$. In general, $\nu$ may depend on various parameters including frequency and also on the spatial
coordinates. Since the resonant absorption or amplification due to mode conversion occurs even in the limit where $\nu\rightarrow 0$, we will choose a sufficiently
small value of $\nu$ in most of our numerical calculations. In such cases, $\nu$ is an artificial damping parameter
introduced to account for the mode conversion and
all the results will be independent of the numerical value of $\nu$ used in the calculations.

The IIM can also be used in calculating the wave function $P=P(z;L)$
inside the inhomogeneous region.
The equation satisfied by $P(z;L)$ is very similar to that for $t$ and takes the form
\begin{eqnarray}
&&\frac{\partial}{\partial l}P(z;l)=ik_{z1}\frac{A(l)}{A_1}P(z;l)\nonumber\\
&&+\frac{i}{2k_{z1}}\left[\frac{C(l)A_1}{D(l)}-\frac{A(l)}{A_1}{k_{z1}}^2\right]\left[1+r(l)\right]P(z;l).
\label{eq:ief}
\end{eqnarray}
This equation is integrated from $l=z$ to $l=L$ using the initial condition $P(z;z)=1+r(z)$
to obtain $P(z;L)$.

\section{Zero $\beta$ limit}
\label{sec_c}

In a zero $\beta$ plasma where the temperature is sufficiently low, we can set $p_0=v_s=v_c=0$ and ignore the pressure field $\tilde p$.
Then the slow magnetosonic mode and the slow resonance are absent and only the fast magnetosonic and transverse Alfv\'en modes and the Alfv\'en resonance remain to occur.
We restrict our interest to the special case where $B_0$ is uniform in the whole space and $U_y=0$ while $U_x\ne 0$. We also assume that $U_x$ is zero in the incident region. Then we can set the pressure field $\tilde p$ to zero in
Eq.~(\ref{eq:ppp}) and rewrite
the wave equation, Eq.~(\ref{eq:wee1}), in a simplified form as
\begin{eqnarray}
\frac{d}{dz}\left(\frac{1}{\epsilon}\frac{d\tilde B_x}{dz}\right)+\left({k_0}^2-\frac{{k_y}^2}{\epsilon}\right)\tilde B_x=0,
\label{eq:weq}
\end{eqnarray}
where
\begin{eqnarray}
&&\epsilon =\frac{\rho_0(z)}{\rho_{01}}\left[1-\frac{U_x(z)}{v_{A1}}\sin\theta\cos\phi\right]^2-\sin^2\theta\cos^2\phi,\nonumber\\
&&k_0=\frac{\omega}{v_{A1}},~~k_y=k_0\sin\theta\sin\phi.
\label{eq:epsp}
\end{eqnarray}
The Alfv\'en resonance occurs at the positions satisfying $\epsilon=0$. The resonance condition can be expressed as
\begin{eqnarray}
&&f_{c1}= 1-\frac{U_x(z)}{v_{A1}}\sin\theta\cos\phi-\sqrt{\frac{\rho_{01}}{\rho_0(z)}}\big\vert\sin\theta\cos\phi\big\vert=0,\nonumber\\
&&f_{c2}= 1-\frac{U_x(z)}{v_{A1}}\sin\theta\cos\phi+\sqrt{\frac{\rho_{01}}{\rho_0(z)}}\big\vert\sin\theta\cos\phi\big\vert=0.\nonumber\\
\label{eq:rescond}
\end{eqnarray}
The singularity at the Alfv\'en resonance can be taken care
of by replacing $\epsilon$ in Eq.~(\ref{eq:epsp}) with
\begin{eqnarray}
&&\epsilon =\frac{\rho_0(z)}{\rho_{01}}\left[1+i\eta-\frac{U_x(z)}{v_{A1}}\sin\theta\cos\phi\right]^2-\sin^2\theta\cos^2\phi,\nonumber\\
&&\eta=\frac{\nu}{\omega},
\label{eq:epsc}
\end{eqnarray}
in the inhomogeneous region. The damping parameter $\eta$ is chosen to be a sufficiently small positive number.
Even though $\eta$ is always positive, the imaginary part of $\epsilon$ becomes negative if $(U_x/v_{A1})\sin\theta\cos\phi>1$, which leads to wave amplification and overreflection.

The invariant imbedding equations for $r$ and $t$ take the simplified forms
\begin{eqnarray}
&&\frac{dr}{dl}=2ik_{z1}\frac{\epsilon}{\epsilon_1}r \nonumber\\ &&~~~~+\frac{i}{2k_{z1}}\left({k_0}^2\epsilon_1-\frac{\epsilon_1}{\epsilon}{k_y}^2-\frac{\epsilon}{\epsilon_1}{k_{z1}}^2\right)(1+r)^2,\nonumber\\
&&\frac{dt}{dl}=ik_{z1}\frac{\epsilon}{\epsilon_1}t \nonumber\\ &&~~~~+\frac{i}{2k_{z1}}\left({k_0}^2\epsilon_1-\frac{\epsilon_1}{\epsilon}{k_y}^2-\frac{\epsilon}{\epsilon_1}{k_{z1}}^2\right)t(1+r),
\label{eq:cpie}
\end{eqnarray}
where
\begin{eqnarray}
\epsilon_1=1-\sin^2\theta\cos^2\phi,~~k_{z1}=k_0\cos\theta.
\end{eqnarray}
The initial conditions are
\begin{eqnarray}
r(0)=\frac{k_{z1}\epsilon_2-k_{z2} \epsilon_1}{k_{z1}\epsilon_2+k_{z2} \epsilon_1},~~t(0)=\frac{2k_{z1}\epsilon_2}{k_{z1}\epsilon_2+k_{z2} \epsilon_1},
\label{eq:maic}
\end{eqnarray}
where
\begin{eqnarray}
\epsilon_2 =\frac{\rho_{02}}{\rho_{01}}\left[1-\frac{U_{x2}}{v_{A1}}\sin\theta\cos\phi\right]^2-\sin^2\theta\cos^2\phi,
\end{eqnarray}
\begin{eqnarray}
k_{z2}=\left\{\begin{array}{ll} \sqrt{{k_0}^2\epsilon_2-{k_y}^2} &~\mbox{if ${k_0}^2\epsilon_2>{k_y}^2$, $\Omega_2>0$} \\
-\sqrt{{k_0}^2\epsilon_2-{k_y}^2} &~\mbox{if ${k_0}^2\epsilon_2>{k_y}^2$, $\Omega_2<0$} \\
i\sqrt{{k_y}^2-{k_0}^2\epsilon_2} &~\mbox{if ${k_0}^2\epsilon_2\le {k_y}^2$} \end{array}\right..
\end{eqnarray}
The quantities $\rho_{02}$ and $U_{x2}$ refer to the mass density and the flow velocity in the transmitted region respectively.
The condition $\Omega_2<0$ is equivalent to $(U_{x2}/v_{A1})\sin\theta\cos\phi>1$.
The transmittance $T$ is obtained using
\begin{eqnarray}
T=\left\{\begin{matrix} \frac{k_{z2} \epsilon_1}{k_{z1}\epsilon_2}\vert t\vert^2 &~~\mbox{if ${k_0}^2\epsilon_2>{k_y}^2$} \\
0 &~~\mbox{if ${k_0}^2\epsilon_2\le {k_y}^2$} \end{matrix}\right..
\end{eqnarray}
When $\Omega_2$ is negative in the propagative case, $T$ becomes negative.
The equation for the field amplitude $\tilde B_x$ [$=\tilde B_x(z;l)$] takes the form
\begin{eqnarray}
\frac{\partial}{\partial l}\tilde B_x(z;l)&=&ik_{z1}\frac{\epsilon(l)}{\epsilon_1}\tilde B_x(z;l)+\frac{i}{2k_{z1}}\bigg[{k_0}^2\epsilon_1
 -\frac{\epsilon_1}{\epsilon(l)}{k_y}^2 \nonumber\\ &&  -\frac{\epsilon(l)}{\epsilon_1}{k_{z1}}^2\bigg][1+r(l)]\tilde B_x(z;l),
\label{eq:fieldb}
\end{eqnarray}
which should be integrated from $l=z$ to $l=L$ using the initial condition $\tilde B_x(z;z)=1+r(z)$
to obtain $\tilde B_x(z;L)$.

\section{Numerical results}
\label{sec_num}

\begin{figure}
\centering\includegraphics[width=8cm]{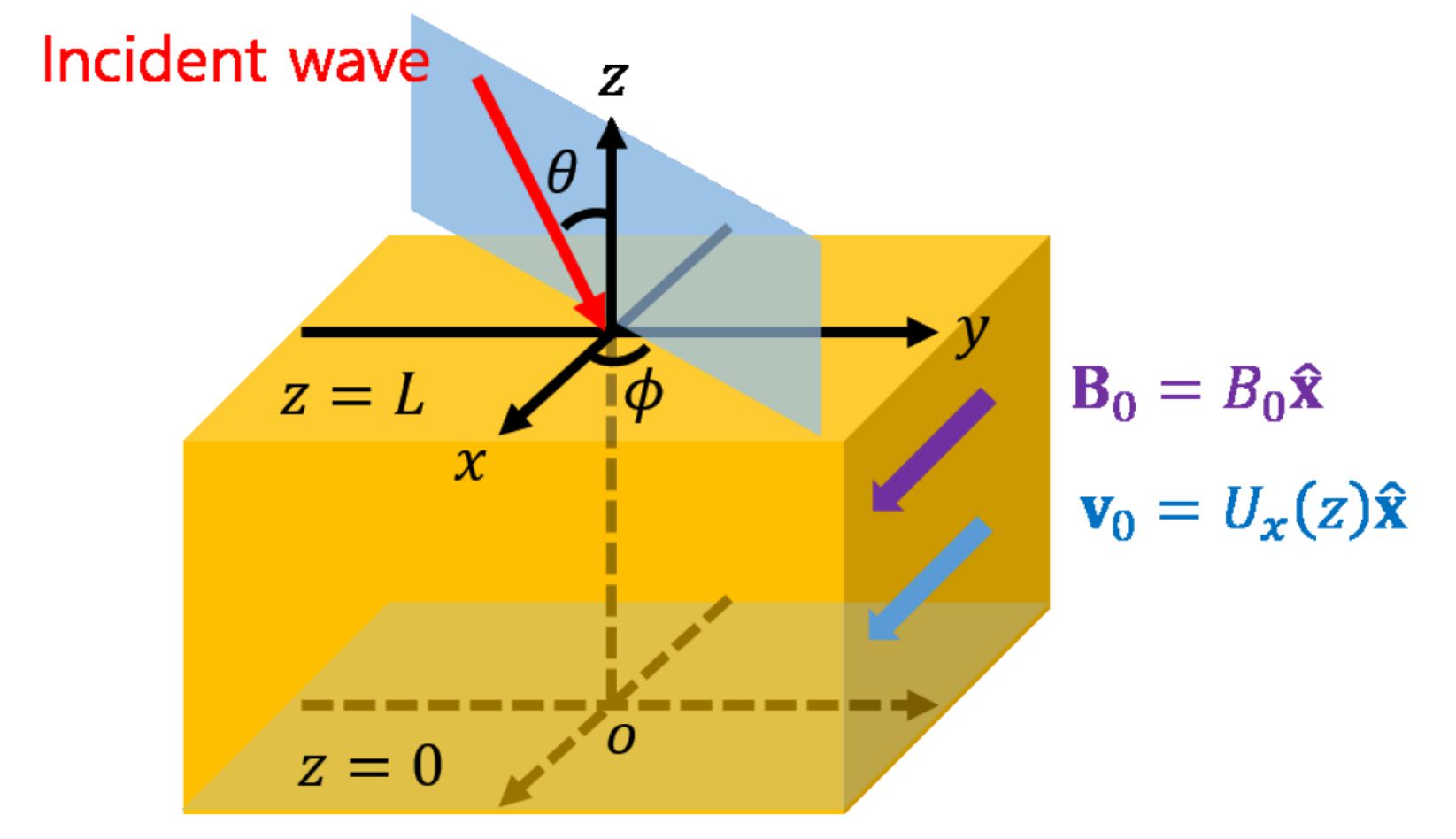}
\caption{Schematic of the configuration considered in this paper.}
\label{figz}
\end{figure}

We restrict our attention to the case where the external magnetic field $B_0$ is uniform throughout the space.
Then the equilibrium pressure $p_0$ and the plasma $\beta$ are also uniform due to Eq.~(\ref{eq:tp}).
From the ideal gas law, we obtain the relationship
\begin{equation}
\rho_0(z) T_0(z)={\rm constant}
\end{equation}
between $\rho_0$ and the temperature $T_0$.
Furthermore, we consider only the case where the flow velocity is parallel or anti-parallel to ${\bf B}_0$.
Therefore $U_y$ is zero, while $U_x$ is nonzero.
In the present situation, $\rho_0(z)$ and $U_x(z)$ can be specified independently by arbitrary functions.
Instead of $\rho_0(z)$, we can equivalently choose to specify $v_A(z)$, $v_s(z)$, or $v_c(z)$.
The ratio of the specific heats $\gamma$ is chosen to be 5/3 which corresponds to an ideal monatomic gas.
In Fig.~\ref{figz}, we show a simple schematic of the configuration considered in the present study.

In order to test our numerical method based on the IIM, we have calculated the absorption coefficient for
the same configurations as those studied in Refs.~\onlinecite{29Csik1998,30Csik2000},
where the Alfv\'en velocity varies linearly while
the shear flow velocity has a step-function profile and
the incident angle has been fixed to a single value. We have reproduced precisely the same results,
confirming that our method is accurate.
In this section, we consider the cases where both the flow velocity
and the Alfv\'en velocity (or the plasma density) vary linearly along the inhomogeneity direction.
We consider both the cases where the plasma density in the incident region is higher and lower than that in the
transmitted region and consider an entire range of the incident angles.

\begin{figure}
\centering\includegraphics[width=8.6cm]{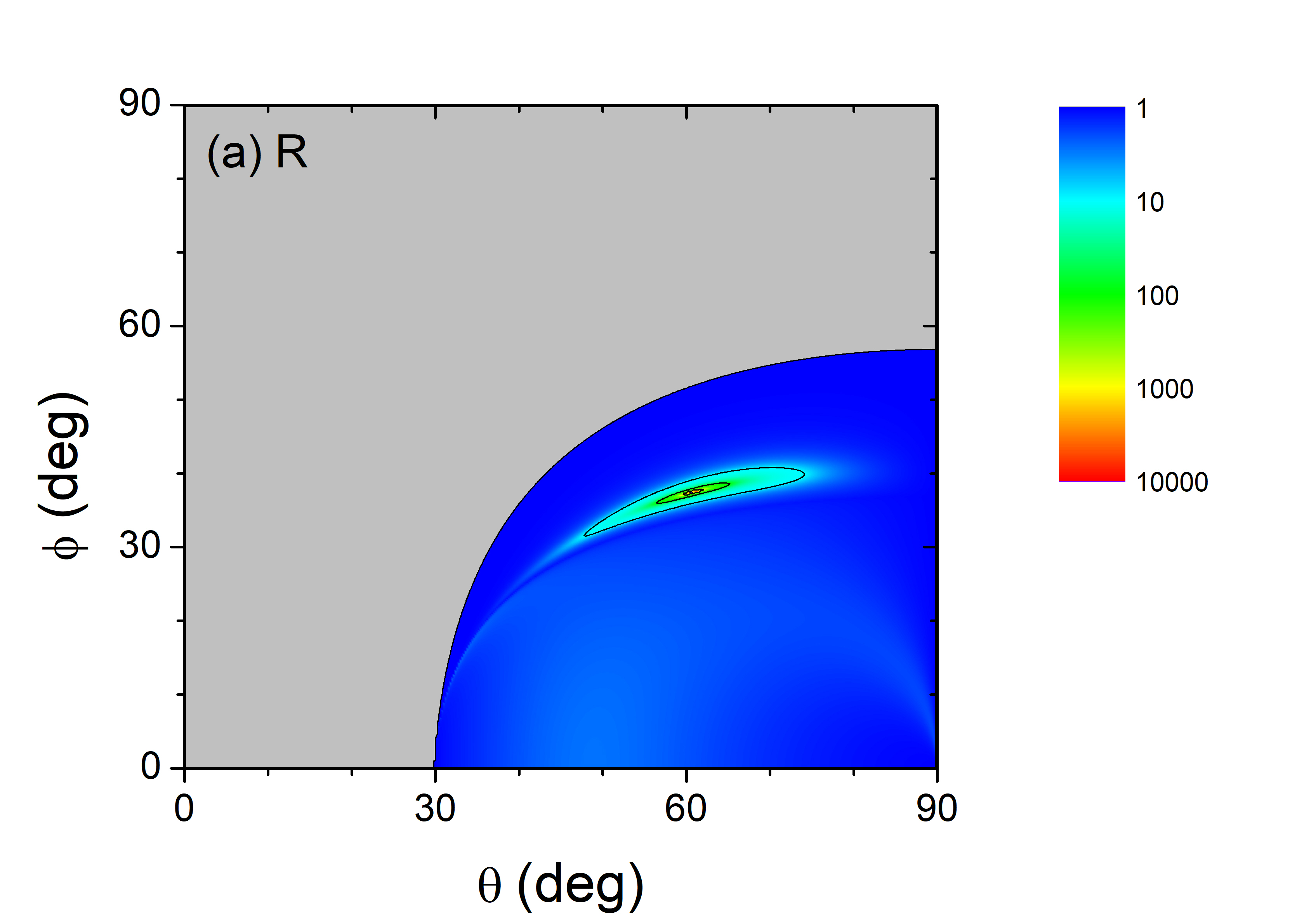}
\centering\includegraphics[width=8.6cm]{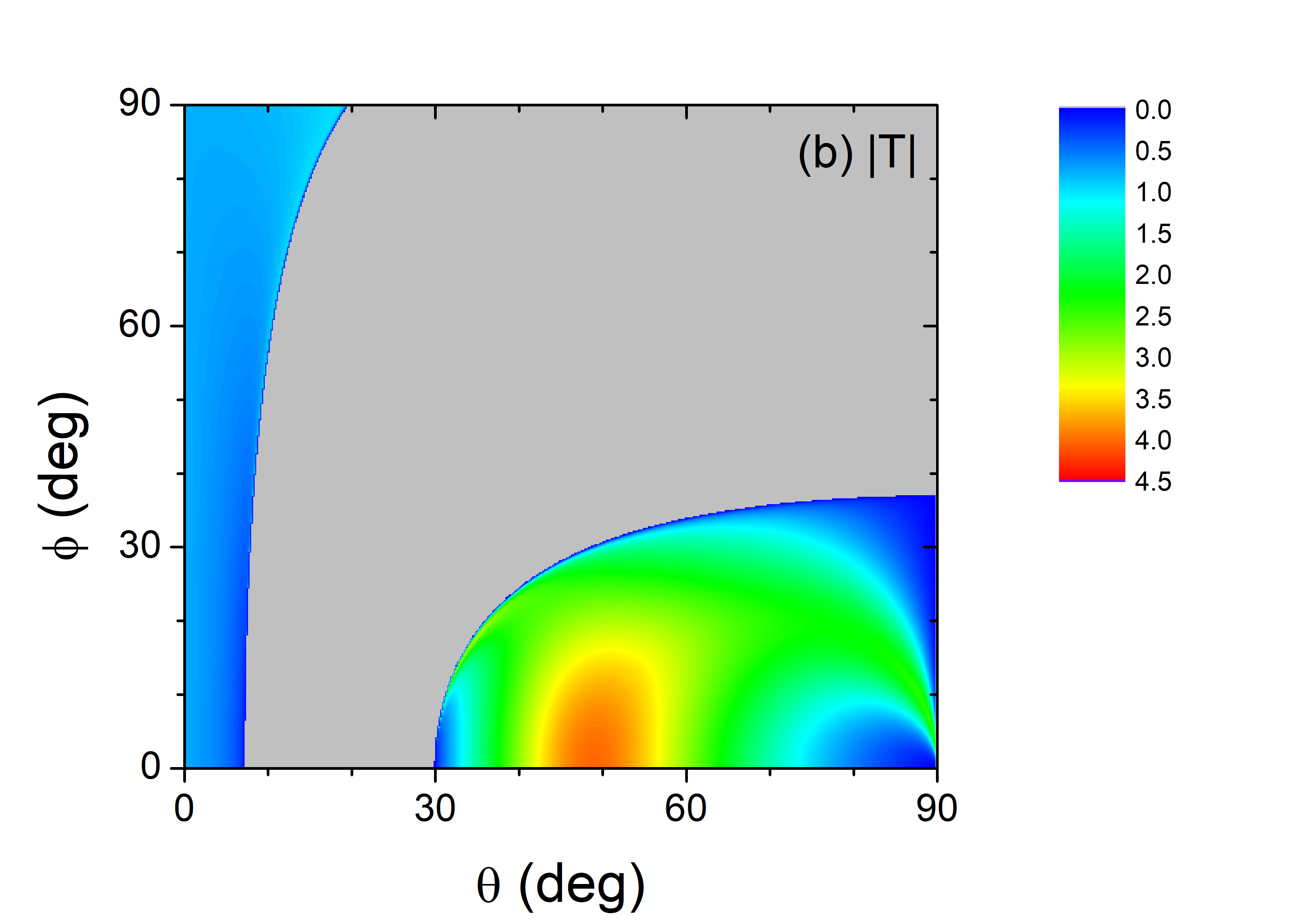}
\caption{Contour plots of (a) the reflectance $R$ and (b) the absolute value of the transmittance $\vert T\vert$
as functions of the incident angles $\theta$ and $\phi$ for the configuration given by Eq.~(\ref{eq:conf1}),
when the frequency $\omega$ of the incident fast wave is equal to $\omega_0=v_{A1}/L$ and the plasma $\beta$ is equal to zero.
In the non-gray region of (a), $R$ is larger than 1. In the gray region of (b), $T$ is strictly zero.}
\label{fig_w}
\end{figure}

\begin{figure}
\centering\includegraphics[width=8.6cm]{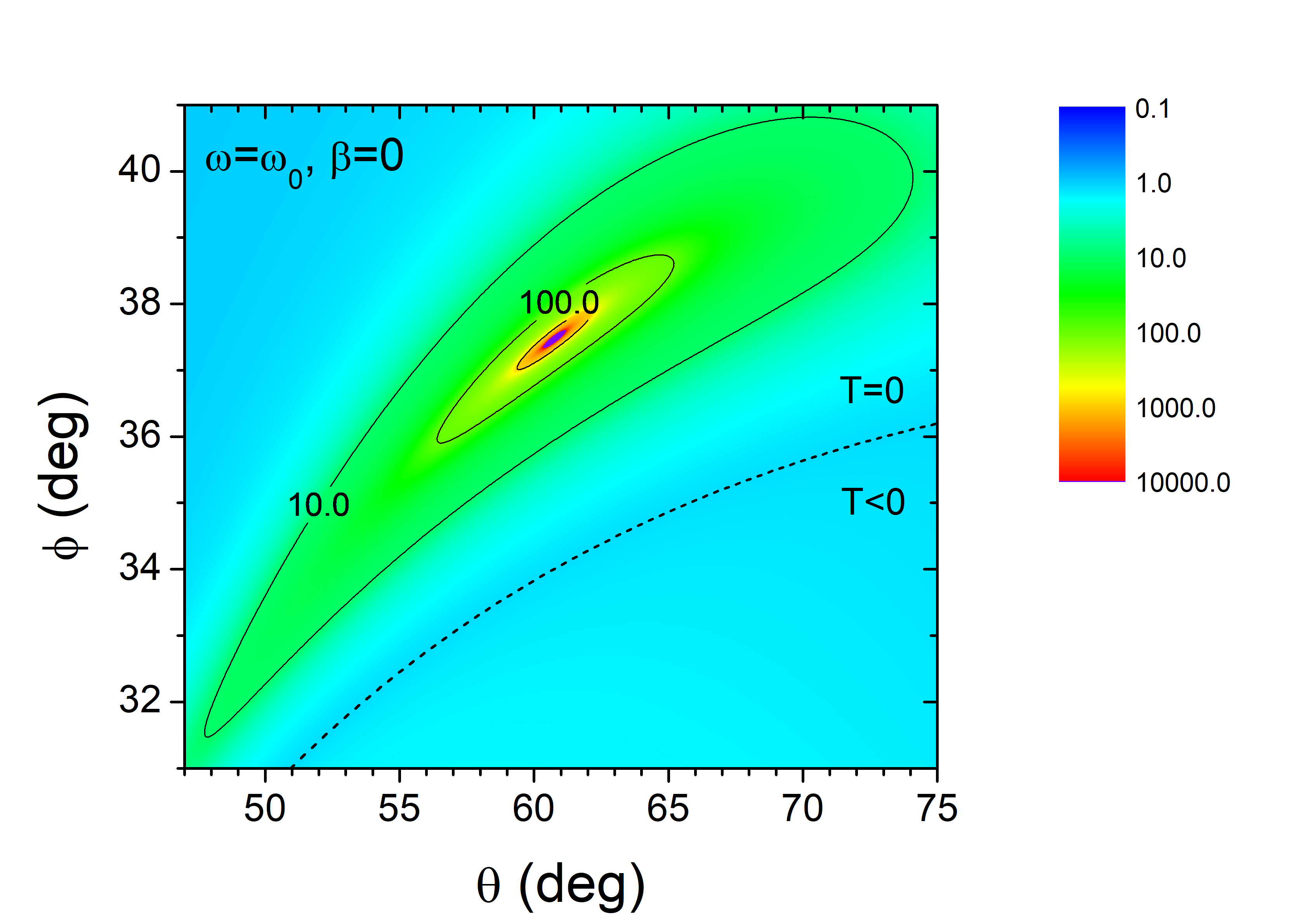}
\caption{Expanded logarithmic contour plot of $R$
as a function of $\theta$ and $\phi$ for the configuration given by Eq.~(\ref{eq:conf1}),
when $\omega=\omega_0$ and $\beta=0$. The region of $(\theta,\phi)$ such that $R>10$ is explicitly displayed. In the narrow purple-colored region, $R$ can be substantially larger than 10,000. The boundary
between the regions where the transmittance is zero and negative is indicated by a dashed line.}
\label{fig_a10}
\end{figure}

We first consider the configuration where both $U_x$ and $v_A$ vary linearly in the region $0\le z\le L$ such that
\begin{eqnarray}
  && \frac{U_x(z)}{v_{A1}}=\left\{ \begin{array}{ll}
  5,&~ \mbox{if } z<0\\
  5\left(1-\frac{z}{L}\right),&~ \mbox{if } 0\le z\le L\\
  0, &~ \mbox{if } z>L
  \end{array} \right.,\nonumber\\
  && \frac{v_A(z)}{v_{A1}}=\left\{ \begin{array}{ll}
  3,&~ \mbox{if } z<0\\
  3-2\frac{z}{L},&~ \mbox{if } 0\le z\le L\\
  1, &~ \mbox{if } z>L
  \end{array} \right..
\label{eq:conf1}
\end{eqnarray}
In this configuration, since $v_A$ is inversely proportional to $\sqrt{\rho_0}$, the wave incident from $z>L$ propagates from the region of higher density to that of lower density.
We remind that we can always choose the flow velocity in the incident region to be zero by using the reference frame moving with that velocity.
We consider the zero $\beta$ plasma case first and set the temperature $T_0$ and the plasma $\beta$ to be zero. Then the slow magnetosonic wave is absent.
We fix the frequency $\omega$ of the incident fast magnetosonic wave to $\omega_0$ ($\equiv v_{A1}/L$) and the
damping parameter $\eta$ to $10^{-8}$. We solve Eq.~(\ref{eq:cpie}) numerically to obtain the reflectance $R$ and the transmittance $T$
as functions of the incident angles $\theta$ ($0\le \theta<\pi/2$) and $\phi$ ($0\le \phi<2\pi$). We note that, due to symmetry, all physical quantities at $\phi=2\pi-\phi_0$ are
equal to those at $\phi=\phi_0$. According to Eq.~(\ref{eq:rescond}), when $U_x(z)$ is positive, amplification is possible only when $\cos\phi$ is positive.
Therefore we restrict the range of $\phi$ to $0\le \phi<\pi/2$. In Fig.~\ref{fig_w}, we show the logarithmic contour plot of $R$ and the linear contour plot of the absolute value of $T$.
In the non-gray region of Fig.~\ref{fig_w}(a), $R$ is larger than 1.
In the gray region of Fig.~\ref{fig_w}(b), $T$ is strictly zero.
We find that overreflection ($R>1$) of the incident wave occurs in a substantial region of the $(\theta,\phi)$ space. We also notice that {\it giant overreflection with $R\gtrsim 10$
occurs predominantly in the parameter region where the incident wave is totally reflected} and the transmittance is identically zero. In the region with $\theta\gtrsim 30^\circ$ and $\phi\lesssim 36^\circ$ in Fig.~\ref{fig_w}(b)
where $\vert T\vert$ is substantially larger
than 1, the transmittance is actually negative
and overreflection arises due to negative energy waves in the transmitted region, not due to the mode conversion
to local resonant MHD modes.
However, the enhancement of the wave energy in this region is much smaller than in the region of giant overreflection shown explicitly in Fig.~\ref{fig_a10}.

In Fig.~\ref{fig_a10}, we show an expanded logarithmic contour plot of $R$ mainly in the region where $R$ is larger than 10. Such a strong overreflection occurs in a fairly wide range of $\theta$
and in a narrower but readily measurable range of $\phi$.
We emphasize again that this region belongs to the gray region in Fig.~\ref{fig_w}(b) where the incident wave is totally reflected and the transmittance is zero.
Inside the narrow purple-colored region, $R$ can be substantially larger than 10,000.
To the best of our knowledge, this type of giant overreflection has never been reported previously.

\begin{figure}
\centering\includegraphics[width=8.6cm]{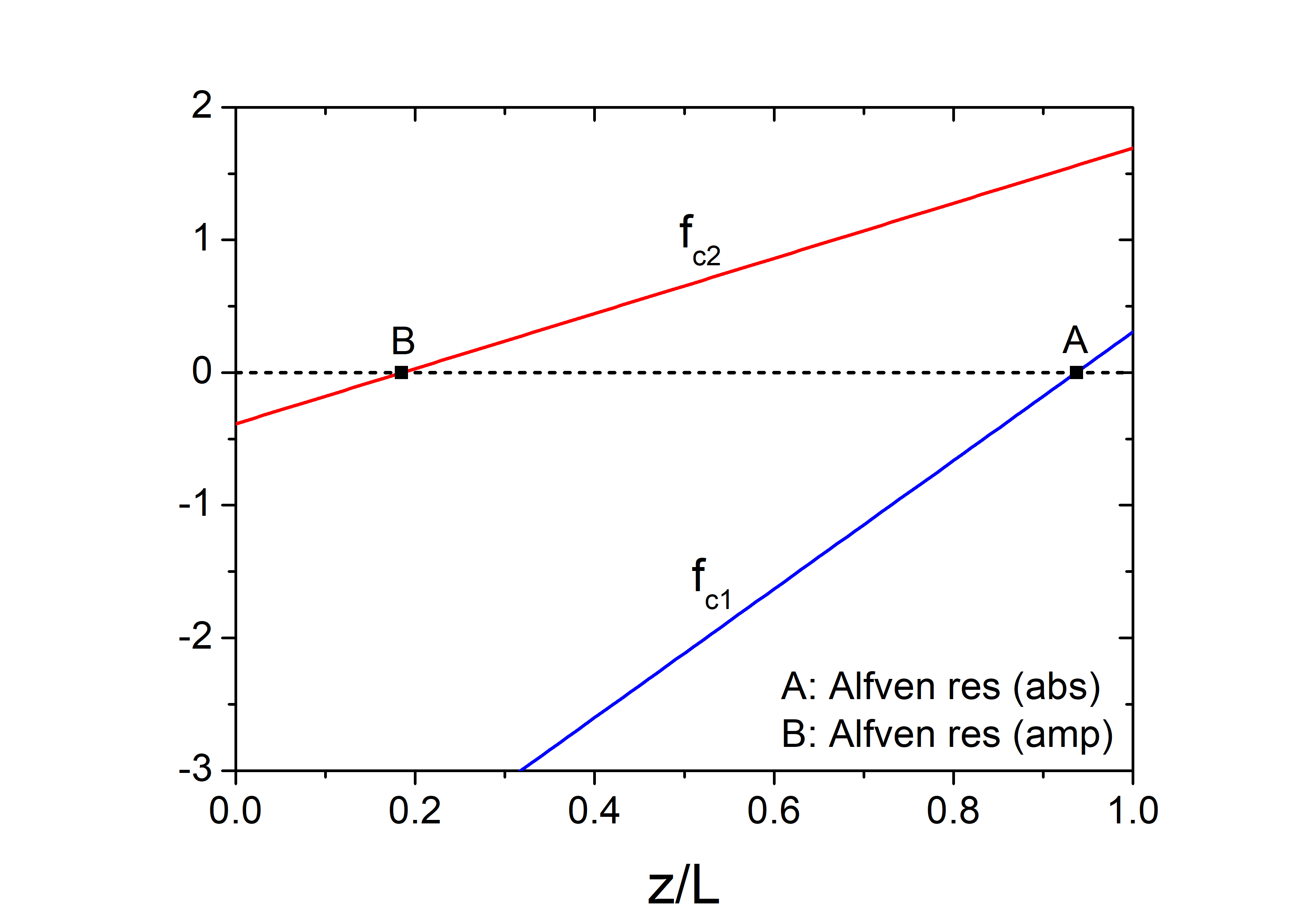}
\caption{The functions $f_{c1}$ and $f_{c2}$ defined by Eq.~(\ref{eq:rescond})
for the configuration given by Eq.~(\ref{eq:conf1}) plotted versus $z/L$
when $\theta=60.8^\circ$ and $\phi=37.5^\circ$, which corresponds
to the maximum value of $R$ in Fig.~\ref{fig_a10}.
The resonant absorption and amplification due to the Alfv\'en resonance occur at
the positions A ($z/L\approx 0.937$) and B ($z/L\approx 0.185$) respectively.}
\label{fig_aa}
\end{figure}

When a strong overreflection with $R\gtrsim 10$ occurs, there should be at least one resonance
plane corresponding to the wave amplification with a negative effective damping parameter. In Fig.~\ref{fig_aa}, we plot
the functions $f_{c1}$ and $f_{c2}$ defined by Eq.~(\ref{eq:rescond})
for the configuration given by Eq.~(\ref{eq:conf1}) versus $z/L$
when $\theta=60.8^\circ$ and $\phi=37.5^\circ$, which corresponds
to the maximum of $R$ in Fig.~\ref{fig_a10}.
The resonant absorption and amplification due to the Alfv\'en resonance occur at
the positions A ($z/L\approx 0.937$) and B ($z/L\approx 0.185$) respectively.
Strong amplification of the wave energy at B is responsible for the giant overreflection.

\begin{figure}
\centering\includegraphics[width=8.6cm]{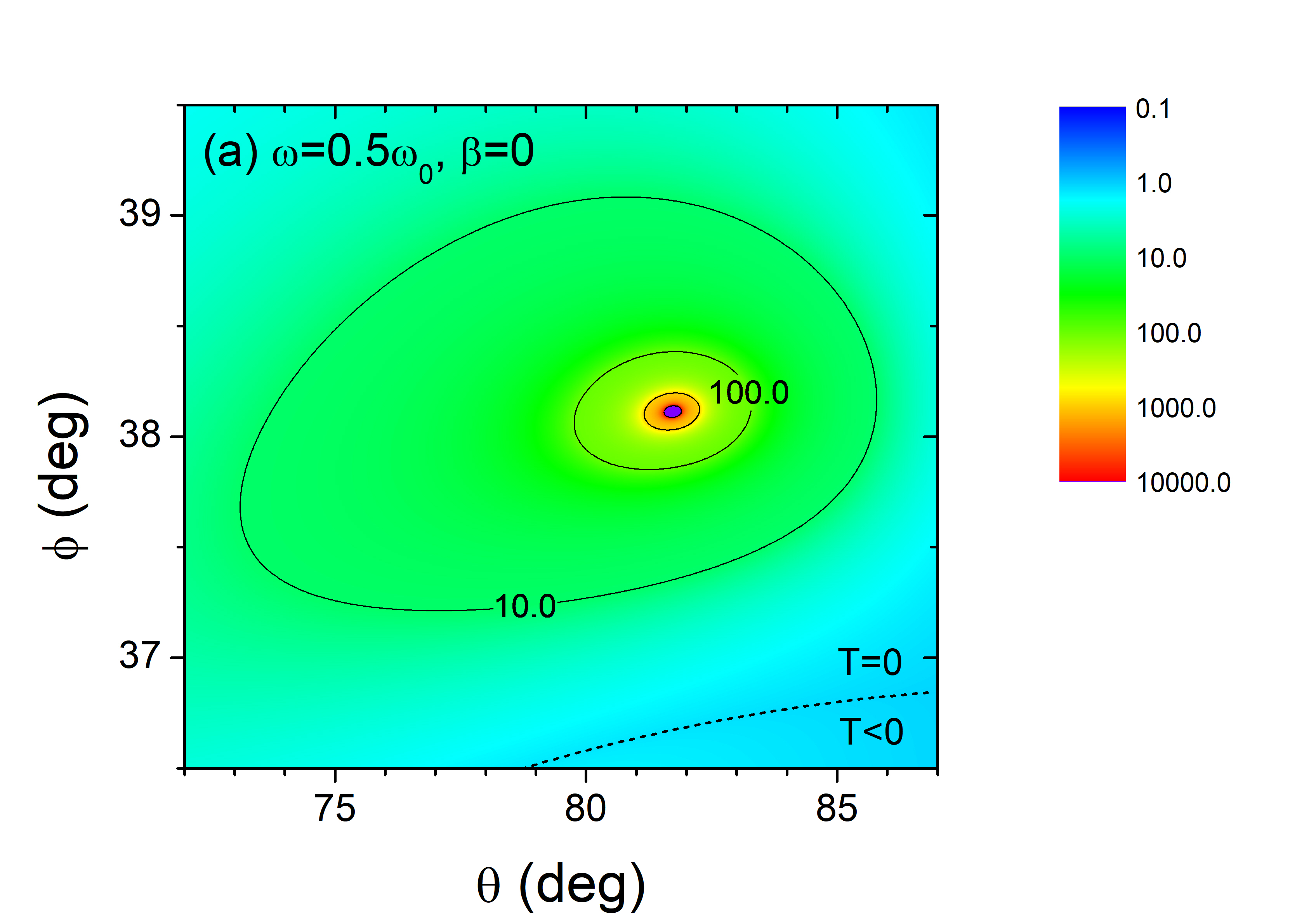}
\centering\includegraphics[width=8.6cm]{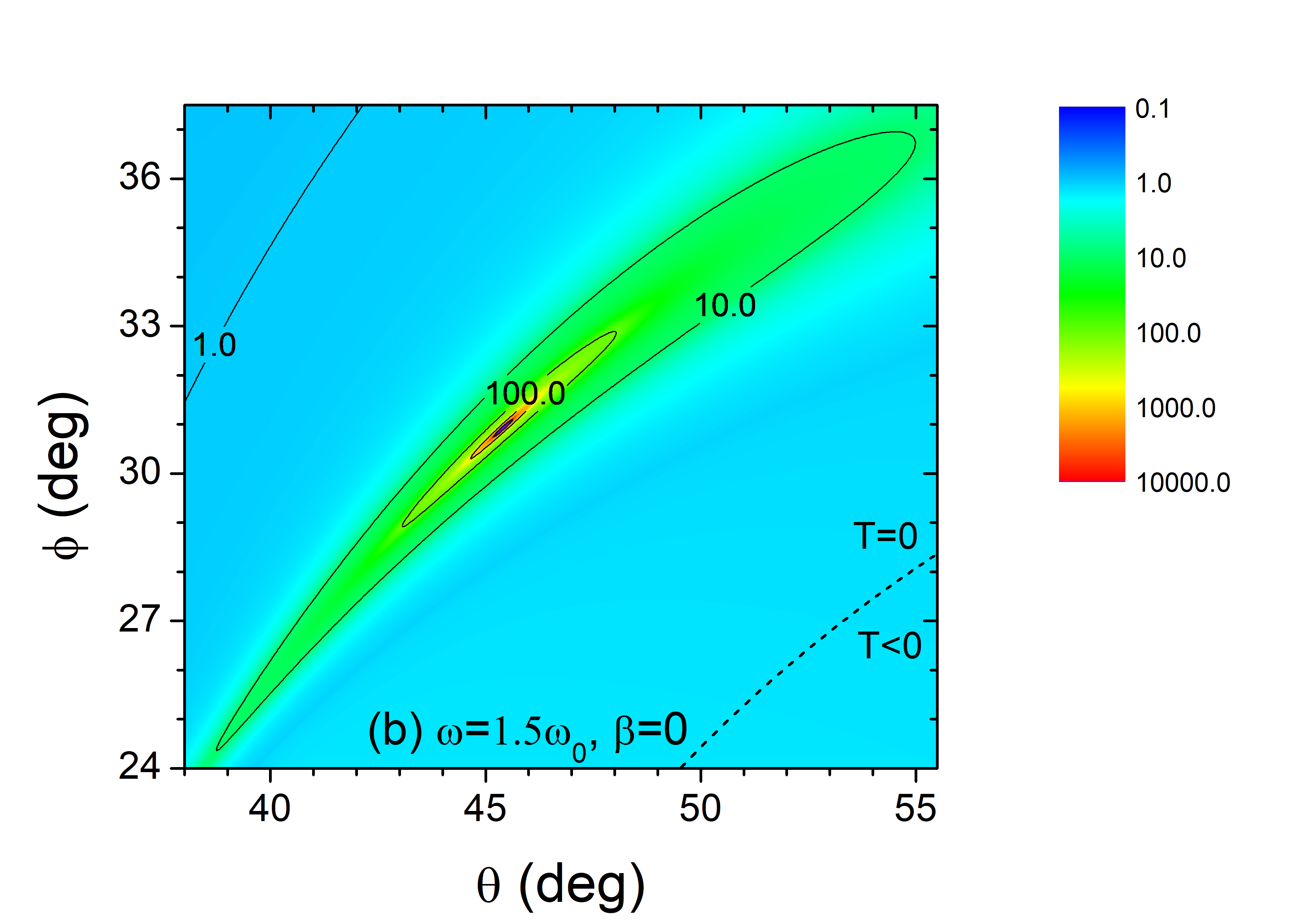}
\caption{Logarithmic contour plots of $R$ versus $\theta$ and $\phi$
for the configuration given by Eq.~(\ref{eq:conf1}), when (a) $\omega=0.5\omega_0$ and (b) $\omega=1.5\omega_0$. The plasma $\beta$ is zero.
The boundary
between the regions where the transmittance is zero and negative is indicated by a dashed line in each figure.}
\label{fig_freq}
\end{figure}

The phenomenon of strong overreflection persists in a wide range of incident wave frequency. In Fig.~\ref{fig_freq}, we show the
logarithmic contour plots of $R$ at $\beta=0$ when $\omega$ is equal to $0.5\omega_0$ and $1.5\omega_0$.
The position and the shape of the region where $R$ is greater than 10 in the $(\theta,\phi)$ space are changed, but its size remains  substantially large as the frequency varies.

\begin{figure}
\centering\includegraphics[width=8.6cm]{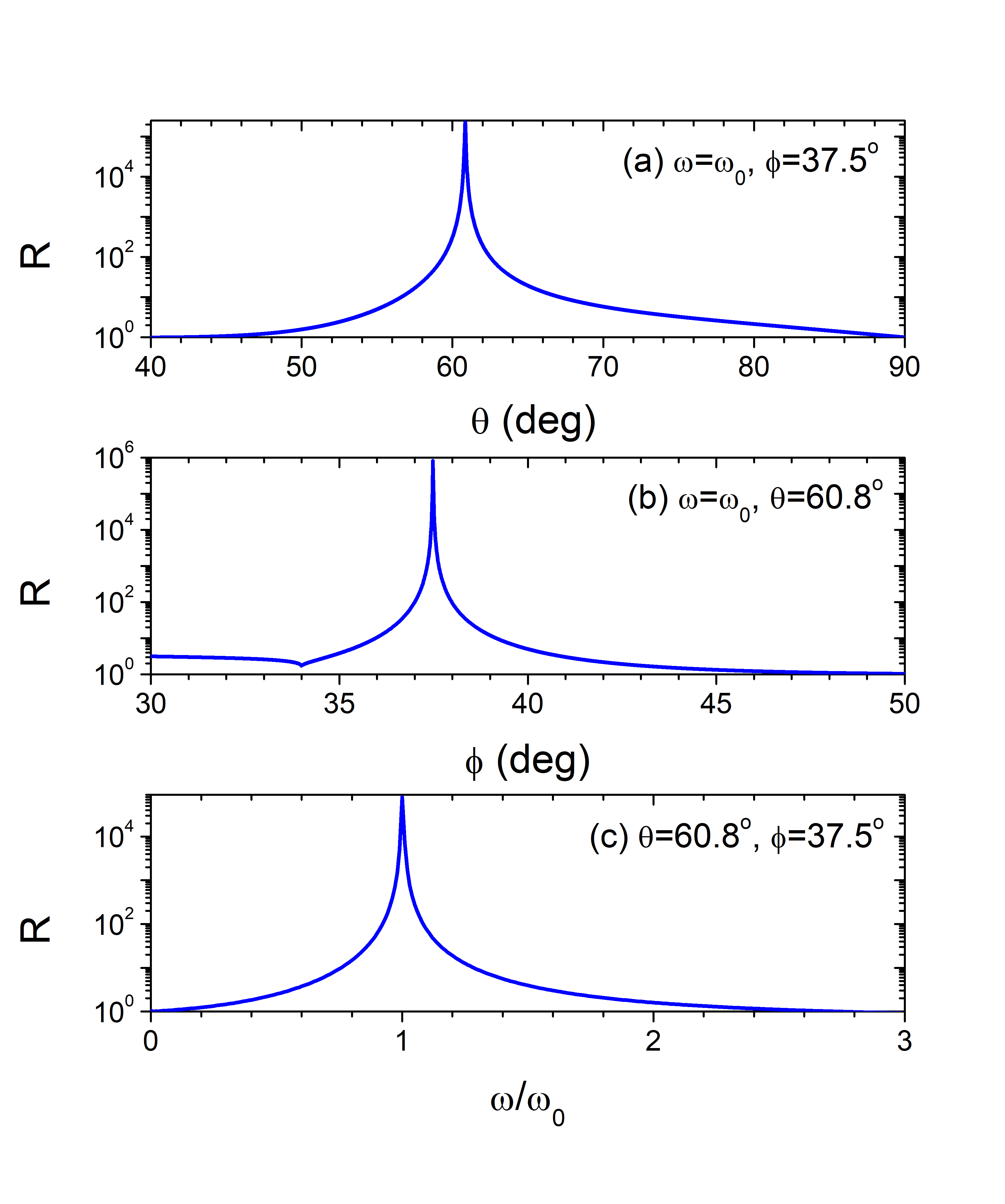}
\caption{Reflectance $R$
plotted (a) versus $\theta$ when $\omega=\omega_0$ and $\phi=37.5^\circ$, (b) versus $\phi$ when $\omega=\omega_0$ and $\theta=60.8^\circ$,
and (c) versus $\omega/\omega_0$ when $\theta=60.8^\circ$ and $\phi=37.5^\circ$ for the configuration given by Eq.~(\ref{eq:conf1}). The plasma $\beta$ is zero.}
\label{fig_tpf}
\end{figure}

In the present configuration, when $\omega=\omega_0$ and $\beta=0$, the reflectance takes an extremely large maximum value at $\theta\approx 60.8^\circ$ and $\phi\approx 37.5^\circ$. The
variations of $R$ versus $\theta$, $\phi$, and $\omega$ close to the maximum are shown in Fig.~\ref{fig_tpf}. In Fig.~\ref{fig_tpf}(a), we observe that
$R$ is larger than 1 in the range $41.18^\circ <\theta <90^\circ$, larger than 2 in $51.54^\circ <\theta< 80.92^\circ$, and larger than 10 in $56.6^\circ <\theta <67.14^\circ$.
In Fig.~\ref{fig_tpf}(b), we find that
$R$ is larger than 1 in $0^\circ <\phi <51.92^\circ$, larger than 2 in $0^\circ <\phi <33.9^\circ$ and
in $34.12^\circ <\phi <42.32^\circ$, and larger than 10 in $35.97^\circ <\phi <39.15^\circ$.
In this figure, the region where $\phi<34^\circ$ corresponds to the region
where overreflection occurs due to negative energy waves in the transmitted region.
In Fig.~\ref{fig_tpf}(c), we find that
$R$ is larger than 1 in $0 <\omega/\omega_0< 2.74$, larger than 2 in $0.43 <\omega/\omega_0< 1.82$, and larger than 10 in $0.76 <\omega/\omega_0< 1.28$.
We notice that the maximum reflectance is well over $10^6$.

\begin{figure}
\centering\includegraphics[width=8.6cm]{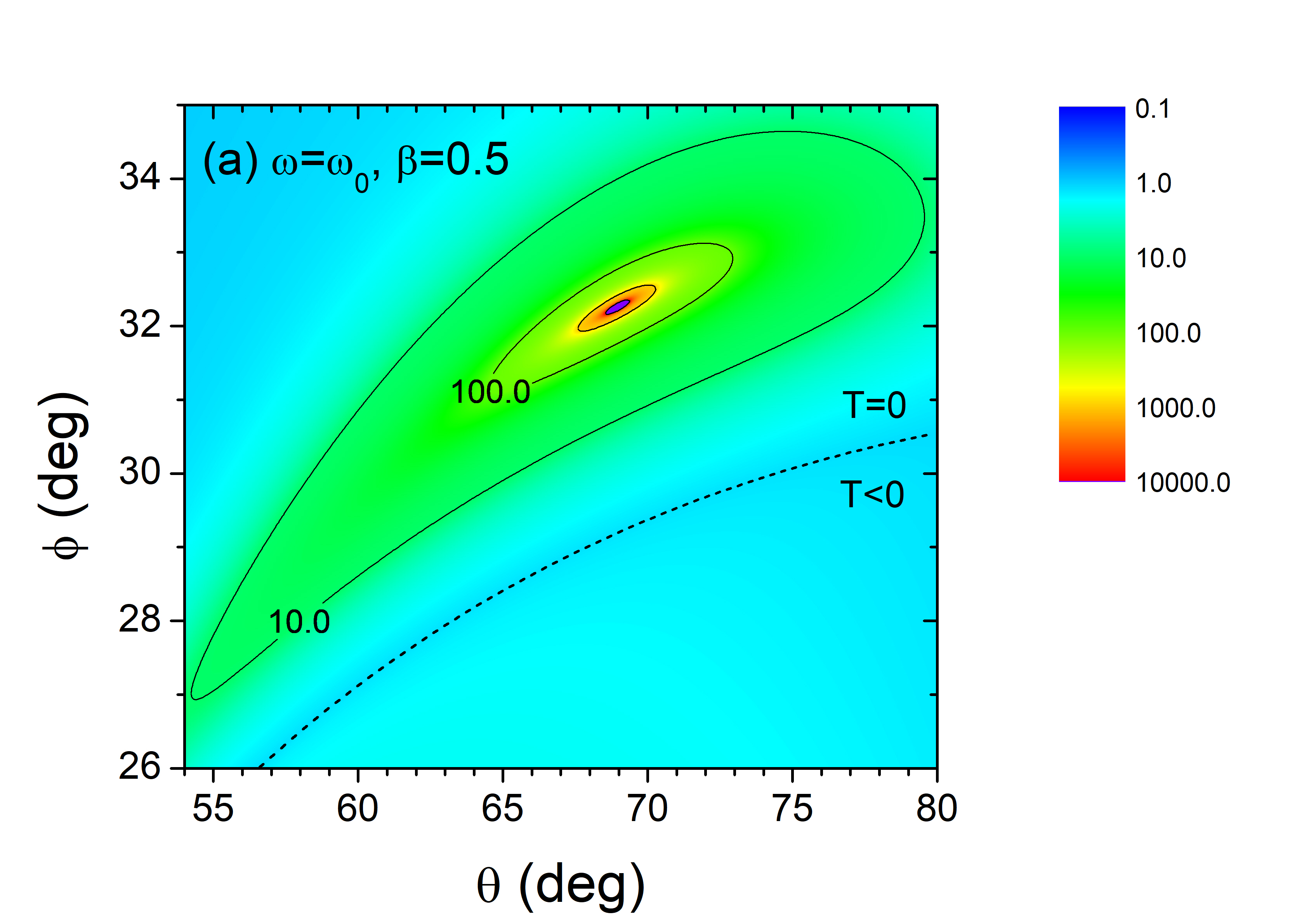}
\centering\includegraphics[width=8.6cm]{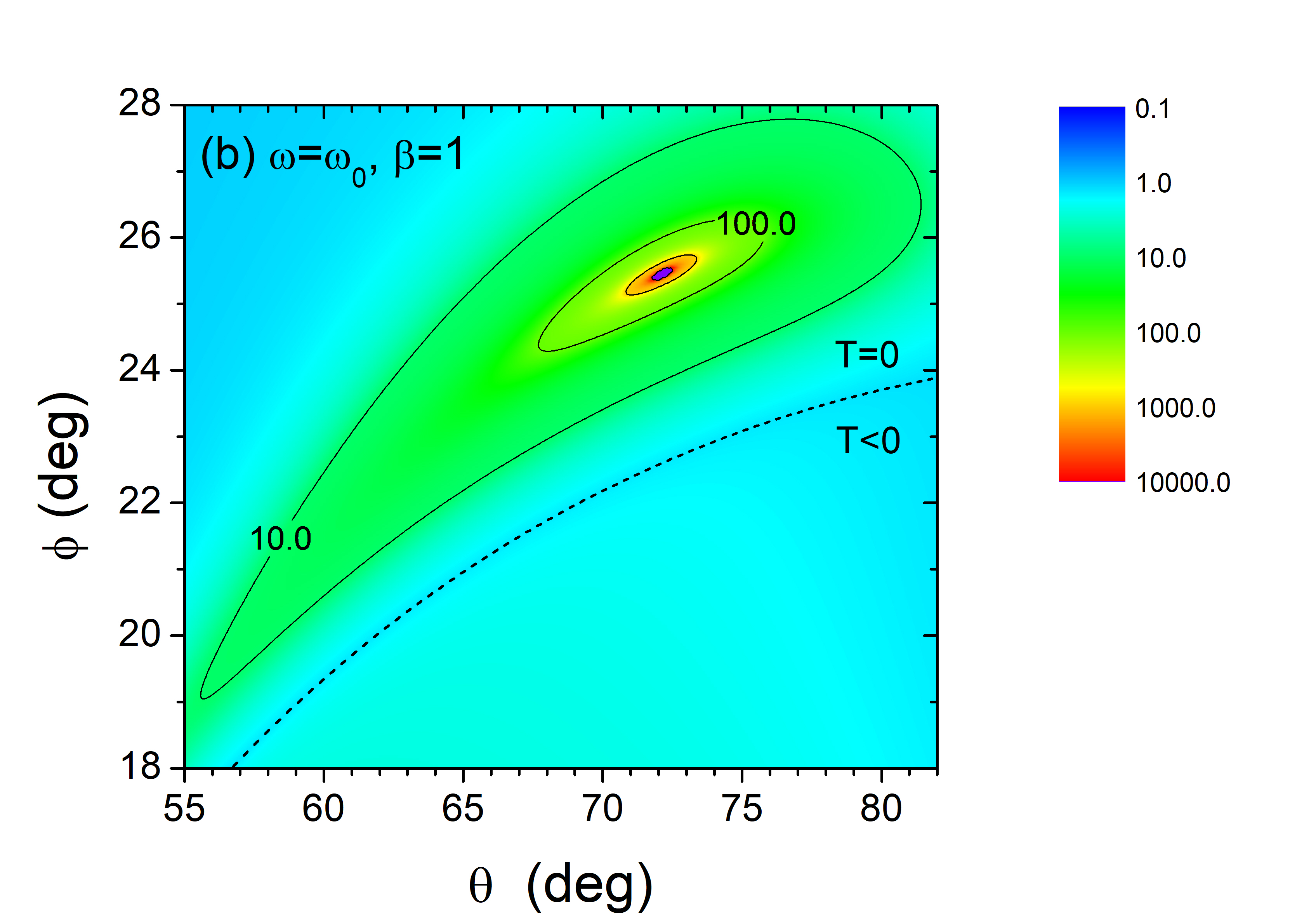}
\caption{Logarithmic contour plots of $R$
for fast waves of frequency $\omega=\omega_0$ versus $\theta$ and $\phi$  for the configuration given by Eq.~(\ref{eq:conf1}), when (a) $\beta=0.5$ and (b) $\beta=1$. The boundary
between the regions where the transmittance is zero and negative is indicated by a dashed line in each figure.}
\label{fig_beta}
\end{figure}

In the case of warm plasmas with finite $\beta$, we can solve Eq.~(\ref{eq:mie}) numerically to obtain $R$ and $T$ for both incident
fast and slow waves. In Fig.~\ref{fig_beta}, we show
the logarithmic contour plots of $R$ for fast waves of frequency $\omega=\omega_0$ when $\beta$ is equal to 0.5 and 1. We find that
the position of the region where $R$ is larger than 10 in the $(\theta,\phi)$ space is shifted, but its size remains approximately the same, as $\beta$ increases. Therefore the finite temperature effect does not destroy the giant overreflection.

\begin{figure}
\centering\includegraphics[width=8.6cm]{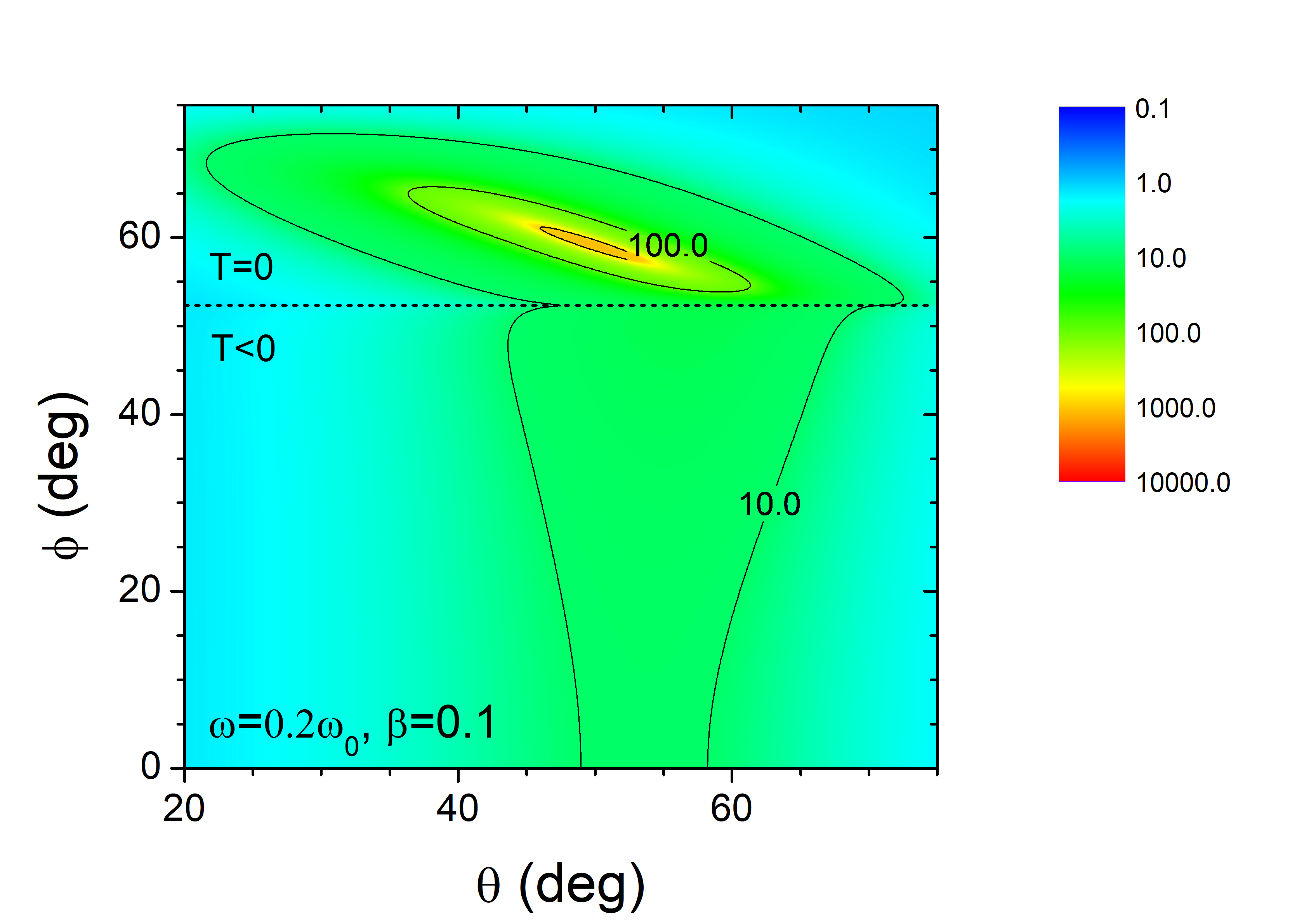}
\caption{Logarithmic contour plot of $R$
for slow waves of frequency $\omega=0.2\omega_0$ versus $\theta$ and $\phi$ for the configuration given by Eq.~(\ref{eq:slow}), when $\beta=0.1$. The boundary
between the regions where the transmittance is zero and negative is indicated by a dashed line.}
\label{fig_slow}
\end{figure}

Giant overreflection also occurs for slow waves, which exist only in finite temperature plasmas.
The parameter values that generate strong overreflection for slow waves are substantially different from those for fast waves.
In Fig.~\ref{fig_slow}, we have considered the configuration given by
\begin{eqnarray}
  && \frac{U_x(z)}{v_{A1}}=\left\{ \begin{array}{ll}
  7,& \mbox{if } z<0\\
  7\left(1-\frac{z}{L}\right),& \mbox{if } 0\le z\le L\\
  0, & \mbox{if } z>L
  \end{array} \right.,\nonumber\\
  && \frac{v_A(z)}{v_{A1}}=\left\{ \begin{array}{ll}
  4,& \mbox{if } z<0\\
  4-3\frac{z}{L},& \mbox{if } 0\le z\le L\\
  1, & \mbox{if } z>L
  \end{array} \right.,
\label{eq:slow}
\end{eqnarray}
and chosen $\beta=0.1$ and $\omega=0.2\omega_0$.
We find that, though the maximum reflectance is only of the order of $10^3$, the region where $R$ is larger than 10 is much wider than that corresponding to fast waves.
This result suggests that even for values of $\beta$ as small as 0.1, slow waves and slow resonances may play
rather important roles in various processes in space plasmas.
The elongated region with $\phi\lesssim 52.3^\circ$ where $R>10$ corresponds to the region where
strong overreflection occurs due to negative energy waves in the transmitted region.

\begin{figure}
\centering\includegraphics[width=8.6cm]{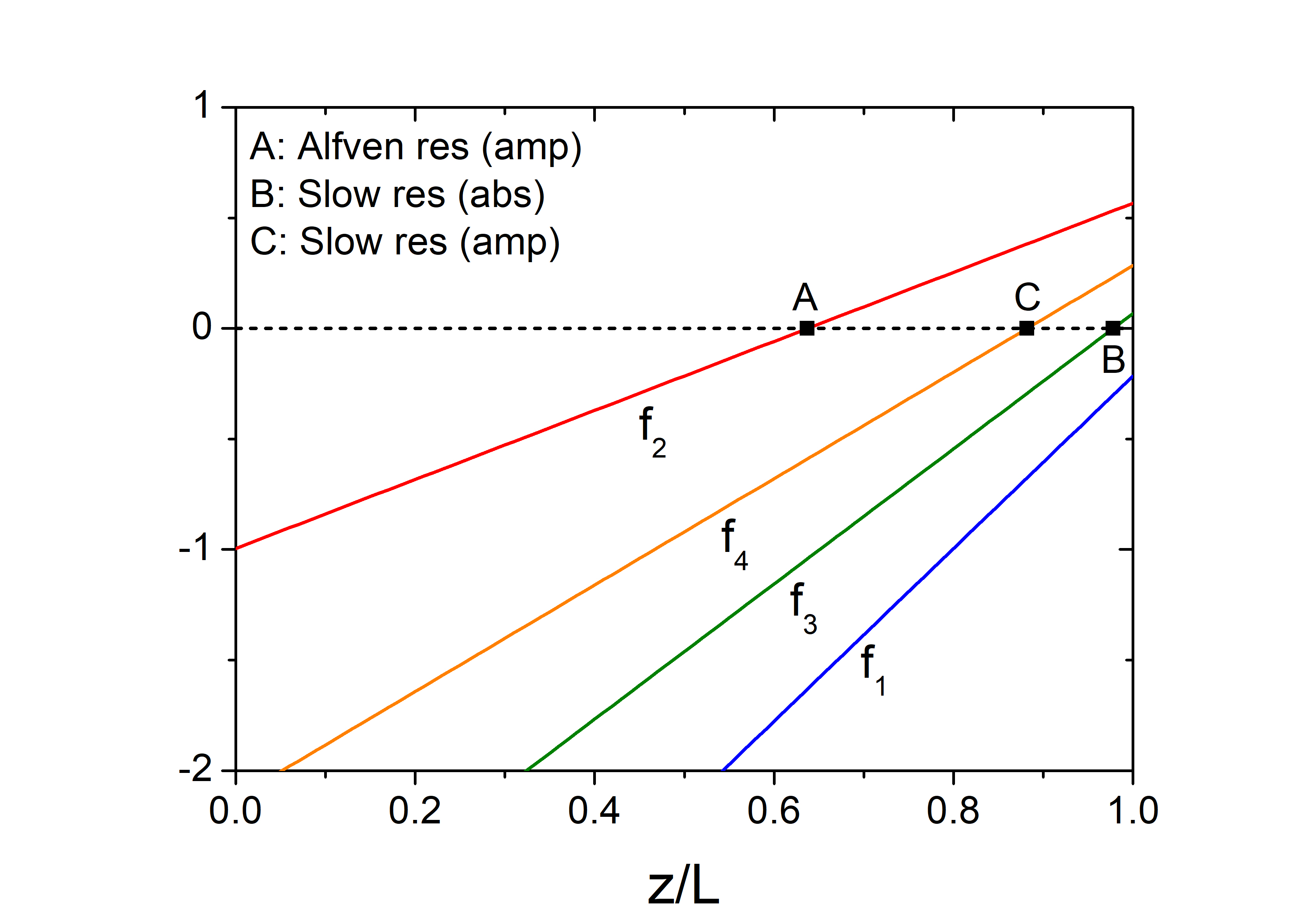}
\caption{The functions $f_{1}$, $f_{2}$, $f_{3}$, and $f_{4}$ defined by Eq.~(\ref{eq:resw})
for the configuration given by Eq.~(\ref{eq:slow}) plotted versus $z/L$
when $\theta=49.9^\circ$ and $\phi=59.3^\circ$, which corresponds
to the maximum value of $R$ in Fig.~\ref{fig_slow}.
The resonant amplification due to the Alfv\'en resonance occurs at
the position A ($z/L\approx 0.637$) and
the resonant absorption and amplification due to the slow resonance occur at
the positions B ($z/L\approx 0.978$) and C ($z/L\approx 0.882$) respectively.}
\label{fig_bb}
\end{figure}

At finite temperatures, both Alfv\'en and slow resonances can occur simultaneously
inside the inhomogeneous plasma. In Fig.~\ref{fig_bb}, we plot the functions $f_{1}$, $f_{2}$, $f_{3}$, and $f_{4}$ defined by Eq.~(\ref{eq:resw})
for the configuration given by Eq.~(\ref{eq:slow}) versus $z/L$
when $\theta=49.9^\circ$ and $\phi=59.3^\circ$, which corresponds
to the maximum of $R$ in Fig.~\ref{fig_slow}.
The resonant amplification due to the Alfv\'en resonance occurs at
the position A ($z/L\approx 0.637$) and
the resonant absorption and amplification due to the slow resonance occur at
the positions B ($z/L\approx 0.978$) and C ($z/L\approx 0.882$) respectively.
We notice that two different types of resonances are responsible for the giant overreflection in this case.

\begin{figure}
\centering\includegraphics[width=8.6cm]{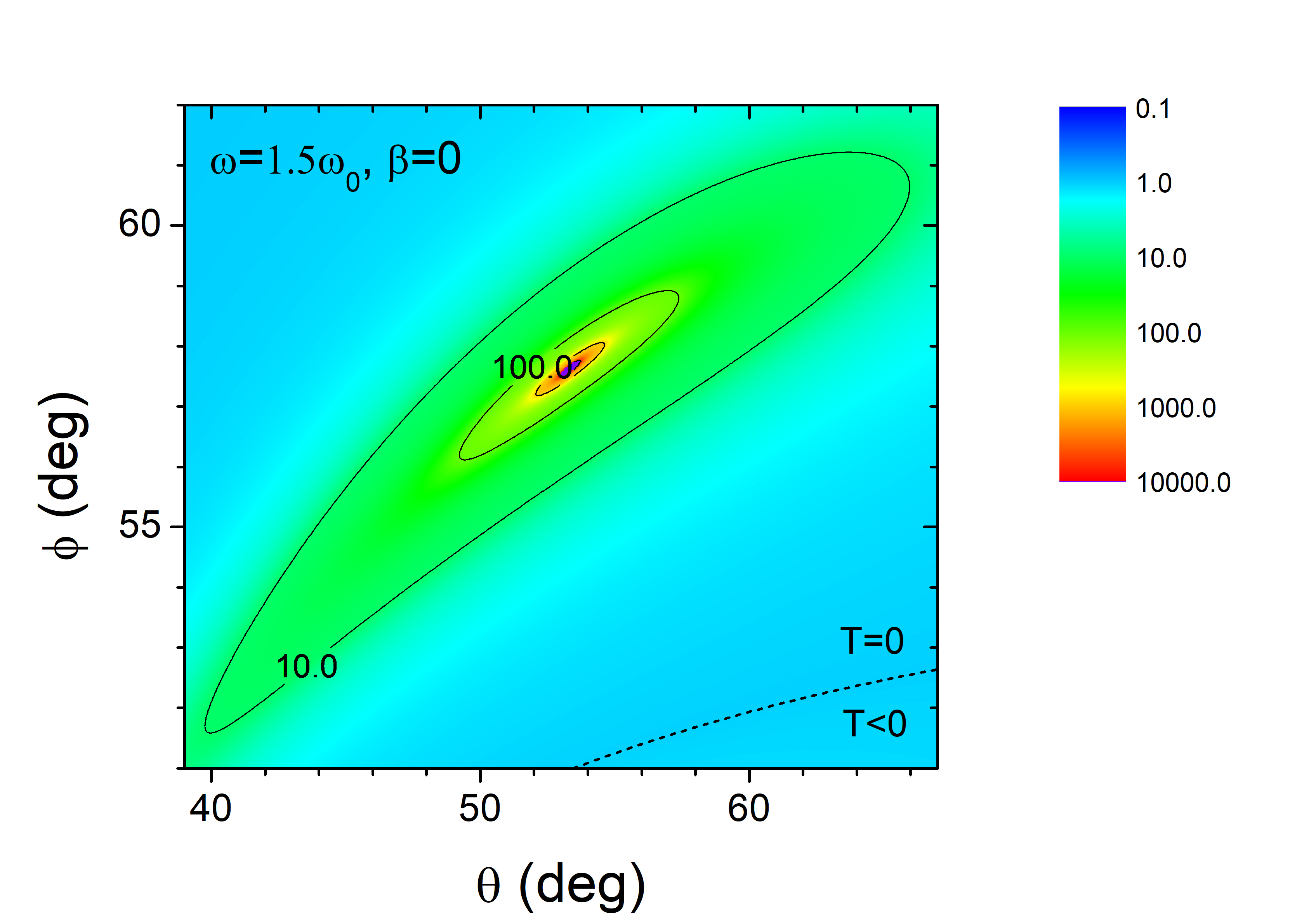}
\caption{Logarithmic contour plot of $R$ versus $\theta$ and $\phi$ for the configuration given by Eq.~(\ref{eq:conf2}), when $\omega=1.5\omega_0$ and $\beta=0$. The boundary
between the regions where the transmittance is zero and negative is indicated by a dashed line.}
\label{fig_c2}
\end{figure}

\begin{figure}
\centering\includegraphics[width=8.6cm]{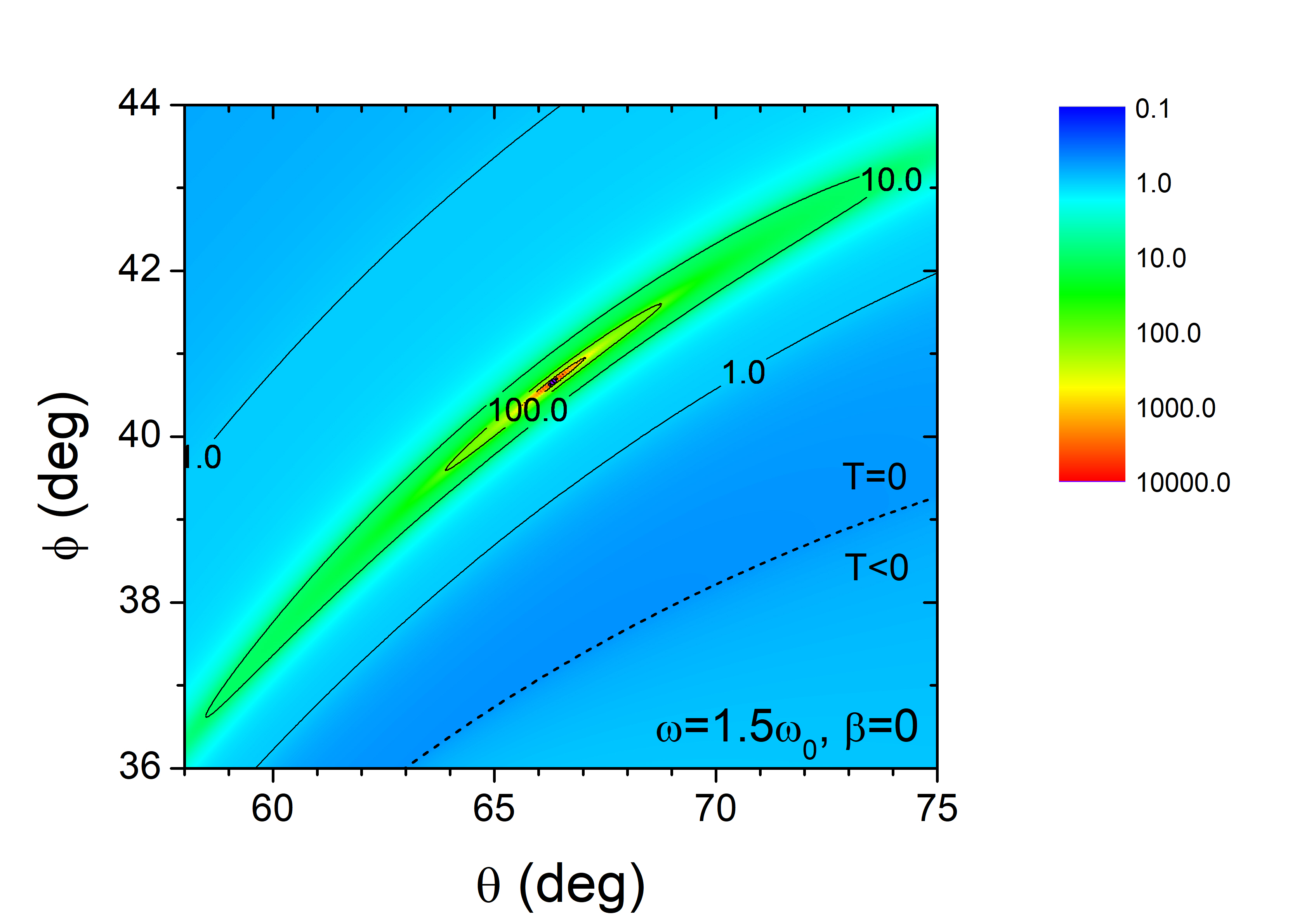}
\caption{Logarithmic contour plot of $R$ versus $\theta$ and $\phi$ for the configuration given by Eq.~(\ref{eq:conf3}), when $\omega=1.5\omega_0$ and $\beta=0$. The boundary
between the regions where the transmittance is zero and negative is indicated by a dashed line.}
\label{fig_c3}
\end{figure}

Strong overreflection and amplification of magnetosonic waves are not limited to specific configurations and occur in a wide range of configurations for the
flow velocity and the plasma density. In Fig.~\ref{fig_c2}, we consider the configuration given by
\begin{eqnarray}
  && \frac{U_x(z)}{v_{A1}}=\left\{ \begin{array}{ll}
  7,& \mbox{if } z<0\\
  7\left(1-\frac{z}{L}\right),& \mbox{if } 0\le z\le L\\
  0, & \mbox{if } z>L
  \end{array} \right.,\nonumber\\
  && \frac{\rho_0(z)}{\rho_{01}}=\left\{ \begin{array}{ll}
  0.1,& \mbox{if } z<0\\
  0.9\frac{z}{L}+0.1,& \mbox{if } 0\le z\le L\\
  1, & \mbox{if } z>L
  \end{array} \right.,
\label{eq:conf2}
\end{eqnarray}
where the plasma density $\rho_0$, instead of the Alfv\'en velocity $v_A$, varies linearly inside the inhomogeneous slab.
We find that giant overreflection occurs in a wide range of $\theta$ and $\phi$ in this case too.

In Fig.~\ref{fig_c3}, we consider the configuration given by
\begin{eqnarray}
  && \frac{U_x(z)}{v_{A1}}=\left\{ \begin{array}{ll}
  2.5,& \mbox{if } z<0\\
  2.5\left(1-\frac{z}{L}\right),& \mbox{if } 0\le z\le L\\
  0, & \mbox{if } z>L
  \end{array} \right.,\nonumber\\
  && \frac{v_A(z)}{v_{A1}}=\left\{ \begin{array}{ll}
  0.9,& \mbox{if } z<0\\
  0.9+0.1\frac{z}{L},& \mbox{if } 0\le z\le L\\
  1, & \mbox{if } z>L
  \end{array} \right.,
\label{eq:conf3}
\end{eqnarray}
where the Alfv\'en velocity decreases (therefore, the plasma density increases) as $z$ decreases from $L$ to 0, in contrast to the cases considered so far.
Since the wave is incident from the region where $z>L$,
it propagates from the region of lower density into that of higher density and gets reflected. In this case,
we find that the region where total reflection occurs and the transmission vanishes
is narrow. Since strong overreflection occurs mostly when total reflection does, the parameter region
where strong overreflection occurs is rather narrow in the present configuration, though it is still in a readily observable range of the $(\theta,\phi)$ space.

We have also performed calculations for much larger values of the damping parameter up to $\eta=1$.
Up to $\eta\approx 0.1$, we have verified that there is no noticeable change in the position and the size of the region where $R$ is strongly amplified in the $(\theta,\phi)$ space.
For $\eta>0.1$, the position of this region is shifted, but its size remains substantially large.
We conclude that the phenomenon of giant overreflection persists even in the presence of realistic damping.

\begin{figure}
\centering\includegraphics[width=8.6cm]{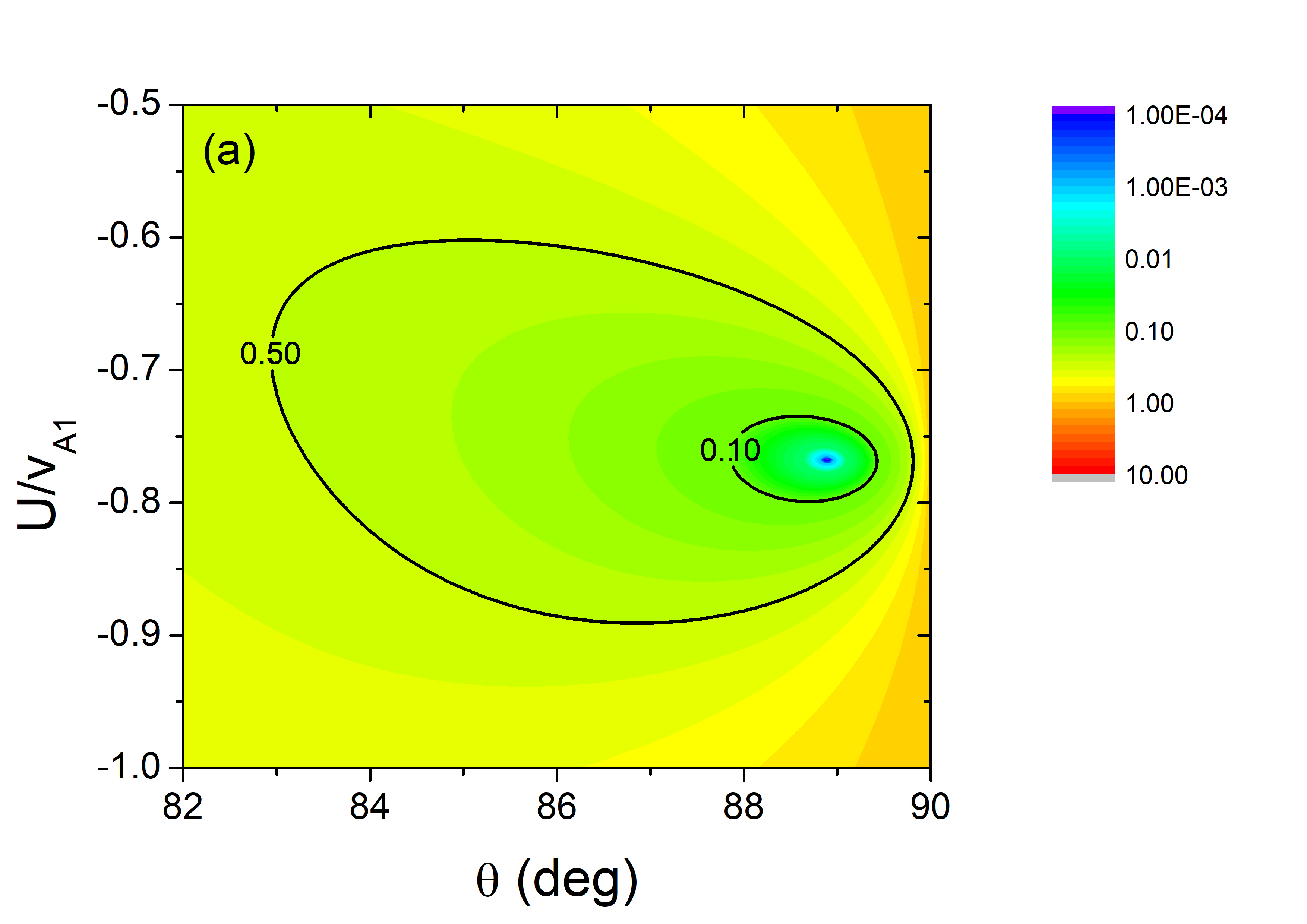}
\centering\includegraphics[width=8.6cm]{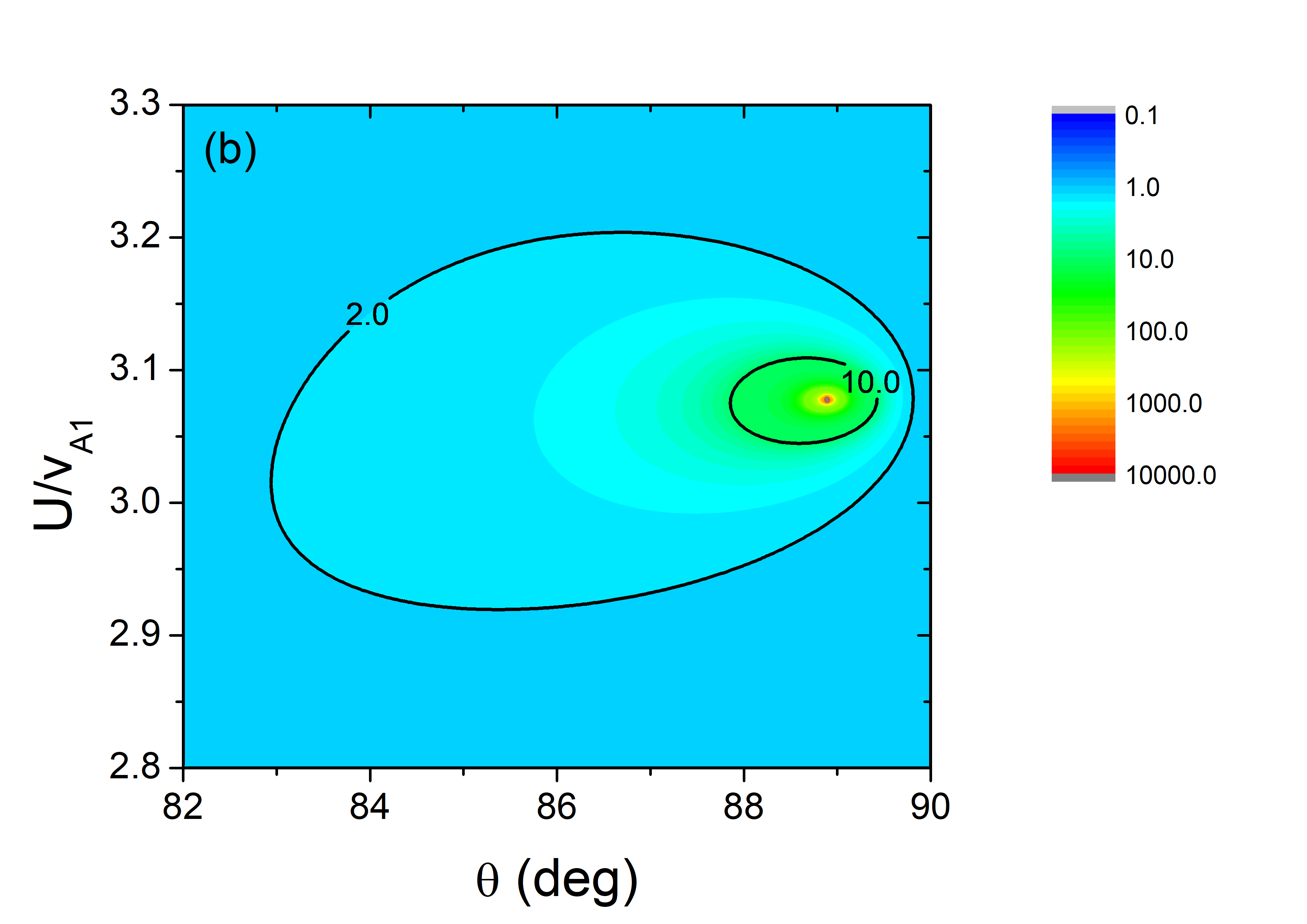}
\caption{Logarithmic contour plots of $R$ versus $\theta$ and $U/v_{A1}$ for the configuration given by Eq.~(\ref{eq:conf0}), when $\phi=30^\circ$, $\omega/\omega_0=\pi$, and $\beta=0$.
The regions where $R$ is (a) smaller than 0.5 and (b) larger than 2 are explicitly displayed.}
\label{fig_c0}
\end{figure}

\section{Mechanism of giant overreflection}
\label{sec_or}

From the results presented so far, we conclude that giant overreflection and amplification of incident magnetosonic waves is a generic and robust phenomenon in inhomogeneous
plasmas with nonuniform shear flows. We now attempt to explain the origin of this phenomenon. To provide a simpler and more transparent explanation,
it is beneficial to consider a configuration where the flow velocity is uniform in the half space, such as given by
\begin{eqnarray}
  && U_x(z)=\left\{ \begin{array}{ll}
  U,& \mbox{if } z\le L\\
  0, & \mbox{if } z>L
  \end{array} \right.,\nonumber\\
  && \frac{\rho_0(z)}{\rho_{01}}=\left\{ \begin{array}{ll}
  0.1,& \mbox{if } z<0\\
  0.9\frac{z}{L}+0.1,& \mbox{if } 0\le z\le L\\
  1, & \mbox{if } z>L
  \end{array} \right..
\label{eq:conf0}
\end{eqnarray}
We fix $\beta=0$, $\omega/\omega_0=\pi$, and $\phi=30^\circ$ and calculate $R$ and $T$ as a function of $U$ and $\theta$. In Fig.~\ref{fig_c0},
we show the logarithmic contour plots in the two regions where $R$ is smaller than 0.5 and larger than 2.
We observe that there is a perfect symmetry between the shapes of these regions.
This symmetry can be understood easily from the complex form of $\epsilon$ given in Eq.~(\ref{eq:epsc}).
There always exist a pair of $U$ values, $U_a$ and $U_b$,
for any configuration of $\rho_0$ and for any values of $\theta$ and $\phi$,
such that the corresponding values of $\epsilon$
are the complex conjugates of each other, that is, $\epsilon^*(U_a)=\epsilon(U_b)$. From Eq.~(\ref{eq:epsc}),
we obtain
\begin{eqnarray}
\frac{U_a}{v_{A1}}+\frac{U_b}{v_{A1}}=\frac{2}{\sin\theta\cos\phi},
\label{eq:symrel}
\end{eqnarray}
which explains the symmetry between Figs.~\ref{fig_c0}(a) and \ref{fig_c0}(b) very well.

The wave scattering coefficients for the medium with the scattering potential $\epsilon$ are closely related to those for the medium with the scattering potential $\epsilon^*$.
These relationships can be derived from scattering theory. For a given medium, one needs to consider the reflection and transmission coefficients, $r_R$ and $t_R$
($r_L$ and $t_L$) for the waves incident from the right (left). We only need the relationship
\begin{eqnarray}
\left[r_R\left(\epsilon^*\right)\right]^*=\frac{r_L\left(\epsilon\right)}{r_R\left(\epsilon\right)r_L\left(\epsilon\right)-t_R\left(\epsilon\right)t_L\left(\epsilon\right)},
\end{eqnarray}
which has been derived in Ref.~\onlinecite{39Rivero2019}.
For the consideration of the giant overreflection,
we are mainly interested in the case where the wave is evanescent in the transmitted region and
the transmission coefficients $t_R$ and $t_L$ vanish. Then we obtain a very simple relationship
between the reflectances for $\epsilon$ and $\epsilon^*$ of the form
\begin{eqnarray}
R\left(\epsilon^*\right)=\frac{1}{R\left(\epsilon\right)}.
\label{eq:refsym}
\end{eqnarray}
We have verified that this relationship is strictly satisfied for $U_a$ and $U_b$ satisfying Eq.~(\ref{eq:symrel}), as can be checked in Fig.~\ref{fig_c0}.

\begin{figure}
\centering\includegraphics[width=8.6cm]{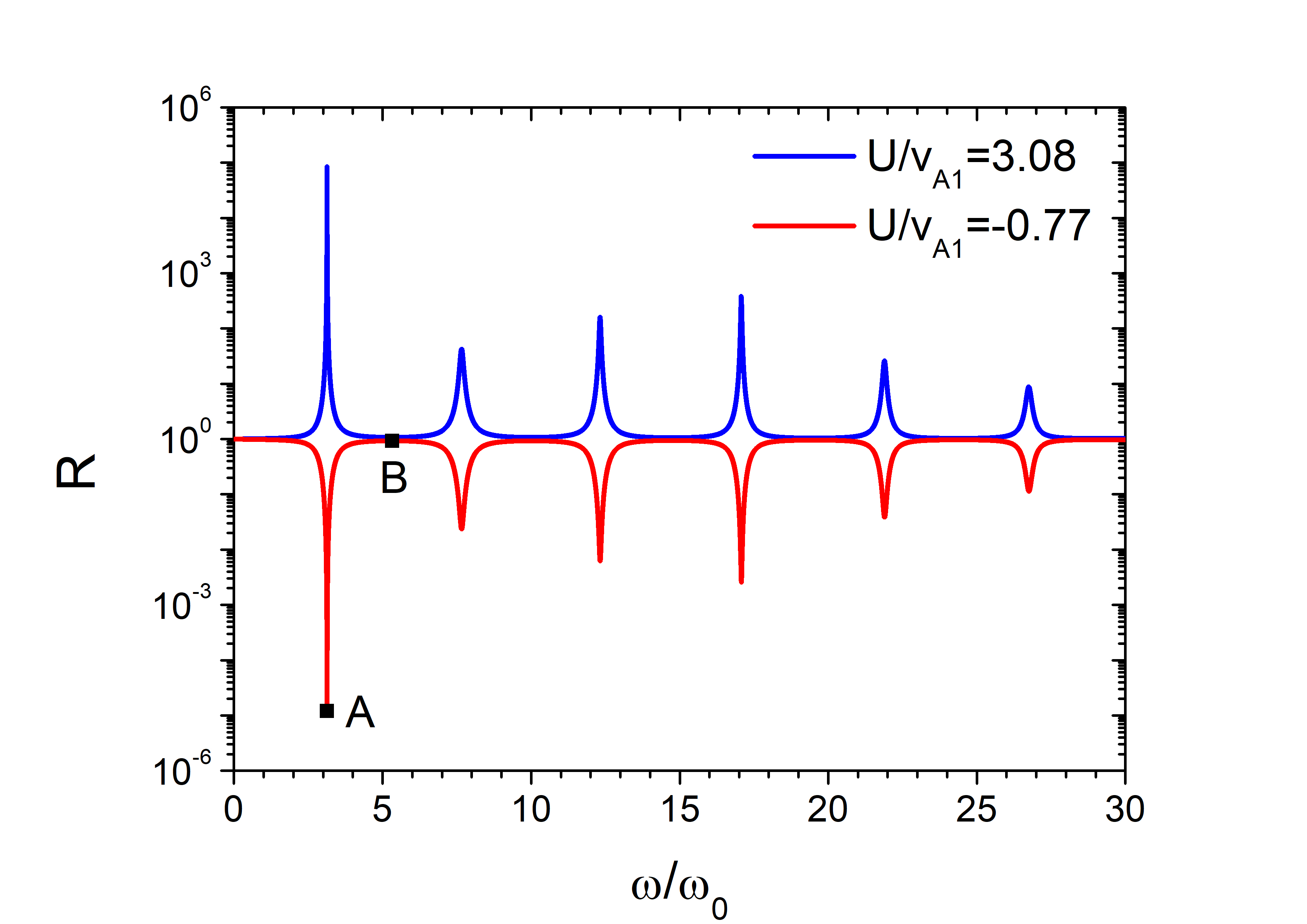}
\caption{Reflectance versus normalized frequency $\omega/\omega_0$ for the configuration given by Eq.~(\ref{eq:conf0}), when $\theta=88.89^\circ$, $\phi=30^\circ$, $\beta=0$, and $U/v_{A1}=3.08$, $-0.77$. The two reflectances are precisely the reciprocal of each other. The points A and B correspond to $\omega/\omega_0=3.13$ and 5.32 respectively.}
\label{fig_sym}
\end{figure}

In the configuration given by Eq.~(\ref{eq:conf0}) and when $\phi=30^\circ$ and $\beta=0$, the reflectance becomes extremely small and the plasma
behaves as a near-perfect absorber when $U/v_{A1}=-0.77$ and $\theta=88.89^\circ$.
For this case, the conjugate value of $U$ obtained from Eq.~(\ref{eq:symrel}) is $U/v_{A1}=3.08$.
In Fig.~\ref{fig_sym}, we show how the reflectance varies as a function of the frequency for these two
conjugate cases. We find that the two reflectances are precisely the reciprocal of each other in excellent agreement with Eq.~(\ref{eq:refsym}). It is intriguing that
{\it the near-perfect absorption and the giant overreflection arise as a pair of conjugate
phenomena}.
Many sharp peaks and dips appear in a regular pattern resembling a Fabry-Perot resonator in Fig.~\ref{fig_sym},
with approximately equal intervals between them.
This suggests clearly that some kind of resonance phenomenon associated with a cavity is taking place.

\begin{figure}
\centering\includegraphics[width=8.6cm]{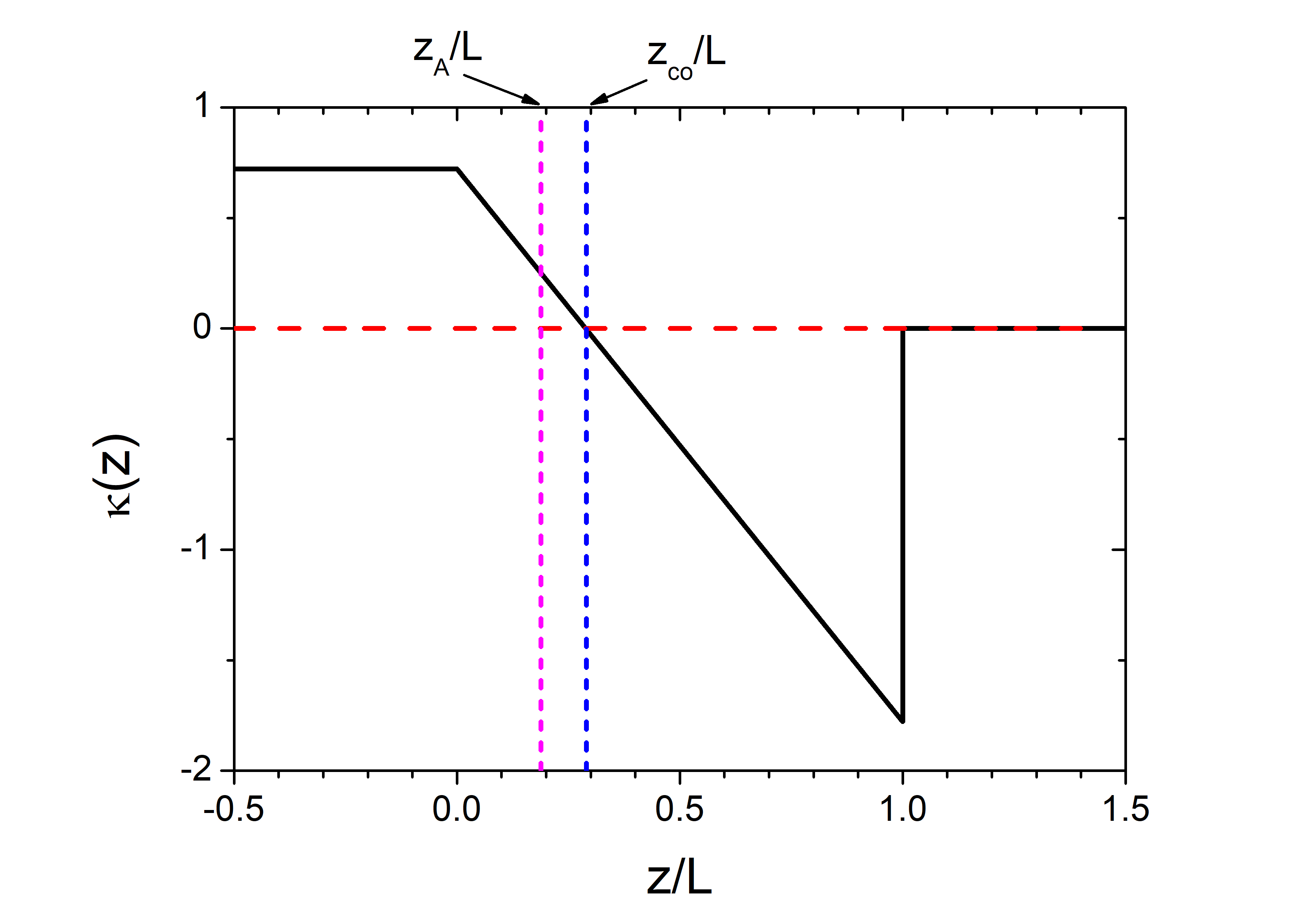}
\caption{The function $\kappa(z)$ [$\equiv\sin^2\theta\sin^2\phi-\epsilon(z)$], where $\epsilon(z)$ is given by Eq.~(\ref{eq:epsp}), plotted versus $z/L$
for the configuration given by Eq.~(\ref{eq:conf0}), when $\theta=88.89^\circ$, $\phi=30^\circ$, and $U/v_{A1}=3.08$ (or $-0.77$). The wave becomes evanescent
in the region $z<z_{\rm co}$ ($\approx 0.29L$) where $\kappa$ is positive.
$z_A$ ($\approx 0.19L$) denotes the position of the Alfv\'en resonance.}
\label{fig_pot}
\end{figure}

From the wave equation, Eq.~(\ref{eq:weq}), we notice that the quantity
\begin{eqnarray}
\kappa(z)\equiv\sin^2\theta\sin^2\phi-\epsilon(z)
\end{eqnarray}
can be considered as the negative square of the $z$ component of the wave vector. The wave becomes evanescent where $\kappa$ is positive.  In Fig.~\ref{fig_pot}, we plot $\kappa$
versus $z/L$ when $\theta=88.89^\circ$, $\phi=30^\circ$, and $U/v_{A1}=3.08$ (or $-0.77$). We note that $\kappa$ is the same for both $U/v_{A1}=3.08$ and $-0.77$. The wave is evanescent in the region $z<z_{\rm co}$ ($\approx 0.29L$). The position of the Alfv\'en resonance $z_A$ ($\approx 0.19L$) is located
inside the evanescent region. We point out that $\kappa$ plays the role of a scattering potential and an effective cavity is formed in the region $0.29<z/L<1$.
Note that this is an {\it open cavity} since the wave can propagate through one end at $z=L$. In the present case, however, a series of semi-bound states are well-formed, since the absolute value of $\kappa$
is very small in $z>L$. The formation of semi-bound states is the cause of both
extremely small reflectance and perfect absorption for $U/v_{A1}=-0.77$ and extremely large reflectance and amplification for $U/v_{A1}=3.08$.

\begin{figure}
\centering\includegraphics[width=8.6cm]{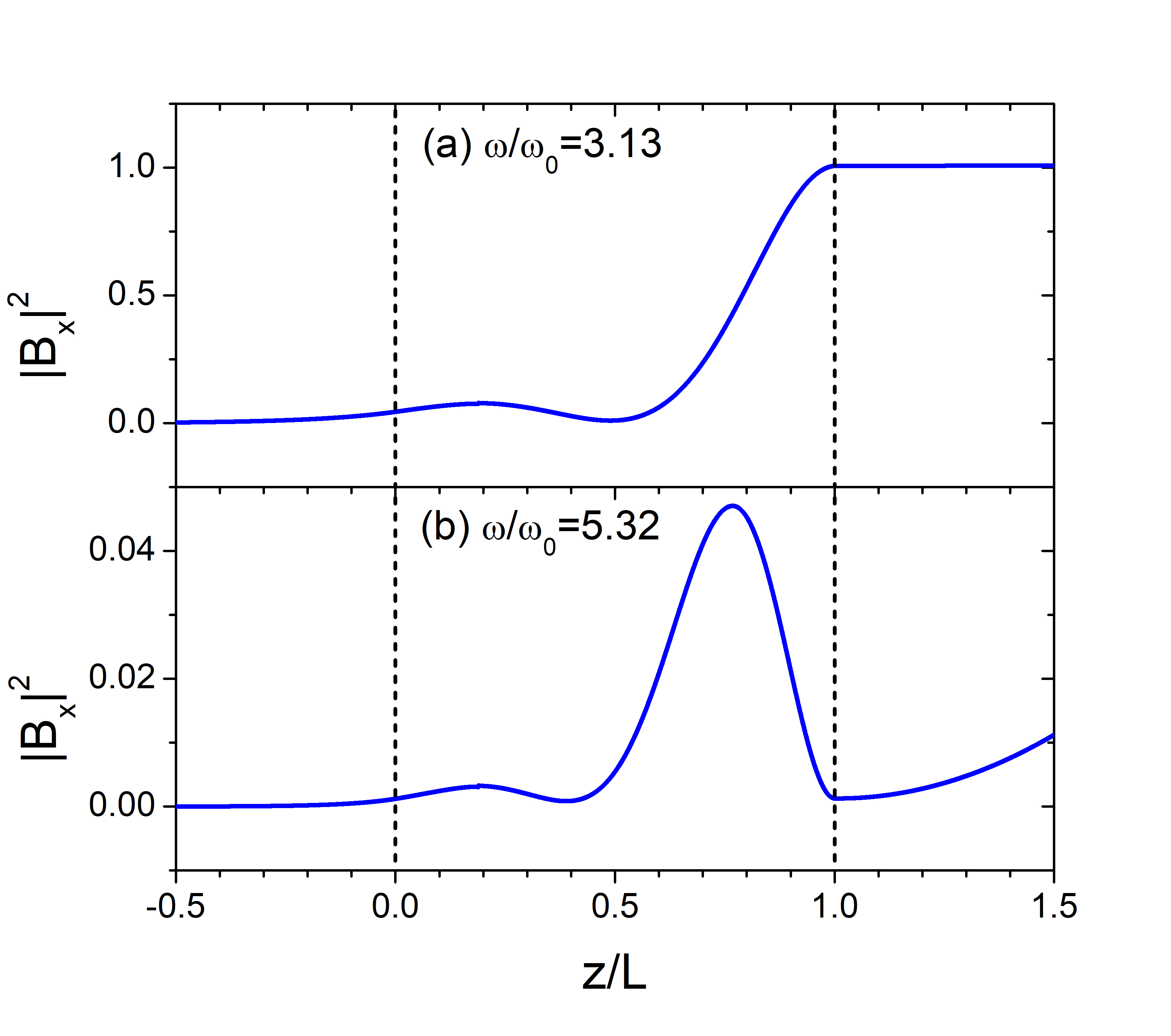}
\caption{Spatial distributions of the intensity of the $x$ component of the magnetic field, $\vert B_x\vert^2$, for the configuration given by Eq.~(\ref{eq:conf0}), when $\theta=88.89^\circ$, $\phi=30^\circ$, $\beta=0$, $U/v_{A1}=-0.77$, and (a) $\omega/\omega_0=3.13$ and (b) $\omega/\omega_0=5.32$,
which correspond to the points A and B in Fig.~\ref{fig_sym} respectively. The wave is assumed to be incident from the right side.
The vertical dashed lines denote the position of the inhomogeneous slab.}
\label{fig_field}
\end{figure}

In Fig.~\ref{fig_field}, we plot the spatial distributions of the intensity of the $x$ component of the magnetic field, $\vert B_x\vert^2$, obtained by solving Eq.~(\ref{eq:fieldb}),
when $\theta=88.89^\circ$, $\phi=30^\circ$, $U/v_{A1}=-0.77$, and $\omega/\omega_0=3.13$ and 5.32,
which correspond to the points A and B denoted in Fig.~\ref{fig_sym} respectively. When $\omega/\omega_0$ is 3.13, corresponding to the minimum reflectance close to zero,
the field profile shows an antinode at $z=L$, whereas, when $\omega/\omega_0$ is 5.32, corresponding to the reflectance close to unity, the field profile shows a node at $z=L$.
This result is fully consistent with the formation of a semi-bound state in an open cavity.

\section{Discussion}
\label{sec_dc}

Let us first estimate the possibility of observing the phenomena predicted in this paper in the terrestrial magnetosphere.
In the configuration given by Eq.~(\ref{eq:conf1}),
the range of the variation and the maximum value of the flow velocity are 2.5 and 5/3 times those of the Alfv\'en velocity respectively
and an incident wave propagates from a higher density region to a lower density region.
The flow speed is slightly super-Alfv\'enic in $0< z/L<2/3$ and sub-Alfv\'enic in $2/3< z/L<1$.
From the numerical results, strong overreflection with the reflectance greater than 10
is found to arise in a broad range of the parameter values when the wave frequency $f$ is roughly in the range of
\begin{eqnarray}
0.5f_0\lesssim f\lesssim 1.5f_0,
\end{eqnarray}
where the unit frequency $f_0$ is defined by
\begin{eqnarray}
f_0=\frac{\omega_0}{2\pi}=\frac{v_{A1}}{2\pi L}.
\end{eqnarray}
As an example, we choose $L$ to be the thickness of the dawnside magnetopause in the Earth's magnetosphere,
which has been reported to be about 1500 km \citep{40Haaland2019}. If we also assume, reasonably, that the Alfv\'en velocity
in the incident region is about 200 km/s, then $f_0$ is about 21 mHz and the frequency range of strong overreflection is between 10 and 32 mHz,
which is well within the region of Pc 3--4 magnetic pulsations.
Strong overreflection occurs also in the case where an incident wave propagates from a lower density region
to a higher density region and for many other different configurations.
This consideration appears to support the proposition that strong overreflection takes place widely in the magnetopause region
and may contribute to the excitation and enhancement
of ULF magnetic pulsations. We also speculate that this phenomenon could be an energy source of giant pulsations (Pg) with large field amplitudes \citep{41Wright2001}.

Similar estimates can be made for other astrophysical plasmas including the solar atmosphere.
In finite $\beta$ plasmas, both slow magnetosonic waves and slow resonances can contribute substantially to strong overreflection.
In the configuration given by Eq.~(\ref{eq:slow}) and considered in Fig.~\ref{fig_slow},
the range of the variation and the maximum value of the flow velocity are 7/3 and 7/4 times those of the Alfv\'en velocity respectively
and an incident wave propagates from a higher density region to a lower density region.
The flow speed is super-Alfv\'enic in $0< z/L<0.75$ and sub-Alfv\'enic in $0.75< z/L<1$.
From Fig.~\ref{fig_slow}, we observe that giant overreflection occurs
for a very wide range of the incident angles when slow magnetosonic waves are incident.
The plasma $\beta$ and the frequency used in the calculation are
$\beta=0.1$, which is a reasonable value for the lower part of the transition layer between the solar chromosphere and the corona, and $f=0.2f_0$. We notice that giant overreflection occurs at a considerably lower frequency range
for slow waves than for fast waves. If, for instance, we assume that $L$ is 500 km and $v_{A1}$ is 100 km/s, then $f_0$ is about
32 mHz. In that case, giant overreflection for slow waves is expected when the frequency is about 6 mHz
and the oscillation period is about 160 seconds.

Finally, we speculate that giant overreflection may provide an alternative route for plasma heating.
Overreflection causes the flow kinetic energy to be converted into the wave energy.
The amplified MHD waves will propagate through various regions of the plasma
and may induce resonant absorption or another overreflection.
Ultimately, the energy of the wave will be converted into heat by some dissipative mechanism.
This can be a multi-step route of plasma heating.

\section{Conclusion}
\label{sec_con}

In this paper,
we have investigated the mode conversion
and resonant overreflection of MHD waves in an inhomogeneous plasma with a nonuniform mass flow theoretically, using the IIM.
We have found that, especially when the parameters are such that the incident waves are totally reflected,
giant overreflection where the reflectance is much larger than 10 arises in a universal manner.
It occurs in a fairly broad range of the incident angles, the frequency,
and the plasma $\beta$
and occurs for many different types of the density and flow velocity profiles.
We have also found that in finite $\beta$ plasmas, slow magnetosonic waves and slow resonances cause
strong overreflection in a broader range of parameters than fast magnetosonic waves.
This result suggests that slow waves may play important roles in various
processes in space plasmas.

In the present work, we have been mainly concerned with the calculation of the reflectance and the transmittance
and presented only limited results of the spatial field distributions. However, it is straightforward
to calculate all the components of various fields using the IIM, which can provide valuable informations about
polarization, compression, vorticity, and other characteristics \cite{42Goossens2020, 43Goossens2021}.
Our method also allows to consider the more general case where the magnetic field as well as the plasma density and the flow velocity is inhomogeneous.
With such a generalization, one can study interesting subjects such as the magnetic reconnection region and
the coronal plume structure \cite{44Provornikova2018, 49Andries2001a}.
Finally, we point out that when a giant overreflection occurs,
some nonlinear effects can occur due to the large enhancement of the field amplitudes.
Future work in those directions will be of great interest.

\begin{acknowledgments}
This research was supported through a National Research Foundation of Korea Grant (NRF-2022R1F1A1074463)
funded by the Korean Government.
It was also supported by the Basic Science Research Program funded by the Ministry of Education (2021R1A6A1A10044950) and by
the Global Frontier Program (2014M3A6B3063708).
\end{acknowledgments}

\section*{Data Availability}
The data that support the findings of this study are available from the corresponding author upon reasonable request.

\appendix*

\section{Invariant imbedding method and the derivation of Eq.~(\ref{eq:mie})}
\label{app:iim}

The IIM is a technique for solving the boundary value problem for an arbitrary number of coupled
ordinary differential equations (ODEs) by transforming it into an equivalent initial value problem.
In the case of wave equations, this method allows to calculate the reflection and transmission coefficients
for an arbitrarily inhomogeneous stratified medium of thickness $L$ when a wave is incident from the outside
homogeneous region. It can also be used to calculate the field amplitudes inside the inhomogeneous medium.
In some sense, the IIM can be considered as a continuum version of the transfer matrix method that has been widely used in
many fields including optics and condensed matter physics.
In the simplest case of the IIM, the original set of equations is transformed to an equivalent set of ODEs where
the independent variable $l$ is the thickness of the inhomogeneous slab rather than the position inside it. For instance, $r(l)$ represents
the reflection coefficient for the medium of thickness $l$ where the region $l<z\le L$ is truncated
from the original medium of thickness $L$. The invariant imbedding equation for $r(l)$
relates it to $r(L)$ through the initial value problem. The initial condition $r(0)$ corresponds to the
situation where there is no inhomogeneous medium and is determined
from the Fresnel formula between the incident and transmitted regions.

Our derivation of the invariant imbedding equations follows closely the presentations given in Refs.~\onlinecite{32Klyatskin2005,33GOLBERG19751,34Kim2016}.
We are interested in solving a boundary value problem of an arbitrary number of
coupled first-order ODEs defined by
\begin{eqnarray}
&&\frac{d}{dz} {\bf u}(z)={\bf F}(z,{\bf u}(z)),~~ z \in [0,L],\label{eq:uuu} \\
&&g{\bf u}(0)+h{\bf u}(L)={\bf w},\label{eq:bc}
\end{eqnarray}
where $\bf u$, $\bf F$, and $\bf w$ are $N$-component vectors and $g$ and $h$ are constant $N\times N$ matrices.
We consider the function $\bf u$ as being parametrically dependent on $L$ and $\bf w$:
\begin{eqnarray}
{\bf u}(z)={\bf u}(z;L,{\bf w})
\end{eqnarray}
and introduce the two vectors
\begin{eqnarray}
{\bf R}(L,{\bf w})={\bf u}(L;L,{\bf w}),~~ {\bf S}(L,{\bf w})={\bf u}(0;L,{\bf w}),
\label{eq:rsd}
\end{eqnarray}
where ${\bf u}(L;L,{\bf w})$ and ${\bf u}(0;L,{\bf w})$ denote the values of $\bf u$ at $z=L$ and $z=0$, respectively,
when the system size is $L$.
These vectors will be related to the reflection and transmission coefficients.
The extra parameters $L$ and $\bf w$ are called imbedding parameters.

We take partial derivatives of Eq.~(\ref{eq:uuu}) with respect to the imbedding parameters $L$ and $\bf w$ and obtain
\begin{eqnarray}
\frac{d}{dz}\frac{\partial u_i(z;L,{\bf w})}{\partial L}&=&\frac{\partial F_i(z,{\bf u})}{\partial u_j}\frac{\partial u_j(z;L,{\bf w})}{\partial L},  \nonumber\\
\frac{d}{dz}\frac{\partial u_i(z;L,{\bf w})}{\partial w_k}&=&\frac{\partial F_i(z,{\bf u})}{\partial u_j}\frac{\partial u_j(z;L,{\bf w})}{\partial w_k},
\end{eqnarray}
where the summation over the repeated index $j$ is assumed.
Since these equations are similar in form to each other, their solutions have to be related by the linear equation
\begin{eqnarray}
\frac{\partial u_i(z;L,{\bf w})}{\partial L}=d_k(L,{\bf w}) \frac{\partial u_i(z;L,{\bf w})}{\partial w_k},
\label{eq:lam}
\end{eqnarray}
where the vector $\bf d$ needs to be determined.
From Eqs.~(\ref{eq:lam}) and (\ref{eq:bc}), we obtain
\begin{eqnarray}
&&g\frac{\partial {\bf u}(0;L,{\bf w})}{\partial L}+h\frac{\partial {\bf u}(z;L,{\bf w})}{\partial L}\bigg\vert_{z=L}\nonumber \\
&&~~~=d_k(L,{\bf w})\frac{\partial}{\partial w_k}\left[g{\bf u}(0;L,{\bf w})+h{\bf u}(L;L,{\bf w})\right] \nonumber \\
&&~~~=d_k(L,{\bf w})\frac{\partial {\bf w}}{\partial w_k}={\mathbf{d}}(L,{\bf w}).
\label{eq:der1}
\end{eqnarray}
Using the definition of ${\bf R}(L,{\bf w})$ in Eq.~(\ref{eq:rsd}), we also get
\begin{eqnarray}
&&\frac{\partial {\bf u}(z;L,{\bf w})}{\partial L}\bigg\vert_{z=L}=\frac{\partial {\bf u}(L;L,{\bf w})}{\partial L}-\frac{\partial {\bf u}(z;L,{\bf w})}{\partial z}\bigg\vert_{z=L} \nonumber \\
&&~~~=\frac{\partial {\bf R}(L,{\bf w})}{\partial L}-{\bf F}(L,{\bf R}(L,{\bf w})).
\label{eq:der2}
\end{eqnarray}
By combining Eqs.~(\ref{eq:der1}), (\ref{eq:der2}), and (\ref{eq:bc}) and using $\partial{\bf w}/\partial L=0$, we obtain
\begin{eqnarray}
{{\mathbf d}(L,{\bf w})}=-h{\bf F}(L,{\bf R}(L,{\bf w})).
\end{eqnarray}
This and Eq.~(\ref{eq:lam}) yield the invariant imbedding equation for the field amplitudes inside the inhomogeneous medium:
\begin{eqnarray}
\frac{\partial u_i(z;l,{\bf w})}{\partial l}=-h_{kj}F_j(l,{\bf R}(l,{\bf w}))\frac{\partial u_i(z;l,{\bf w})}{\partial w_k},
\label{eq:iif}
\end{eqnarray}
which needs to be integrated from $l=z$ to $l=L$
using the initial condition
\begin{equation}
{\bf u}(z;z,{\bf w})={\bf R}(z,{\bf w}).
\end{equation}
In the above equation, the first variable $z$ of $\bf u$ denotes the spatial coordinate, while its second $z$ variable
and the $z$ variable of $\bf R$ denote the thickness of the medium.

In order to solve Eq.~(\ref{eq:iif}) and obtain the fields for $0\le z\le L$, we need to
calculate the function ${\bf R}(l,{\bf w})$ for $0\le l \le L$ in advance to provide the necessary initial conditions.
From the identity
\begin{eqnarray}
\frac{\partial {\bf R}(L,{\bf w})}{\partial L}
=\frac{\partial {\bf u}(z;L,{\bf w})}{\partial z}\bigg\vert_{z=L}+\frac{\partial {\bf u}(z;L,{\bf w})}{\partial L}\bigg\vert_{z=L},
\end{eqnarray}
we obtain the invariant imbedding equation for $\bf R$:
\begin{eqnarray}
\frac{\partial {\bf R}(l,{\bf w})}{\partial l}
&=&{\bf F}(l,{\bf R}(l,{\bf w})) \nonumber\\
&&-h_{kj}F_j(l,{\bf R}(l,{\bf w}))\frac{\partial {\bf R}(l,{\bf w})}{\partial w_k}
\label{eq:iir}
\end{eqnarray}
in a straightforward manner using Eqs.~(\ref{eq:uuu}) and (\ref{eq:iif}).
This equation is integrated from $l=0$ to $l=L$
using the initial condition
\begin{equation}
{\bf R}(0,{\bf w})=(g+h)^{-1}{\bf w},
\end{equation}
which is obtained from Eq.~(\ref{eq:bc}) by setting $L=0$.
In a similar way, we obtain the invariant imbedding equation for $\bf S$:
\begin{eqnarray}
\frac{\partial {\bf S}(l,{\bf w})}{\partial l}=-h_{kj}F_j(l,{\bf R}(l,{\bf w}))\frac{\partial {\bf S}(l,{\bf w})}{\partial w_k}
\label{eq:iis}
\end{eqnarray}
with the initial condition
\begin{equation}
{\bf S}(0,{\bf w})=(g+h)^{-1}{\bf w}.
\end{equation}
This equation can be solved jointly with (\ref{eq:iir}).

Next we specialize to the case where the function ${\bf F}(z,{\bf u})$ is linear in $\bf u$ such that
\begin{eqnarray}
F_i(z,{\bf u}(z))=M_{ij}(z)u_j(z),
\end{eqnarray}
which covers a very broad range of linear wave equations.
Then it is self-consistent to assume that the function ${\bf u}$ is linear in $\bf w$:
\begin{eqnarray}
{\bf u}(z;l,{\bf w})={\tilde U}(z;l){\bf w},
\label{eq:iieu}
\end{eqnarray}
where $\tilde U$ is an $N\times N$ matrix.
From this, we also get
\begin{eqnarray}
&&{\bf R}(l,{\bf w})={\bf u}(l;l,{\bf w})={\tilde U}(l;l){\bf w}\equiv {\tilde R}(l){\bf w},\nonumber\\
&&{\bf S}(l,{\bf w})={\bf u}(0;l,{\bf w})={\tilde U}(0;l){\bf w}\equiv {\tilde S}(l){\bf w}.
\end{eqnarray}
Substituting Eq.~(\ref{eq:iieu}) into Eq.~(\ref{eq:iif}),
we obtain a differential equation for the matrix function ${\tilde U}$:
\begin{eqnarray}
\frac{\partial}{\partial l}{\tilde U}(z;l)=-{\tilde U}(z;l)hM(l){\tilde R}(l)
\label{eq:ieuu}
\end{eqnarray}
supplemented by the initial condition
\begin{eqnarray}
{\tilde U}(z;z)={\tilde R}(z).
\end{eqnarray}
The invariant imbedding equations for the matrices ${\tilde R}$ and $\tilde S$ are obtained similarly:
\begin{eqnarray}
\frac{d}{dl}{\tilde R}(l)&=&M(l){\tilde R}(l)-{\tilde R}(l)h M(l){\tilde R}(l),\nonumber\\
\frac{d}{dl}{\tilde S}(l)&=&-{\tilde S}(l)h M(l){\tilde R}(l),
\label{eq:imbed1}
\end{eqnarray}
together with the initial conditions
\begin{eqnarray}
{\tilde R}(0)={\tilde S}(0)=(g+h)^{-1}.
\label{eq:ic}
\end{eqnarray}
The invariant imbedding equations, Eqs.~(\ref{eq:ieuu}) and (\ref{eq:imbed1}), are the basis for solving a whole range of linear wave equations.

We now apply the IIM to Eq.~(\ref{eq:wee1}).
We are mainly interested in calculating the reflection and transmission coefficients $r$ and $t$ defined by Eq.~(\ref{eq:rt}).
We define ${\bf u}=(u_1,u_2)^{\rm T}$ so that
\begin{eqnarray}
u_1(z;L)=P(z;L),~~u_2(z;L)=\frac{1}{A(z)}\frac{d}{dz}u_1(z;L).
\end{eqnarray}
Then the wave equation, Eq.~(\ref{eq:wee1}), takes the form
\begin{eqnarray}
\frac{d{\bf u}}{dz}=M(z){\bf u},
\end{eqnarray}
where the $2\times 2$ matrix $M$ is given by
\begin{eqnarray}
M(z)=\begin{pmatrix} 0 & A(z) \\
     -\frac{C(z)}{D(z)} & 0
 \end{pmatrix}.
 \label{eq:mm}
\end{eqnarray}
We define the value of $A$ in the incident region ($z>L$) as $A_1$ and that in the transmitted region ($z<0$) as $A_2$.
At the boundaries of the inhomogeneous medium, we have
\begin{eqnarray}
u_1(0;L)&=&t(L),~~~u_1(L;L)=1+r(L),\nonumber\\
u_2(0;L)&=&\frac{-ik_{z2}t(L)}{A_2} =\frac{-ik_{z2}}{A_2} u_1(0;L),\nonumber\\
u_2(L;L)&=&\frac{ik_{z1}[r(L)-1]}{A_1}= \frac{ik_{z1}}{A_1} [u_1(L;L)-2].
\label{eq:bcw}
\end{eqnarray}
By rearranging Eq.~(\ref{eq:bcw}), we obtain
\begin{eqnarray}
g{\bf{S}}+h{\bf{R}}={\bf w},
\end{eqnarray}
where
\begin{eqnarray}
&&{\bf{S}}=\begin{pmatrix}u_1(0;L)\\u_2(0;L)\end{pmatrix},~~{\bf{R}}=\begin{pmatrix}u_1(L;L)\\u_2(L;L)\end{pmatrix},\nonumber\\
&&{\bf w}=\begin{pmatrix} 0\\w_2\end{pmatrix}=\begin{pmatrix} 0\\2ik_{z1}\end{pmatrix},\nonumber\\
&&g=\begin{pmatrix}ik_{z2} & A_2\\ 0&0\end{pmatrix},~~h=\begin{pmatrix} 0 & 0\\ ik_{z1} &-A_1\end{pmatrix}.
\label{eq:dh}
\end{eqnarray}
From the definitions of $\tilde R$ and $\tilde S$ and the form of $\bf w$,
we have
\begin{eqnarray}
&&\tilde R_{12}w_2=1+r(L),~~\tilde S_{12}w_2=t(L),\nonumber\\
&&\tilde R_{22}w_2=
\frac{ik_{z1}[r(L)-1]}{A_1}.
\label{eq:rtd}
\end{eqnarray}
The differential equations satisfied by $r$ and $t$ are obtained from
\begin{eqnarray}
\frac{dr}{dl}=\frac{d\tilde R_{12}}{dl}w_2,~~\frac{dt}{dl}=\frac{d\tilde S_{12}}{dl}w_2,
\end{eqnarray}
in which we use Eq.~(\ref{eq:imbed1}) and the expressions for $M$, $h$, and $w_2$ given by Eqs.~(\ref{eq:mm}) and (\ref{eq:dh}).
After tedious but straightforward calculations, we derive
the invariant imbedding equations for $r$ and $t$, Eq.~(\ref{eq:mie}).
The initial conditions for $r$ and $t$, Eq.~(\ref{eq:maic}), can be derived easily from Eq.~(\ref{eq:ic}) using
\begin{eqnarray}
\left[\left(g+h\right)^{-1}\right]_{12}w_2=t(0)=r(0)+1.
\end{eqnarray}
Similarly, the differential equation satisfied by the field $P(z;l)$ is obtained from
\begin{eqnarray}
\frac{\partial P(z;l)}{\partial l}=\frac{\partial\tilde U_{12}(z;l)}{\partial l}w_2,
\end{eqnarray}
in which we use Eq.~(\ref{eq:ieuu}) to derive the invariant imbedding equation for the field, Eq.~(\ref{eq:ief}).

\nocite{*}
\bibliography{MHD}

\begin{thebibliography}{51}%
\makeatletter
\providecommand \@ifxundefined [1]{%
 \@ifx{#1\undefined}
}%
\providecommand \@ifnum [1]{%
 \ifnum #1\expandafter \@firstoftwo
 \else \expandafter \@secondoftwo
 \fi
}%
\providecommand \@ifx [1]{%
 \ifx #1\expandafter \@firstoftwo
 \else \expandafter \@secondoftwo
 \fi
}%
\providecommand \natexlab [1]{#1}%
\providecommand \enquote  [1]{``#1''}%
\providecommand \bibnamefont  [1]{#1}%
\providecommand \bibfnamefont [1]{#1}%
\providecommand \citenamefont [1]{#1}%
\providecommand \href@noop [0]{\@secondoftwo}%
\providecommand \href [0]{\begingroup \@sanitize@url \@href}%
\providecommand \@href[1]{\@@startlink{#1}\@@href}%
\providecommand \@@href[1]{\endgroup#1\@@endlink}%
\providecommand \@sanitize@url [0]{\catcode `\\12\catcode `\$12\catcode
  `\&12\catcode `\#12\catcode `\^12\catcode `\_12\catcode `\%12\relax}%
\providecommand \@@startlink[1]{}%
\providecommand \@@endlink[0]{}%
\providecommand \url  [0]{\begingroup\@sanitize@url \@url }%
\providecommand \@url [1]{\endgroup\@href {#1}{\urlprefix }}%
\providecommand \urlprefix  [0]{URL }%
\providecommand \Eprint [0]{\href }%
\providecommand \doibase [0]{https://doi.org/}%
\providecommand \selectlanguage [0]{\@gobble}%
\providecommand \bibinfo  [0]{\@secondoftwo}%
\providecommand \bibfield  [0]{\@secondoftwo}%
\providecommand \translation [1]{[#1]}%
\providecommand \BibitemOpen [0]{}%
\providecommand \bibitemStop [0]{}%
\providecommand \bibitemNoStop [0]{.\EOS\space}%
\providecommand \EOS [0]{\spacefactor3000\relax}%
\providecommand \BibitemShut  [1]{\csname bibitem#1\endcsname}%
\let\auto@bib@innerbib\@empty
\bibitem [{\citenamefont {{Keiling}}, \citenamefont {{Lee}},\ and\
  \citenamefont {{Nakariakov}}(2016)}]{1Keiling2016}%
  \BibitemOpen
  \bibinfo {editor} {\bibfnamefont {A.}~\bibnamefont {{Keiling}}}, \bibinfo
  {editor} {\bibfnamefont {D.-H.}\ \bibnamefont {{Lee}}},\ and\ \bibinfo
  {editor} {\bibfnamefont {V.}~\bibnamefont {{Nakariakov}}},\ eds.,\ \href@noop
  {} {\emph {\bibinfo {title} {Low-Frequency Waves in Space Plasmas}}}\
  (\bibinfo  {publisher} {American Geophysical Union, Wiley},\ \bibinfo
  {address} {Washignton, D.C., Hoboken, NJ},\ \bibinfo {year}
  {2016})\BibitemShut {NoStop}%
\bibitem [{\citenamefont {Walker}(2005)}]{2Walker2005}%
  \BibitemOpen
  \bibfield  {author} {\bibinfo {author} {\bibfnamefont {A.~D.~M.}\
  \bibnamefont {Walker}},\ }\href@noop {} {\emph {\bibinfo {title}
  {Magnetohydrodynamic Waves in Geospace}}}\ (\bibinfo  {publisher} {Institute
  of Physics Publishing},\ \bibinfo {address} {Bristol},\ \bibinfo {year}
  {2005})\BibitemShut {NoStop}%
\bibitem [{\citenamefont {Roberts}(2019)}]{3Roberts2019}%
  \BibitemOpen
  \bibfield  {author} {\bibinfo {author} {\bibfnamefont {B.}~\bibnamefont
  {Roberts}},\ }\href@noop {} {\emph {\bibinfo {title} {MHD Waves in the Solar
  Atmosphere}}}\ (\bibinfo  {publisher} {Cambridge Univ. Press},\ \bibinfo
  {address} {Cambridge},\ \bibinfo {year} {2019})\BibitemShut {NoStop}%
\bibitem [{\citenamefont {Swanson}(1998)}]{4Swanson1998}%
  \BibitemOpen
  \bibfield  {author} {\bibinfo {author} {\bibfnamefont {D.~G.}\ \bibnamefont
  {Swanson}},\ }\href@noop {} {\emph {\bibinfo {title} {Theory of Mode
  Conversion and Tunneling in Inhomogeneous Plasmas}}}\ (\bibinfo  {publisher}
  {Wiley},\ \bibinfo {address} {New York},\ \bibinfo {year} {1998})\BibitemShut
  {NoStop}%
\bibitem [{\citenamefont {Forslund}\ \emph {et~al.}(1975)\citenamefont
  {Forslund}, \citenamefont {Kindel}, \citenamefont {Lee}, \citenamefont
  {Lindman},\ and\ \citenamefont {Morse}}]{5Forslund1975}%
  \BibitemOpen
  \bibfield  {author} {\bibinfo {author} {\bibfnamefont {D.~W.}\ \bibnamefont
  {Forslund}}, \bibinfo {author} {\bibfnamefont {J.~M.}\ \bibnamefont
  {Kindel}}, \bibinfo {author} {\bibfnamefont {K.}~\bibnamefont {Lee}},
  \bibinfo {author} {\bibfnamefont {E.~L.}\ \bibnamefont {Lindman}},\ and\
  \bibinfo {author} {\bibfnamefont {R.~L.}\ \bibnamefont {Morse}},\ }\bibfield
  {title} {\enquote {\bibinfo {title} {Theory and simulation of resonant
  absorption in a hot plasma},}\ }\href
  {https://doi.org/10.1103/PhysRevA.11.679} {\bibfield  {journal} {\bibinfo
  {journal} {Phys. Rev. A}\ }\textbf {\bibinfo {volume} {11}},\ \bibinfo
  {pages} {679--683} (\bibinfo {year} {1975})}\BibitemShut {NoStop}%
\bibitem [{\citenamefont {{Mj{\o}lhus}}(1990)}]{6Mjolhus1990}%
  \BibitemOpen
  \bibfield  {author} {\bibinfo {author} {\bibfnamefont {E.}~\bibnamefont
  {{Mj{\o}lhus}}},\ }\bibfield  {title} {\enquote {\bibinfo {title} {{On linear
  conversion in a magnetized plasma}},}\ }\href
  {https://doi.org/10.1029/RS025i006p01321} {\bibfield  {journal} {\bibinfo
  {journal} {Radio Sci.}\ }\textbf {\bibinfo {volume} {25}},\ \bibinfo {pages}
  {1321--1339} (\bibinfo {year} {1990})}\BibitemShut {NoStop}%
\bibitem [{\citenamefont {{Hinkel-Lipsker}}, \citenamefont {{Fried}},\ and\
  \citenamefont {{Morales}}(1992)}]{7Hinkel1992}%
  \BibitemOpen
  \bibfield  {author} {\bibinfo {author} {\bibfnamefont {D.~E.}\ \bibnamefont
  {{Hinkel-Lipsker}}}, \bibinfo {author} {\bibfnamefont {B.~D.}\ \bibnamefont
  {{Fried}}},\ and\ \bibinfo {author} {\bibfnamefont {G.~J.}\ \bibnamefont
  {{Morales}}},\ }\bibfield  {title} {\enquote {\bibinfo {title} {{Analytic
  expressions for mode conversion in a plasma with a linear density
  profile}},}\ }\href {https://doi.org/10.1063/1.860255} {\bibfield  {journal}
  {\bibinfo  {journal} {Phys. Fluids B}\ }\textbf {\bibinfo {volume} {4}},\
  \bibinfo {pages} {559--575} (\bibinfo {year} {1992})}\BibitemShut {NoStop}%
\bibitem [{\citenamefont {{Kim}}\ and\ \citenamefont {{Lee}}(2005)}]{8Kim2005}%
  \BibitemOpen
  \bibfield  {author} {\bibinfo {author} {\bibfnamefont {K.}~\bibnamefont
  {{Kim}}}\ and\ \bibinfo {author} {\bibfnamefont {D.-H.}\ \bibnamefont
  {{Lee}}},\ }\bibfield  {title} {\enquote {\bibinfo {title} {{Invariant
  imbedding theory of mode conversion in inhomogeneous plasmas. I. Exact
  calculation of the mode conversion coefficient in cold, unmagnetized
  plasmas}},}\ }\href {https://doi.org/10.1063/1.1914536} {\bibfield  {journal}
  {\bibinfo  {journal} {Phys. Plasmas}\ }\textbf {\bibinfo {volume} {12}},\
  \bibinfo {eid} {062101} (\bibinfo {year} {2005})}\BibitemShut {NoStop}%
\bibitem [{\citenamefont {{Kim}}\ and\ \citenamefont {{Lee}}(2006)}]{9Kim2006}%
  \BibitemOpen
  \bibfield  {author} {\bibinfo {author} {\bibfnamefont {K.}~\bibnamefont
  {{Kim}}}\ and\ \bibinfo {author} {\bibfnamefont {D.-H.}\ \bibnamefont
  {{Lee}}},\ }\bibfield  {title} {\enquote {\bibinfo {title} {{Invariant
  imbedding theory of mode conversion in inhomogeneous plasmas. II. Mode
  conversion in cold, magnetized plasmas with perpendicular inhomogeneity}},}\
  }\href {https://doi.org/10.1063/1.2186529} {\bibfield  {journal} {\bibinfo
  {journal} {Phys. Plasmas}\ }\textbf {\bibinfo {volume} {13}},\ \bibinfo {eid}
  {042103} (\bibinfo {year} {2006})}\BibitemShut {NoStop}%
\bibitem [{\citenamefont {Kim}, \citenamefont {Cairns},\ and\ \citenamefont
  {Robinson}(2007)}]{10Kim2007}%
  \BibitemOpen
  \bibfield  {author} {\bibinfo {author} {\bibfnamefont {E.-H.}\ \bibnamefont
  {Kim}}, \bibinfo {author} {\bibfnamefont {I.~H.}\ \bibnamefont {Cairns}},\
  and\ \bibinfo {author} {\bibfnamefont {P.~A.}\ \bibnamefont {Robinson}},\
  }\bibfield  {title} {\enquote {\bibinfo {title} {Extraordinary-mode radiation
  produced by linear-mode conversion of {Langmuir} waves},}\ }\href
  {https://doi.org/10.1103/PhysRevLett.99.015003} {\bibfield  {journal}
  {\bibinfo  {journal} {Phys. Rev. Lett.}\ }\textbf {\bibinfo {volume} {99}},\
  \bibinfo {pages} {015003} (\bibinfo {year} {2007})}\BibitemShut {NoStop}%
\bibitem [{\citenamefont {{McDougall}}\ and\ \citenamefont
  {{Hood}}(2007)}]{11McDougall2007}%
  \BibitemOpen
  \bibfield  {author} {\bibinfo {author} {\bibfnamefont {A.~M.~D.}\
  \bibnamefont {{McDougall}}}\ and\ \bibinfo {author} {\bibfnamefont {A.~W.}\
  \bibnamefont {{Hood}}},\ }\bibfield  {title} {\enquote {\bibinfo {title} {{A
  new look at mode conversion in a stratified isothermal atmosphere}},}\ }\href
  {https://doi.org/10.1007/s11207-007-0393-5} {\bibfield  {journal} {\bibinfo
  {journal} {Sol. Phys.}\ }\textbf {\bibinfo {volume} {246}},\ \bibinfo {pages}
  {259--271} (\bibinfo {year} {2007})}\BibitemShut {NoStop}%
\bibitem [{\citenamefont {{Lee}}\ \emph {et~al.}(2008)\citenamefont {{Lee}},
  \citenamefont {{Johnson}}, \citenamefont {{Kim}},\ and\ \citenamefont
  {{Kim}}}]{12Lee2008}%
  \BibitemOpen
  \bibfield  {author} {\bibinfo {author} {\bibfnamefont {D.-H.}\ \bibnamefont
  {{Lee}}}, \bibinfo {author} {\bibfnamefont {J.~R.}\ \bibnamefont
  {{Johnson}}}, \bibinfo {author} {\bibfnamefont {K.}~\bibnamefont {{Kim}}},\
  and\ \bibinfo {author} {\bibfnamefont {K.~S.}\ \bibnamefont {{Kim}}},\
  }\bibfield  {title} {\enquote {\bibinfo {title} {{Effects of heavy ions on
  ULF wave resonances near the equatorial region}},}\ }\href
  {https://doi.org/10.1029/2008JA013088} {\bibfield  {journal} {\bibinfo
  {journal} {J. Geophys. Res.}\ }\textbf {\bibinfo {volume} {113}},\ \bibinfo
  {eid} {A11212} (\bibinfo {year} {2008})}\BibitemShut {NoStop}%
\bibitem [{\citenamefont {{Yu}}\ and\ \citenamefont {{Van
  Doorsselaere}}(2016)}]{13Yu2016}%
  \BibitemOpen
  \bibfield  {author} {\bibinfo {author} {\bibfnamefont {D.~J.}\ \bibnamefont
  {{Yu}}}\ and\ \bibinfo {author} {\bibfnamefont {T.}~\bibnamefont {{Van
  Doorsselaere}}},\ }\bibfield  {title} {\enquote {\bibinfo {title} {{A study
  on the excitation and resonant absorption of coronal loop kink
  oscillations}},}\ }\href {https://doi.org/10.3847/0004-637X/831/1/30}
  {\bibfield  {journal} {\bibinfo  {journal} {Astrophys. J.}\ }\textbf
  {\bibinfo {volume} {831}},\ \bibinfo {eid} {30} (\bibinfo {year}
  {2016})}\BibitemShut {NoStop}%
\bibitem [{\citenamefont {{Chen}}\ and\ \citenamefont
  {{Hasegawa}}(1974{\natexlab{a}})}]{14Chen1974}%
  \BibitemOpen
  \bibfield  {author} {\bibinfo {author} {\bibfnamefont {L.}~\bibnamefont
  {{Chen}}}\ and\ \bibinfo {author} {\bibfnamefont {A.}~\bibnamefont
  {{Hasegawa}}},\ }\bibfield  {title} {\enquote {\bibinfo {title} {{Plasma
  heating by spatial resonance of Alfv{\'e}n wave}},}\ }\href
  {https://doi.org/10.1063/1.1694904} {\bibfield  {journal} {\bibinfo
  {journal} {Phys. Fluids}\ }\textbf {\bibinfo {volume} {17}},\ \bibinfo
  {pages} {1399--1403} (\bibinfo {year} {1974}{\natexlab{a}})}\BibitemShut
  {NoStop}%
\bibitem [{\citenamefont {{Southwood}}(1974)}]{15Southwood1974}%
  \BibitemOpen
  \bibfield  {author} {\bibinfo {author} {\bibfnamefont {D.~J.}\ \bibnamefont
  {{Southwood}}},\ }\bibfield  {title} {\enquote {\bibinfo {title} {{Some
  features of field line resonances in the magnetosphere}},}\ }\href
  {https://doi.org/10.1016/0032-0633(74)90078-6} {\bibfield  {journal}
  {\bibinfo  {journal} {Planet. Space Sci.}\ }\textbf {\bibinfo {volume}
  {22}},\ \bibinfo {pages} {483--491} (\bibinfo {year} {1974})}\BibitemShut
  {NoStop}%
\bibitem [{\citenamefont {{Leonovich}}\ and\ \citenamefont
  {{Kozlov}}(2013)}]{16Leonovich2013}%
  \BibitemOpen
  \bibfield  {author} {\bibinfo {author} {\bibfnamefont {A.~S.}\ \bibnamefont
  {{Leonovich}}}\ and\ \bibinfo {author} {\bibfnamefont {D.~A.}\ \bibnamefont
  {{Kozlov}}},\ }\bibfield  {title} {\enquote {\bibinfo {title} {{Magnetosonic
  resonances in the magnetospheric plasma}},}\ }\href
  {https://doi.org/10.5047/eps.2012.07.002} {\bibfield  {journal} {\bibinfo
  {journal} {Earth Planets Space}\ }\textbf {\bibinfo {volume} {65}},\ \bibinfo
  {pages} {369--384} (\bibinfo {year} {2013})}\BibitemShut {NoStop}%
\bibitem [{\citenamefont {{Lee}}\ \emph {et~al.}(2002)\citenamefont {{Lee}},
  \citenamefont {{Hudson}}, \citenamefont {{Kim}}, \citenamefont {{Lysak}},\
  and\ \citenamefont {{Song}}}]{17Lee2002}%
  \BibitemOpen
  \bibfield  {author} {\bibinfo {author} {\bibfnamefont {D.-H.}\ \bibnamefont
  {{Lee}}}, \bibinfo {author} {\bibfnamefont {M.~K.}\ \bibnamefont {{Hudson}}},
  \bibinfo {author} {\bibfnamefont {K.}~\bibnamefont {{Kim}}}, \bibinfo
  {author} {\bibfnamefont {R.~L.}\ \bibnamefont {{Lysak}}},\ and\ \bibinfo
  {author} {\bibfnamefont {Y.}~\bibnamefont {{Song}}},\ }\bibfield  {title}
  {\enquote {\bibinfo {title} {{Compressional MHD wave transport in the
  magnetosphere 1. Reflection and transmission across the plasmapause}},}\
  }\href {https://doi.org/10.1029/2002JA009239} {\bibfield  {journal} {\bibinfo
   {journal} {J. Geophys. Res.}\ }\textbf {\bibinfo {volume} {107}},\ \bibinfo
  {eid} {1307} (\bibinfo {year} {2002})}\BibitemShut {NoStop}%
\bibitem [{\citenamefont {{Lysak}}(2022)}]{18Lysak2022}%
  \BibitemOpen
  \bibfield  {author} {\bibinfo {author} {\bibfnamefont {R.~L.}\ \bibnamefont
  {{Lysak}}},\ }\bibfield  {title} {\enquote {\bibinfo {title} {{Field line
  resonances and cavity modes at Earth and Jupiter}},}\ }\href
  {https://doi.org/10.3389/fspas.2022.913554} {\bibfield  {journal} {\bibinfo
  {journal} {Front. Astron. Space Sci.}\ }\textbf {\bibinfo {volume} {9}},\
  \bibinfo {eid} {913554} (\bibinfo {year} {2022})}\BibitemShut {NoStop}%
\bibitem [{\citenamefont {{Chen}}\ and\ \citenamefont
  {{Hasegawa}}(1974{\natexlab{b}})}]{19Chen1974}%
  \BibitemOpen
  \bibfield  {author} {\bibinfo {author} {\bibfnamefont {L.}~\bibnamefont
  {{Chen}}}\ and\ \bibinfo {author} {\bibfnamefont {A.}~\bibnamefont
  {{Hasegawa}}},\ }\bibfield  {title} {\enquote {\bibinfo {title} {{A theory of
  long-period magnetic pulsations: 1. Steady state excitation of field line
  resonance}},}\ }\href {https://doi.org/10.1029/JA079i007p01024} {\bibfield
  {journal} {\bibinfo  {journal} {J. Geophys. Res.}\ }\textbf {\bibinfo
  {volume} {79}},\ \bibinfo {pages} {1024--1032} (\bibinfo {year}
  {1974}{\natexlab{b}})}\BibitemShut {NoStop}%
\bibitem [{\citenamefont {{Nakariakov}}\ \emph {et~al.}(2016)\citenamefont
  {{Nakariakov}}, \citenamefont {{Pilipenko}}, \citenamefont {{Heilig}},
  \citenamefont {{Jel{\'\i}nek}}, \citenamefont {{Karlick{\'y}}}, \citenamefont
  {{Klimushkin}}, \citenamefont {{Kolotkov}}, \citenamefont {{Lee}},
  \citenamefont {{Nistic{\`o}}}, \citenamefont {{Van Doorsselaere}},
  \citenamefont {{Verth}},\ and\ \citenamefont
  {{Zimovets}}}]{20Nakariakov2016}%
  \BibitemOpen
  \bibfield  {author} {\bibinfo {author} {\bibfnamefont {V.~M.}\ \bibnamefont
  {{Nakariakov}}}, \bibinfo {author} {\bibfnamefont {V.}~\bibnamefont
  {{Pilipenko}}}, \bibinfo {author} {\bibfnamefont {B.}~\bibnamefont
  {{Heilig}}}, \bibinfo {author} {\bibfnamefont {P.}~\bibnamefont
  {{Jel{\'\i}nek}}}, \bibinfo {author} {\bibfnamefont {M.}~\bibnamefont
  {{Karlick{\'y}}}}, \bibinfo {author} {\bibfnamefont {D.~Y.}\ \bibnamefont
  {{Klimushkin}}}, \bibinfo {author} {\bibfnamefont {D.~Y.}\ \bibnamefont
  {{Kolotkov}}}, \bibinfo {author} {\bibfnamefont {D.-H.}\ \bibnamefont
  {{Lee}}}, \bibinfo {author} {\bibfnamefont {G.}~\bibnamefont
  {{Nistic{\`o}}}}, \bibinfo {author} {\bibfnamefont {T.}~\bibnamefont {{Van
  Doorsselaere}}}, \bibinfo {author} {\bibfnamefont {G.}~\bibnamefont
  {{Verth}}},\ and\ \bibinfo {author} {\bibfnamefont {I.~V.}\ \bibnamefont
  {{Zimovets}}},\ }\bibfield  {title} {\enquote {\bibinfo {title}
  {{Magnetohydrodynamic oscillations in the solar corona and Earth's
  magnetosphere: Towards consolidated understanding}},}\ }\href
  {https://doi.org/10.1007/s11214-015-0233-0} {\bibfield  {journal} {\bibinfo
  {journal} {Space Sci. Rev.}\ }\textbf {\bibinfo {volume} {200}},\ \bibinfo
  {pages} {75--203} (\bibinfo {year} {2016})}\BibitemShut {NoStop}%
\bibitem [{\citenamefont {{Nakariakov}}\ and\ \citenamefont
  {{Kolotkov}}(2020)}]{21Nakariakov2020}%
  \BibitemOpen
  \bibfield  {author} {\bibinfo {author} {\bibfnamefont {V.~M.}\ \bibnamefont
  {{Nakariakov}}}\ and\ \bibinfo {author} {\bibfnamefont {D.~Y.}\ \bibnamefont
  {{Kolotkov}}},\ }\bibfield  {title} {\enquote {\bibinfo {title}
  {{Magnetohydrodynamic waves in the solar corona}},}\ }\href
  {https://doi.org/10.1146/annurev-astro-032320-042940} {\bibfield  {journal}
  {\bibinfo  {journal} {Annu. Rev. Astron. Astrophys.}\ }\textbf {\bibinfo
  {volume} {58}},\ \bibinfo {pages} {441--481} (\bibinfo {year}
  {2020})}\BibitemShut {NoStop}%
\bibitem [{\citenamefont {{Van Doorsselaere}}\ \emph
  {et~al.}(2020)\citenamefont {{Van Doorsselaere}}, \citenamefont
  {{Srivastava}}, \citenamefont {{Antolin}}, \citenamefont {{Magyar}},
  \citenamefont {{Vasheghani Farahani}}, \citenamefont {{Tian}}, \citenamefont
  {{Kolotkov}}, \citenamefont {{Ofman}}, \citenamefont {{Guo}}, \citenamefont
  {{Arregui}}, \citenamefont {{De Moortel}},\ and\ \citenamefont
  {{Pascoe}}}]{22VanDoorsselaere2020}%
  \BibitemOpen
  \bibfield  {author} {\bibinfo {author} {\bibfnamefont {T.}~\bibnamefont {{Van
  Doorsselaere}}}, \bibinfo {author} {\bibfnamefont {A.~K.}\ \bibnamefont
  {{Srivastava}}}, \bibinfo {author} {\bibfnamefont {P.}~\bibnamefont
  {{Antolin}}}, \bibinfo {author} {\bibfnamefont {N.}~\bibnamefont {{Magyar}}},
  \bibinfo {author} {\bibfnamefont {S.}~\bibnamefont {{Vasheghani Farahani}}},
  \bibinfo {author} {\bibfnamefont {H.}~\bibnamefont {{Tian}}}, \bibinfo
  {author} {\bibfnamefont {D.}~\bibnamefont {{Kolotkov}}}, \bibinfo {author}
  {\bibfnamefont {L.}~\bibnamefont {{Ofman}}}, \bibinfo {author} {\bibfnamefont
  {M.}~\bibnamefont {{Guo}}}, \bibinfo {author} {\bibfnamefont
  {I.}~\bibnamefont {{Arregui}}}, \bibinfo {author} {\bibfnamefont
  {I.}~\bibnamefont {{De Moortel}}},\ and\ \bibinfo {author} {\bibfnamefont
  {D.}~\bibnamefont {{Pascoe}}},\ }\bibfield  {title} {\enquote {\bibinfo
  {title} {{Coronal heating by MHD Waves}},}\ }\href
  {https://doi.org/10.1007/s11214-020-00770-y} {\bibfield  {journal} {\bibinfo
  {journal} {Space Sci. Rev.}\ }\textbf {\bibinfo {volume} {216}},\ \bibinfo
  {eid} {140} (\bibinfo {year} {2020})}\BibitemShut {NoStop}%
\bibitem [{\citenamefont {{Walker}}(1998)}]{23Walker1998}%
  \BibitemOpen
  \bibfield  {author} {\bibinfo {author} {\bibfnamefont {A.~D.~M.}\
  \bibnamefont {{Walker}}},\ }\bibfield  {title} {\enquote {\bibinfo {title}
  {{Excitation of magnetohydrodynamic cavities in the magnetosphere}},}\ }\href
  {https://doi.org/10.1016/S1364-6826(98)00077-7} {\bibfield  {journal}
  {\bibinfo  {journal} {J. Atmos. Sol. Terr. Phys.}\ }\textbf {\bibinfo
  {volume} {60}},\ \bibinfo {pages} {1279--1293} (\bibinfo {year}
  {1998})}\BibitemShut {NoStop}%
\bibitem [{\citenamefont {{Walker}}(2000)}]{24Walker2000}%
  \BibitemOpen
  \bibfield  {author} {\bibinfo {author} {\bibfnamefont {A.~D.~M.}\
  \bibnamefont {{Walker}}},\ }\bibfield  {title} {\enquote {\bibinfo {title}
  {{Coupling between waveguide modes and field line resonances}},}\ }\href
  {https://doi.org/10.1016/S1364-6826(00)00046-8} {\bibfield  {journal}
  {\bibinfo  {journal} {J. Atmos. Sol. Terr. Phys.}\ }\textbf {\bibinfo
  {volume} {62}},\ \bibinfo {pages} {799--813} (\bibinfo {year}
  {2000})}\BibitemShut {NoStop}%
\bibitem [{\citenamefont {{Mazur}}\ and\ \citenamefont
  {{Chuiko}}(2013)}]{25Mazur2013}%
  \BibitemOpen
  \bibfield  {author} {\bibinfo {author} {\bibfnamefont {V.~A.}\ \bibnamefont
  {{Mazur}}}\ and\ \bibinfo {author} {\bibfnamefont {D.~A.}\ \bibnamefont
  {{Chuiko}}},\ }\bibfield  {title} {\enquote {\bibinfo {title} {{Influence of
  the outer-magnetospheric magnetohydrodynamic waveguide on the reflection of
  hydromagnetic waves from a shear flow at the magnetopause}},}\ }\href
  {https://doi.org/10.1134/S1063780X13120064} {\bibfield  {journal} {\bibinfo
  {journal} {Plasma Phys. Rep.}\ }\textbf {\bibinfo {volume} {39}},\ \bibinfo
  {pages} {959--975} (\bibinfo {year} {2013})}\BibitemShut {NoStop}%
\bibitem [{\citenamefont {{McKenzie}}(1970)}]{26McKenzie1970}%
  \BibitemOpen
  \bibfield  {author} {\bibinfo {author} {\bibfnamefont {J.~F.}\ \bibnamefont
  {{McKenzie}}},\ }\bibfield  {title} {\enquote {\bibinfo {title}
  {{Hydromagnetic wave interaction with the magnetopause and the bow shock}},}\
  }\href {https://doi.org/10.1016/0032-0633(70)90063-2} {\bibfield  {journal}
  {\bibinfo  {journal} {Planet. Space Sci.}\ }\textbf {\bibinfo {volume}
  {18}},\ \bibinfo {pages} {1--23} (\bibinfo {year} {1970})}\BibitemShut
  {NoStop}%
\bibitem [{\citenamefont {{Mann}}\ \emph {et~al.}(1999)\citenamefont {{Mann}},
  \citenamefont {{Wright}}, \citenamefont {{Mills}},\ and\ \citenamefont
  {{Nakariakov}}}]{28Mann1999}%
  \BibitemOpen
  \bibfield  {author} {\bibinfo {author} {\bibfnamefont {I.~R.}\ \bibnamefont
  {{Mann}}}, \bibinfo {author} {\bibfnamefont {A.~N.}\ \bibnamefont
  {{Wright}}}, \bibinfo {author} {\bibfnamefont {K.~J.}\ \bibnamefont
  {{Mills}}},\ and\ \bibinfo {author} {\bibfnamefont {V.~M.}\ \bibnamefont
  {{Nakariakov}}},\ }\bibfield  {title} {\enquote {\bibinfo {title}
  {{Excitation of magnetospheric waveguide modes by magnetosheath flows}},}\
  }\href {https://doi.org/10.1029/1998JA900026} {\bibfield  {journal} {\bibinfo
   {journal} {J. Geophys. Res.}\ }\textbf {\bibinfo {volume} {104}},\ \bibinfo
  {pages} {333--354} (\bibinfo {year} {1999})}\BibitemShut {NoStop}%
\bibitem [{\citenamefont {{Goossens}}, \citenamefont {{Hollweg}},\ and\
  \citenamefont {{Sakurai}}(1992)}]{48Goossens1992}%
  \BibitemOpen
  \bibfield  {author} {\bibinfo {author} {\bibfnamefont {M.}~\bibnamefont
  {{Goossens}}}, \bibinfo {author} {\bibfnamefont {J.~V.}\ \bibnamefont
  {{Hollweg}}},\ and\ \bibinfo {author} {\bibfnamefont {T.}~\bibnamefont
  {{Sakurai}}},\ }\bibfield  {title} {\enquote {\bibinfo {title} {{Resonant
  behaviour of magnetohydrodynamic waves on magnetic flux tubes - Part
  Three}},}\ }\href {https://doi.org/10.1007/BF00151914} {\bibfield  {journal}
  {\bibinfo  {journal} {Sol. Phys.}\ }\textbf {\bibinfo {volume} {138}},\
  \bibinfo {pages} {233--255} (\bibinfo {year} {1992})}\BibitemShut {NoStop}%
\bibitem [{\citenamefont {{Cs{\'\i}k}}, \citenamefont {{{\v{C}}ade{\v{z}}}},\
  and\ \citenamefont {{Goossens}}(1998)}]{29Csik1998}%
  \BibitemOpen
  \bibfield  {author} {\bibinfo {author} {\bibfnamefont {{\'A}.~T.}\
  \bibnamefont {{Cs{\'\i}k}}}, \bibinfo {author} {\bibfnamefont {V.~M.}\
  \bibnamefont {{{\v{C}}ade{\v{z}}}}},\ and\ \bibinfo {author} {\bibfnamefont
  {M.}~\bibnamefont {{Goossens}}},\ }\bibfield  {title} {\enquote {\bibinfo
  {title} {{Effects of mass flow on resonant absorption and on over-reflection
  of magnetosonic waves in low $\beta$ solar plasmas}},}\ }\href@noop {}
  {\bibfield  {journal} {\bibinfo  {journal} {Astron. Astrophys.}\ }\textbf
  {\bibinfo {volume} {339}},\ \bibinfo {pages} {215--224} (\bibinfo {year}
  {1998})}\BibitemShut {NoStop}%
\bibitem [{\citenamefont {{Cs{\'\i}k}}, \citenamefont {{{\v{C}}ade{\v{z}}}},\
  and\ \citenamefont {{Goossens}}(2000)}]{30Csik2000}%
  \BibitemOpen
  \bibfield  {author} {\bibinfo {author} {\bibfnamefont {{\'A}.~T.}\
  \bibnamefont {{Cs{\'\i}k}}}, \bibinfo {author} {\bibfnamefont {V.~M.}\
  \bibnamefont {{{\v{C}}ade{\v{z}}}}},\ and\ \bibinfo {author} {\bibfnamefont
  {M.}~\bibnamefont {{Goossens}}},\ }\bibfield  {title} {\enquote {\bibinfo
  {title} {{Frequency dependence of resonant absorption and over-reflection of
  magnetosonic waves in nonuniform structures with shear mass flow}},}\
  }\href@noop {} {\bibfield  {journal} {\bibinfo  {journal} {Astron.
  Astrophys.}\ }\textbf {\bibinfo {volume} {358}},\ \bibinfo {pages}
  {1090--1096} (\bibinfo {year} {2000})}\BibitemShut {NoStop}%
\bibitem [{\citenamefont {{Andries}}, \citenamefont {{Tirry}},\ and\
  \citenamefont {{Goossens}}(2000)}]{46Andries2000}%
  \BibitemOpen
  \bibfield  {author} {\bibinfo {author} {\bibfnamefont {J.}~\bibnamefont
  {{Andries}}}, \bibinfo {author} {\bibfnamefont {W.~J.}\ \bibnamefont
  {{Tirry}}},\ and\ \bibinfo {author} {\bibfnamefont {M.}~\bibnamefont
  {{Goossens}}},\ }\bibfield  {title} {\enquote {\bibinfo {title} {{Modified
  Kelvin-Helmholtz instabilities and resonant flow instabilities in a
  one-dimensional coronal plume model: Results for plasma
  {\ensuremath{\beta}}=0}},}\ }\href {https://doi.org/10.1086/308430}
  {\bibfield  {journal} {\bibinfo  {journal} {Astrophys. J.}\ }\textbf
  {\bibinfo {volume} {531}},\ \bibinfo {pages} {561--570} (\bibinfo {year}
  {2000})}\BibitemShut {NoStop}%
\bibitem [{\citenamefont {{Andries}}\ and\ \citenamefont
  {{Goossens}}(2001{\natexlab{a}})}]{49Andries2001a}%
  \BibitemOpen
  \bibfield  {author} {\bibinfo {author} {\bibfnamefont {J.}~\bibnamefont
  {{Andries}}}\ and\ \bibinfo {author} {\bibfnamefont {M.}~\bibnamefont
  {{Goossens}}},\ }\bibfield  {title} {\enquote {\bibinfo {title}
  {Kelvin-{H}elmholtz instabilities and resonant flow instabilities for a
  coronal plume model with plasma pressure},}\ }\href
  {https://doi.org/10.1051/0004-6361:20010050} {\bibfield  {journal} {\bibinfo
  {journal} {Astron. Astrophys.}\ }\textbf {\bibinfo {volume} {368}},\ \bibinfo
  {pages} {1083--1094} (\bibinfo {year} {2001}{\natexlab{a}})}\BibitemShut
  {NoStop}%
\bibitem [{\citenamefont {{Andries}}\ and\ \citenamefont
  {{Goossens}}(2001{\natexlab{b}})}]{50Andries2001b}%
  \BibitemOpen
  \bibfield  {author} {\bibinfo {author} {\bibfnamefont {J.}~\bibnamefont
  {{Andries}}}\ and\ \bibinfo {author} {\bibfnamefont {M.}~\bibnamefont
  {{Goossens}}},\ }\bibfield  {title} {\enquote {\bibinfo {title} {The
  influence of resonant {MHD} wave coupling in the boundary layer on the
  reflection and transmission process},}\ }\href
  {https://doi.org/10.1051/0004-6361:20010854} {\bibfield  {journal} {\bibinfo
  {journal} {Astron. Astrophys.}\ }\textbf {\bibinfo {volume} {375}},\ \bibinfo
  {pages} {1100--1110} (\bibinfo {year} {2001}{\natexlab{b}})}\BibitemShut
  {NoStop}%
\bibitem [{\citenamefont {Turkakin}, \citenamefont {Rankin},\ and\
  \citenamefont {Mann}(2013)}]{47Turkakin2013}%
  \BibitemOpen
  \bibfield  {author} {\bibinfo {author} {\bibfnamefont {H.}~\bibnamefont
  {Turkakin}}, \bibinfo {author} {\bibfnamefont {R.}~\bibnamefont {Rankin}},\
  and\ \bibinfo {author} {\bibfnamefont {I.~R.}\ \bibnamefont {Mann}},\
  }\bibfield  {title} {\enquote {\bibinfo {title} {Primary and secondary
  compressible {K}elvin-{H}elmholtz surface wave instabilities on the {E}arth's
  magnetopause},}\ }\href {https://doi.org/https://doi.org/10.1002/jgra.50394}
  {\bibfield  {journal} {\bibinfo  {journal} {J. Geophys. Res. Space Phys.}\
  }\textbf {\bibinfo {volume} {118}},\ \bibinfo {pages} {4161--4175} (\bibinfo
  {year} {2013})}\BibitemShut {NoStop}%
\bibitem [{\citenamefont {{Delamere}}\ \emph {et~al.}(2021)\citenamefont
  {{Delamere}}, \citenamefont {{Ng}}, \citenamefont {{Damiano}}, \citenamefont
  {{Neupane}}, \citenamefont {{Johnson}}, \citenamefont {{Burkholder}},
  \citenamefont {{Ma}},\ and\ \citenamefont {{Nykyri}}}]{delamere}%
  \BibitemOpen
  \bibfield  {author} {\bibinfo {author} {\bibfnamefont {P.~A.}\ \bibnamefont
  {{Delamere}}}, \bibinfo {author} {\bibfnamefont {C.~S.}\ \bibnamefont
  {{Ng}}}, \bibinfo {author} {\bibfnamefont {P.~A.}\ \bibnamefont {{Damiano}}},
  \bibinfo {author} {\bibfnamefont {B.~R.}\ \bibnamefont {{Neupane}}}, \bibinfo
  {author} {\bibfnamefont {J.~R.}\ \bibnamefont {{Johnson}}}, \bibinfo {author}
  {\bibfnamefont {B.}~\bibnamefont {{Burkholder}}}, \bibinfo {author}
  {\bibfnamefont {X.}~\bibnamefont {{Ma}}},\ and\ \bibinfo {author}
  {\bibfnamefont {K.}~\bibnamefont {{Nykyri}}},\ }\bibfield  {title} {\enquote
  {\bibinfo {title} {Kelvin–{H}elmholtz-related turbulent heating at
  {S}aturn's magnetopause boundary},}\ }\href
  {https://doi.org/https://doi.org/10.1029/2020JA028479} {\bibfield  {journal}
  {\bibinfo  {journal} {J. Geophys. Res. Space Phys.}\ }\textbf {\bibinfo
  {volume} {126}},\ \bibinfo {pages} {e2020JA028479} (\bibinfo {year}
  {2021})}\BibitemShut {NoStop}%
\bibitem [{\citenamefont {{Kim}}, \citenamefont {{Johnson}},\ and\
  \citenamefont {{Nykyri}}(2022)}]{45Kim2022}%
  \BibitemOpen
  \bibfield  {author} {\bibinfo {author} {\bibfnamefont {E.-H.}\ \bibnamefont
  {{Kim}}}, \bibinfo {author} {\bibfnamefont {J.~R.}\ \bibnamefont
  {{Johnson}}},\ and\ \bibinfo {author} {\bibfnamefont {K.}~\bibnamefont
  {{Nykyri}}},\ }\bibfield  {title} {\enquote {\bibinfo {title} {{Coupling
  between Alfvén wave and Kelvin-Helmholtz waves in the low latitude boundary
  layer}},}\ }\href {https://doi.org/10.3389/fspas.2021.785413} {\bibfield
  {journal} {\bibinfo  {journal} {Front. Astron. Space Sci.}\ }\textbf
  {\bibinfo {volume} {8}},\ \bibinfo {eid} {228} (\bibinfo {year}
  {2022})}\BibitemShut {NoStop}%
\bibitem [{\citenamefont {Bellman}\ and\ \citenamefont
  {Wing}(1976)}]{31Bellman1976}%
  \BibitemOpen
  \bibfield  {author} {\bibinfo {author} {\bibfnamefont {R.}~\bibnamefont
  {Bellman}}\ and\ \bibinfo {author} {\bibfnamefont {G.~M.}\ \bibnamefont
  {Wing}},\ }\href@noop {} {\emph {\bibinfo {title} {An Introduction to
  Invariant Imbedding}}}\ (\bibinfo  {publisher} {Wiley},\ \bibinfo {address}
  {New York},\ \bibinfo {year} {1976})\BibitemShut {NoStop}%
\bibitem [{\citenamefont {Klyatskin}(2005)}]{32Klyatskin2005}%
  \BibitemOpen
  \bibfield  {author} {\bibinfo {author} {\bibfnamefont {V.~I.}\ \bibnamefont
  {Klyatskin}},\ }\href@noop {} {\emph {\bibinfo {title} {Stochastic Equations
  through the Eye of the Physicist}}}\ (\bibinfo  {publisher} {Elsevier},\
  \bibinfo {address} {Amsterdam},\ \bibinfo {year} {2005})\BibitemShut
  {NoStop}%
\bibitem [{\citenamefont {Golberg}(1975)}]{33GOLBERG19751}%
  \BibitemOpen
  \bibfield  {author} {\bibinfo {author} {\bibfnamefont {M.~A.}\ \bibnamefont
  {Golberg}},\ }\bibfield  {title} {\enquote {\bibinfo {title} {Invariant
  imbedding and riccati transformations},}\ }\href
  {https://doi.org/https://doi.org/10.1016/0096-3003(75)90027-2} {\bibfield
  {journal} {\bibinfo  {journal} {Appl. Math. Comput.}\ }\textbf {\bibinfo
  {volume} {1}},\ \bibinfo {pages} {1--24} (\bibinfo {year}
  {1975})}\BibitemShut {NoStop}%
\bibitem [{\citenamefont {{Kim}}\ and\ \citenamefont
  {{Kim}}(2016{\natexlab{a}})}]{34Kim2016}%
  \BibitemOpen
  \bibfield  {author} {\bibinfo {author} {\bibfnamefont {S.}~\bibnamefont
  {{Kim}}}\ and\ \bibinfo {author} {\bibfnamefont {K.}~\bibnamefont {{Kim}}},\
  }\bibfield  {title} {\enquote {\bibinfo {title} {{Invariant imbedding theory
  of wave propagation in arbitrarily inhomogeneous stratified bi-isotropic
  media}},}\ }\href {https://doi.org/10.1088/2040-8978/18/6/065605} {\bibfield
  {journal} {\bibinfo  {journal} {J. Opt.}\ }\textbf {\bibinfo {volume} {18}},\
  \bibinfo {eid} {065605} (\bibinfo {year} {2016}{\natexlab{a}})}\BibitemShut
  {NoStop}%
\bibitem [{\citenamefont {{Cairns}}(1979)}]{27Cairns1979}%
  \BibitemOpen
  \bibfield  {author} {\bibinfo {author} {\bibfnamefont {R.~A.}\ \bibnamefont
  {{Cairns}}},\ }\bibfield  {title} {\enquote {\bibinfo {title} {{The role of
  negative energy waves in some instabilities of parallel flows}},}\ }\href
  {https://doi.org/10.1017/S0022112079000495} {\bibfield  {journal} {\bibinfo
  {journal} {J. Fluid Mech.}\ }\textbf {\bibinfo {volume} {92}},\ \bibinfo
  {pages} {1--14} (\bibinfo {year} {1979})}\BibitemShut {NoStop}%
\bibitem [{\citenamefont {{Ostrovski{\u{i}}}}, \citenamefont {{Rybak}},\ and\
  \citenamefont {{Sh.~Tsimring}}(1986)}]{37Ostrovskii1986}%
  \BibitemOpen
  \bibfield  {author} {\bibinfo {author} {\bibfnamefont {L.~A.}\ \bibnamefont
  {{Ostrovski{\u{i}}}}}, \bibinfo {author} {\bibfnamefont {S.~A.}\ \bibnamefont
  {{Rybak}}},\ and\ \bibinfo {author} {\bibfnamefont {L.}~\bibnamefont
  {{Sh.~Tsimring}}},\ }\bibfield  {title} {\enquote {\bibinfo {title}
  {{Negative energy waves in hydrodynamics}},}\ }\href
  {https://doi.org/10.1070/PU1986v029n11ABEH003538} {\bibfield  {journal}
  {\bibinfo  {journal} {Sov. Phys. Usp.}\ }\textbf {\bibinfo {volume} {29}},\
  \bibinfo {pages} {1040--1052} (\bibinfo {year} {1986})}\BibitemShut {NoStop}%
\bibitem [{\citenamefont {{Joarder}}, \citenamefont {{Nakariakov}},\ and\
  \citenamefont {{Roberts}}(1997)}]{38Joarder1997}%
  \BibitemOpen
  \bibfield  {author} {\bibinfo {author} {\bibfnamefont {P.~S.}\ \bibnamefont
  {{Joarder}}}, \bibinfo {author} {\bibfnamefont {V.~M.}\ \bibnamefont
  {{Nakariakov}}},\ and\ \bibinfo {author} {\bibfnamefont {B.}~\bibnamefont
  {{Roberts}}},\ }\bibfield  {title} {\enquote {\bibinfo {title} {{A
  manifestation of negative energy waves in the solar atmosphere}},}\ }\href
  {https://doi.org/10.1023/A:1004977928351} {\bibfield  {journal} {\bibinfo
  {journal} {Sol. Phys.}\ }\textbf {\bibinfo {volume} {176}},\ \bibinfo {pages}
  {285--297} (\bibinfo {year} {1997})}\BibitemShut {NoStop}%
\bibitem [{\citenamefont {{Kim}}\ and\ \citenamefont
  {{Kim}}(2016{\natexlab{b}})}]{35Kim2016}%
  \BibitemOpen
  \bibfield  {author} {\bibinfo {author} {\bibfnamefont {S.}~\bibnamefont
  {{Kim}}}\ and\ \bibinfo {author} {\bibfnamefont {K.}~\bibnamefont {{Kim}}},\
  }\bibfield  {title} {\enquote {\bibinfo {title} {{Resonant absorption and
  amplification of circularly-polarized waves in inhomogeneous chiral
  media}},}\ }\href {https://doi.org/10.1364/OE.24.001794} {\bibfield
  {journal} {\bibinfo  {journal} {Opt. Express}\ }\textbf {\bibinfo {volume}
  {24}},\ \bibinfo {pages} {1794--1803} (\bibinfo {year}
  {2016}{\natexlab{b}})}\BibitemShut {NoStop}%
\bibitem [{\citenamefont {{\v{C}ade\v{z}}}\ \emph {et~al.}(1997)\citenamefont
  {{\v{C}ade\v{z}}}, \citenamefont {{Cs\'{i}k}}, \citenamefont
  {{Erd\'{e}lyi}},\ and\ \citenamefont {{Goossens}}}]{36Cadez1997}%
  \BibitemOpen
  \bibfield  {author} {\bibinfo {author} {\bibfnamefont {V.~M.}\ \bibnamefont
  {{\v{C}ade\v{z}}}}, \bibinfo {author} {\bibfnamefont {{\'{A}}.}~\bibnamefont
  {{Cs\'{i}k}}}, \bibinfo {author} {\bibfnamefont {R.}~\bibnamefont
  {{Erd\'{e}lyi}}},\ and\ \bibinfo {author} {\bibfnamefont {M.}~\bibnamefont
  {{Goossens}}},\ }\bibfield  {title} {\enquote {\bibinfo {title} {{Absorption
  of magnetosonic waves in presence of resonant slow waves in the solar
  atmosphere}},}\ }\href@noop {} {\bibfield  {journal} {\bibinfo  {journal}
  {Astron. Astrophys.}\ }\textbf {\bibinfo {volume} {326}},\ \bibinfo {pages}
  {1241--1251} (\bibinfo {year} {1997})}\BibitemShut {NoStop}%
\bibitem [{\citenamefont {{Rivero}}\ and\ \citenamefont
  {{Ge}}(2019)}]{39Rivero2019}%
  \BibitemOpen
  \bibfield  {author} {\bibinfo {author} {\bibfnamefont {J.~D.~H.}\
  \bibnamefont {{Rivero}}}\ and\ \bibinfo {author} {\bibfnamefont
  {L.}~\bibnamefont {{Ge}}},\ }\bibfield  {title} {\enquote {\bibinfo {title}
  {{Time-reversal-invariant scaling of light propagation in one-dimensional
  non-Hermitian systems}},}\ }\href
  {https://doi.org/10.1103/PhysRevA.100.023819} {\bibfield  {journal} {\bibinfo
   {journal} {Phys. Rev. E}\ }\textbf {\bibinfo {volume} {100}},\ \bibinfo
  {eid} {023819} (\bibinfo {year} {2019})}\BibitemShut {NoStop}%
\bibitem [{\citenamefont {{Haaland}}\ \emph {et~al.}(2019)\citenamefont
  {{Haaland}}, \citenamefont {{Runov}}, \citenamefont {{Artemyev}},\ and\
  \citenamefont {{Angelopoulos}}}]{40Haaland2019}%
  \BibitemOpen
  \bibfield  {author} {\bibinfo {author} {\bibfnamefont {S.}~\bibnamefont
  {{Haaland}}}, \bibinfo {author} {\bibfnamefont {A.}~\bibnamefont {{Runov}}},
  \bibinfo {author} {\bibfnamefont {A.}~\bibnamefont {{Artemyev}}},\ and\
  \bibinfo {author} {\bibfnamefont {V.}~\bibnamefont {{Angelopoulos}}},\
  }\bibfield  {title} {\enquote {\bibinfo {title} {{Characteristics of the
  flank magnetopause: THEMIS observations}},}\ }\href
  {https://doi.org/10.1029/2019JA026459} {\bibfield  {journal} {\bibinfo
  {journal} {J. Geophys. Res. Space Phys.}\ }\textbf {\bibinfo {volume}
  {124}},\ \bibinfo {pages} {3421--3435} (\bibinfo {year} {2019})}\BibitemShut
  {NoStop}%
\bibitem [{\citenamefont {{Wright}}\ \emph {et~al.}(2001)\citenamefont
  {{Wright}}, \citenamefont {{Yeoman}}, \citenamefont {{Rae}}, \citenamefont
  {{Storey}}, \citenamefont {{Stockton-Chalk}}, \citenamefont {{Roeder}},\ and\
  \citenamefont {{Trattner}}}]{41Wright2001}%
  \BibitemOpen
  \bibfield  {author} {\bibinfo {author} {\bibfnamefont {D.~M.}\ \bibnamefont
  {{Wright}}}, \bibinfo {author} {\bibfnamefont {T.~K.}\ \bibnamefont
  {{Yeoman}}}, \bibinfo {author} {\bibfnamefont {I.~J.}\ \bibnamefont {{Rae}}},
  \bibinfo {author} {\bibfnamefont {J.}~\bibnamefont {{Storey}}}, \bibinfo
  {author} {\bibfnamefont {A.~B.}\ \bibnamefont {{Stockton-Chalk}}}, \bibinfo
  {author} {\bibfnamefont {J.~L.}\ \bibnamefont {{Roeder}}},\ and\ \bibinfo
  {author} {\bibfnamefont {K.~J.}\ \bibnamefont {{Trattner}}},\ }\bibfield
  {title} {\enquote {\bibinfo {title} {{Ground-based and polar spacecraft
  observations of a giant (Pg) pulsation and its associated source
  mechanism}},}\ }\href {https://doi.org/10.1029/2001JA900022} {\bibfield
  {journal} {\bibinfo  {journal} {J. Geophys. Res.}\ }\textbf {\bibinfo
  {volume} {106}},\ \bibinfo {pages} {10837--10852} (\bibinfo {year}
  {2001})}\BibitemShut {NoStop}%
\bibitem [{\citenamefont {{Goossens}}\ \emph {et~al.}(2020)\citenamefont
  {{Goossens}}, \citenamefont {{Arregui}}, \citenamefont {{Soler}},\ and\
  \citenamefont {{Van Doorsselaere}}}]{42Goossens2020}%
  \BibitemOpen
  \bibfield  {author} {\bibinfo {author} {\bibfnamefont {M.}~\bibnamefont
  {{Goossens}}}, \bibinfo {author} {\bibfnamefont {I.}~\bibnamefont
  {{Arregui}}}, \bibinfo {author} {\bibfnamefont {R.}~\bibnamefont {{Soler}}},\
  and\ \bibinfo {author} {\bibfnamefont {T.}~\bibnamefont {{Van
  Doorsselaere}}},\ }\bibfield  {title} {\enquote {\bibinfo {title} {{Resonant
  absorption: Transformation of compressive motions into vortical motions}},}\
  }\href {https://doi.org/10.1051/0004-6361/202038394} {\bibfield  {journal}
  {\bibinfo  {journal} {Astron. Astrophys.}\ }\textbf {\bibinfo {volume}
  {641}},\ \bibinfo {eid} {A106} (\bibinfo {year} {2020})}\BibitemShut
  {NoStop}%
\bibitem [{\citenamefont {{Goossens}}\ \emph {et~al.}(2021)\citenamefont
  {{Goossens}}, \citenamefont {{Chen}}, \citenamefont {{Geeraerts}},
  \citenamefont {{Li}},\ and\ \citenamefont {{Van
  Doorsselaere}}}]{43Goossens2021}%
  \BibitemOpen
  \bibfield  {author} {\bibinfo {author} {\bibfnamefont {M.}~\bibnamefont
  {{Goossens}}}, \bibinfo {author} {\bibfnamefont {S.~X.}\ \bibnamefont
  {{Chen}}}, \bibinfo {author} {\bibfnamefont {M.}~\bibnamefont {{Geeraerts}}},
  \bibinfo {author} {\bibfnamefont {B.}~\bibnamefont {{Li}}},\ and\ \bibinfo
  {author} {\bibfnamefont {T.}~\bibnamefont {{Van Doorsselaere}}},\ }\bibfield
  {title} {\enquote {\bibinfo {title} {{Mixed properties of magnetohydrodynamic
  waves undergoing resonant absorption in the cusp continuum}},}\ }\href
  {https://doi.org/10.1051/0004-6361/202039780} {\bibfield  {journal} {\bibinfo
   {journal} {Astron. Astrophys.}\ }\textbf {\bibinfo {volume} {646}},\
  \bibinfo {eid} {A86} (\bibinfo {year} {2021})}\BibitemShut {NoStop}%
\bibitem [{\citenamefont {{Provornikova}}, \citenamefont {{Laming}},\ and\
  \citenamefont {{Lukin}}(2018)}]{44Provornikova2018}%
  \BibitemOpen
  \bibfield  {author} {\bibinfo {author} {\bibfnamefont {E.}~\bibnamefont
  {{Provornikova}}}, \bibinfo {author} {\bibfnamefont {J.~M.}\ \bibnamefont
  {{Laming}}},\ and\ \bibinfo {author} {\bibfnamefont {V.~S.}\ \bibnamefont
  {{Lukin}}},\ }\bibfield  {title} {\enquote {\bibinfo {title} {{Reflection of
  fast magnetosonic waves near a magnetic reconnection region}},}\ }\href
  {https://doi.org/10.3847/1538-4357/aac1c1} {\bibfield  {journal} {\bibinfo
  {journal} {Astrophys. J.}\ }\textbf {\bibinfo {volume} {860}},\ \bibinfo
  {eid} {138} (\bibinfo {year} {2018})}\BibitemShut {NoStop}%
\end{thebibliography}%

\end{document}